\PassOptionsToPackage{prologue,dvipsnames,svgnames}{xcolor}
\documentclass[screen,acmsmall,nonacm,natbib=false]{acmart}
\usepackage[
  datamodel=acmdatamodel,
  style=acmnumeric,
  ]{biblatex}

\usepackage{stmaryrd}
\usepackage{mathtools}
\usepackage{listings}
\usepackage{tikz}
\usepackage{tikz-cd}
\tikzset{>={Straight Barb[scale=0.8]}}
\tikzcdset{
arrow style=tikz
}

\usepackage{csquotes}

\usepackage{macros}
\usepackage[capitalize]{cleveref}
\setcopyright{cc}
\setcctype{by}
\acmDOI{10.1145/3776664}
\acmYear{2026}
\acmJournal{PACMPL}
\acmVolume{10}
\acmNumber{POPL}
\acmArticle{22}
\acmMonth{1}
\received{2025-07-10}
\received[accepted]{2025-11-06}

\title{AdapTT: Functoriality for Dependent Type Casts}

\author{Arthur Adjedj}
\orcid{0009-0007-3683-6239}
\affiliation{%
  \institution{ENS Paris-Saclay - Université Paris-Saclay}
  \city{Gif-sur-Yvette}
  \country{France}
}
\email{arthur.adjedj@ens-paris-saclay.fr}

\author{Meven Lennon-Bertrand}
\orcid{0000-0002-7079-8826}
\affiliation{%
  \institution{Université Paris Cité, INRIA, CNRS, IRIF}
  \city{Paris}
  \country{France}
}
\email{meven.bertrand@inria.fr}

\author{Thibaut Benjamin}
\orcid{0000-0002-9481-1896}
\affiliation{%
  \institution{Université Paris-Saclay, CNRS, ENS Paris-Saclay, LMF}
  \city{Gif-sur-Yvette}
  \country{France}
}
\email{thibaut.benjamin@universite-paris-saclay.fr}

\author{Kenji Maillard}
\orcid{0000-0001-5554-3203}
\affiliation{%
  \institution{Nantes Université, École Centrale Nantes, CNRS, INRIA, LS2N, UMR 6004}
  \city{Rennes}
  \country{France}
}
\email{kenji.maillard@inria.fr}

\begin{abstract}
  The ability to \emph{cast} values between related types is a leitmotiv of many flavors of dependent type theory, such as observational type theories,
  subtyping, or cast calculi for gradual typing. These casts all exhibit a common structural behavior that boils down to the pervasive \emph{functoriality} of type formers. We propose and extensively study a type theory, called \AdapTT{}, which makes systematic and precise this idea of functorial type formers, with respect to an abstract notion of \emph{adapters} relating types.
  Leveraging descriptions for functorial inductive types in \AdapTT, we derive structural laws for type casts on general inductive type formers.
\end{abstract}

\ccsdesc[500]{Theory of computation~Type theory}
\ccsdesc[300]{Theory of computation~Categorical semantics}
\ccsdesc[300]{Theory of computation~Type structures}

\keywords{Dependent types, Natural models, Inductive types}

\newcommand\castTTobs{\operatorname{cast}}

\begin{document}

\maketitle

\section{Introduction}
\label{sec:intro}

\emph{Type casting} is a fundamental operation in typed programming languages,
turning a value of one type into a value of another to mediate between
superficially different types.
In the setting of proof assistants and
dependent type systems, where types are extremely rich and precise, the need for
such mediation is particularly dire.
% where annotations can be used to direct elaboration through unification or provide witnesses pertinent for deciding typechecking.
%
Proof assistants thus offer many type casting mechanisms: coercions~\cite{Saibi1997} are heavily employed in \Rocq~\cite{rocq} and \Lean~\cite{deMouraUllrich2021}, in particular to deal with hierarchies of structures~\cite{Garillot2009,CohenST20}; predicate subtyping~\cite{Shankar2000} is a central tool for verifying properties in PVS~\cite{PVS}, the Russel sub-language of \Rocq~\cite{Sozeau2007}, and \Fstar~\cite{SwamyHKRDFBFSKZ16}; and cumulativity~\cite{Luo1989}, present by default in \Rocq and optionally in \Agda\cite{agdaCumulativity}, greatly simplifies the management of universe levels.
A particularly attractive feature of type casting is to be \emph{structural},
meaning that type casts can be systematically lifted through type formers.
Structural type casts  can be applied deeply in types, rather than be confined
to the top level, which makes them much more powerful and modular.
% since they propagate in any context, they can be substituted arbitrarily while retaining their expected behavior.
%
And indeed, failures of typecast to properly lift through type formers
manifest as painful errors that can be hard to debug~\cite[p.35]{laurent:tel-05144931}
or expressivity limitations~\cite{fstar65}.
%
% \tb{I think this needs a bit more motivation, why is structural type casting good?}\km{better ?}
% of the theory.
%
Structural type casting is an important feature in recent work such as
Observational Type Theory (\TTobs)~\cite{Altenkirch2007, Pujet2022}, dependent type theories with coercive subtyping (\MLTTcoe)~\cite{Luo2008,Subtyping2024}, or cast calculi for gradual dependent types (\CastCIC)~\cite{LennonBertrand2022}.
\Cref{fig:casts-comparison} shows the structural type casts between
function types ($\Pi$) in these three systems.
Although the exact mechanisms differ, in all cases the cast of a
function $f$ proceeds by first casting its argument before passing it to $f$,
and finally casting the result back.
Such striking similarities can be observed with structural casts for other type formers too.
In this paper, we argue that this common core is no coincidence, but emerges
from a deeper structure:
\begin{center}
  \textbf{Structural type casts arise from functorial type formers.}
\end{center}

\begin{figure}
\begin{small}
\begin{mathpar}
\inferdef
  {
    \typing{\Gamma}{e}[\P x : A.B = \P x : A'.B'] \\
    \typing{\Gamma}{f}[\P x : A.B] \\\\
    \typing{\Gamma}{a'}[A'] \\
    a \coloneq \castTTobs{}(A',A,\operatorname{fst}(e),a')
  }
  {\conv{\Gamma}{
      \castTTobs{}(\P x : A.B,\P x : A'.B',e,f)~a'
    }{
      \castTTobs{}(\subs{B}{a},\subs{B'}{a'},\operatorname{snd}(e),f\,a)
    }[
      \subs{B}{a}
    ]
  } \and
\inferdef
  {
    \Gamma \vdash A' \sub A \\
    \Gamma, x : A' \vdash \subs{B}{\coe[A'][A]{x}} \sub B' \\\\
    \typing{\Gamma}{f}[\P x : A.B] \\
    \typing{\Gamma}{a'}[A'] \\
    a \coloneq \coe[A'][A]{a'}
  }
  {\conv{\Gamma}{(\coe[\P x : A.B][\P x : A'.B']{f})~a'}{
  \coe[\subs{B}{a}][\subs{B'}{a'}]{(f\,a)}
  }[\subs{B'}{a'}]}
\and
\inferdef
  {
    \Gamma \vdash A \\ \Gamma \vdash A' \\
    \Gamma, x : A \vdash B \\ \Gamma, x : A' \vdash B' \\\\
    \typing{\Gamma}{f}[\P x : A.B] \\
    \typing{\Gamma}{a'}[A'] \\
    a \coloneq \langle A \Leftarrow A'\rangle\, a'
  }
  {\conv{\Gamma}{
    \left(\langle \P x : A'.B' \Leftarrow \P x : A.B\rangle\,f\right)~a'
  }{
    \langle \subs{B'}{a'} \Leftarrow \subs{B}{a} \rangle \left( f~a \right)
  }[\subs{B'}{a'}]}
\end{mathpar}
\end{small}
\caption[Casts between \(\P\) types in observational equality,
subtyping and cast calculi for gradual typing.]
{Casts between \(\P\) types in observational equality,
subtyping and cast calculi for gradual typing.\protect\footnotemark}
\label{fig:casts-comparison}
\Description{We see that the three rules are very similar.}
\end{figure}
\footnotetext{
  To highlight their similarity, we express these rules as definitional equalities,
  which are inter-derivable with the original rules from
  \textcite{Pujet2022a,Subtyping2024,LennonBertrand2022}
  in presence of β and η for functions.
}

Indeed, the three examples of~\cref{fig:casts-comparison} all rely on a functor structure on the $\P$ type former, a structure independently described in any model of
\MLTT by \textcite[Lemma 4.8]{Castellan2017}.
In each case, this functorial structure applies to different %but related
notions of morphisms between types: observational equalities in \TTobs, subtyping coercions in \MLTTcoe and arbitrary cast in \CastCIC.
This approach extends far beyond \(\P\): \textcite{Subtyping2024} observe that many usual type formers ($\P, \Sig, \Id, \W, +, \List,\dots$) can be equipped with a functorial
action on functions that definitionally satisfies functoriality equations,
while keeping decidability of definitional equality and typing.
% and show that $\List$ can be made definitionally functorial in an equational extension of \MLTT dubbed \MLTTmap.

\paragraph{Challenges of structural type casting.} % of functorial types formers
Multiple difficulties arise if one tries to unify diverse instances of structural type casts.
First, type casting can apply to varied notions of morphism between types,
as shown in the previous examples, and a unifying framework ought to allow for this diversity.
Second, type formers can introduce contravariant type casts, as exemplified by the domain of function types, and mixing variance and type dependency quickly becomes a subtle matter.
Third, \textcite{Subtyping2024} observed a strong obstruction to derive the functoriality of general inductive types
encoded via the composition of atomic type formers (\(\Pi, \Sigma, \W, \ldots\)).
% the definitional functoriality properties needed to define a general theory of type casts is not compositional as it relies on properties of the conversion of the underlying type theory. \mlb{I don't understand this sentence}
%
The last two difficulties hamper the design of a general schema for inductive type formers
---as can be found in day-to-day proof assistants (\Rocq, \Lean, \Agda)--- that moreover supports structural type cast.
%
% \km{Add a small but intricate example where the domain category is not evident ?}

\paragraph{Introducing \AdapTT{}}
To attack these challenges, we take a step back and % take a clean-slate approach to
provide a solid and general framework to understand cast operations,
functoriality of type formers, and the relationship between the two. The
dependent type theory we develop, \AdapTT{}, comes
equipped with a primitive notion of type morphisms,
\emph{adapters}~\cite{McBride2021}, which capture the data necessary to construct a cast.
The action of adapters on terms ---type casting--- is functorial with respect to the categorical structure adapters give to types. As a design principle, any type former in \AdapTT{} should respect this structure as well, \ie be functorial.
We show that \AdapTT{} is able to encompass the above
examples of type theories with structural type casts.
We provide a corresponding notion of model, a variant of natural models~\cite{Awodey2018} dubbed \NMDO{}s, and exhibit multiple instances of this structure.
Moreover, we establish a correspondence with the works of \textcite{Coraglia2024a,Najmaei2025} that develop related semantic notions based on comprehension categories~\cite{Jacobs1993}.

\begin{comment}
To substantiate this claim, we must first clarify what are the categories between
which a type former is a functor. For this, we develop an extension of categories with
families (CwF), which we dub \NMDO{}s, where types in a context form a category.
Morphisms in this category, which we call \emph{adapters}~\cite{McBride2021}, capture the data
necessary to construct a cast. The operation mapping a type to its set of terms is
a functor on this category, which captures the coercion operation itself.
\textcite{Coraglia2024a,Najmaei2025} developed independently a similar
definition formulated in terms of comprehension categories rather than CwF, and
we explore the relationship between both approaches.
\AdapTT{} is the syntax corresponding to this semantic structure.
We extend the standard CwF of sets to a \NMDO, and observe that it faithfully embeds into a \NMDO on categories.
\mlb{Should mention the other instances (subtyping/\TTobs/GCIC), too?}
% Just as CwFs have a “standard model” in sets, our \NMDO{}s has
% a standard model in categories, which we describe.
\end{comment}

\paragraph{Functorial type formers with \AdapTTt}
While \AdapTT{} provides an adequate framework to state that a given type former is functorial
on a case-by-case basis, it does not provide the means to generically represent
the categories with respect to which a type former is functorial,
nor to derive that complex, composite types are structurally functorial.
%
% \km{The rhetoric of the following paragraph needs to be changed: the point
%   should be on how we can effectively define functorial type formers and how the
% 2-dimensional \NMDO can help us in this endeavour}
% \km{Mention that in \AdapTTt type formers will automagically be functorial}
% Using \NMDO, we can already formulate a very general definition of (functorial) type former.
% However, this definition is semantic, very general,
% and %in some sense too general. In particular, it is
% not really helpful for developing of a (syntactic) description of functorial types, in
% particular inductive types.
We thus extend \AdapTT{} with type variables which provides an internalization of type formers, including their action on adapters.
% This leads us to design a type theory \AdapTTt{} equipped with type variables
% which let us capture the source category of many type formers of interest, in a syntactic way.
Contexts in this extension, \AdapTTt{}, naturally exhibit a 2-categorical structure, which harmoniously interacts with our category of types and adapters.
% The \NMDO structure on categories
% naturally extends to account for this extra 2-categorical structure and type variables.
% \mlb{Mention the standard model in Cat?}

\AdapTTt{} gives us the tools we need to construct a theory of signatures in the style of Kaposi, Kovács et al.~\cite{KovacsPhD,Kaposi2020a,Kaposi2019a} to describe general inductive type formers.
An inductive type described with such a signature is by construction functorial with respect to its parameters, described as a context in \AdapTTt{}, entailing
a well-behaved notion of structural casts.
We illustrate the expressivity of our signatures by describing many standard examples,
and their functorial structure:
lists, sum types, W types, or the inductive identity type.

% Finally, we sketch how the examples of observational equality~\cite{Altenkirch2007,Pujet2022},
% subtyping~\cite{Luo1999,Subtyping2024} and cast calculi for gradual typing
% \cite{LennonBertrand2022} can be easily recovered as special instances of the general
% framework. As this would lead us too far, we
% do not attack the technical details of how to precisely integrate some
% of the most complex examples of subtyping, such as cumulativity or subset types,
% but believe these could be relatively easily integrated in our framework.

% \mlb{More limitations? Like, no normalization, no implementation/decidable type-checking?
%   Or should we mention that only at the end?}

\paragraph{Contributions} We bring the following to the study of type casts in dependent type theory:
% \mlb{TODO, can we try not to repeat the above?}\km{We should balance between the
% two, but I would rather make the motivation paragraphs more to the point,
% leaving the technical buzzword to this paragraph}
\begin{itemize}
\item In \cref{sec:cwf-cat}, we introduce \AdapTT{}, a type theory with type casts,
  and \NMDO, a categorical structure in which it interprets, which we relate to
  previous work~\cite{Coraglia2024a,Najmaei2025};
\item \cref{sec:type-vars} extends \AdapTT{} to a 2-dimensional type theory, \AdapTTt{}, giving the ability to internalize type formers using type variables, and
exhibiting important constraints on the allowed interactions of type variance and dependency;
\item We moreover show in \cref{sec:back-adaptt} that the $2$-category of ($\Cat$-valued) presheaves on such a \NMDO $\mathcal{C}$ naturally interprets \AdapTTt{}, giving rise to functorial type former on $\mathcal{C}$.
  %from any description of an inductive type in the signature.
\item Finally, \Cref{sec:functorial-types} describes a variance-aware theory of signatures,
deriving structural type casts for inductive types through their inherent functorial action on adapters.
% \item \km{Agda formalization of the theory of signatures ?}
\end{itemize}
% \Cref{sec:related-work} discusses related work and \cref{sec:future-work} future research directions. \mlb{meh}\km{Specialize}

We have type checked most%
\footnote{
We managed to check most rules in isolation, but dire performance issues
prevented us from checking them all together.
}
rules of the paper in \Agda \cite{Formalisation}.
The point of this formalisation was not to verify properties
of the system, rather, \Agda was used as a guide to ensure the
numerous rules were type-correct.
This only provides partial support for our work,
but we still believe \Agda's roles was valuable enough to be acknowledged.

This is the long version, with appendices, of an article \cite{POPLVersion} published
at POPL '26.

\section{A Category of Types}
\label{sec:cwf-cat}

This section introduces \AdapTT{}, a dependent type theory with extra structure
for type casts.
\Cref{sec:adaptt-syntax} presents its syntax
while \cref{sec:adapters} defines its categorical models, with examples.
\Cref{sec:other-models} relates \AdapTT{} to
an existing notion of model for (relevant) subtyping: \GCwF~\cite{Coraglia2024a}.

\subsection{AdapTT, Syntactically}
\label{sec:adaptt-syntax}

\begin{figure}
\begin{small}

  \[
    \begin{array}{clcl}
      \ctxty{\Box} & \text{well-formed context} &
      \typing{\Gamma}{\Box}[\Delta] & \text{well-formed substitution in } \ctxty{\Gamma}, \ctxty{\Delta} \\
      \typing{\Gamma}{\Box} & \text{well-formed type in } \ctxty{\Gamma} &
      \typing{\Gamma}{\Box}[\adso{A}{B}] & \text{well-formed adapter in } \typing{\Gamma}{A}, \typing{\Gamma}{B} \\
      \typing{\Gamma}{\Box}[A] & \text{well-formed term in }\typing{\Gamma}{A} && \text{\textcolor{gray}{definitional equality judgments omitted}}
    \end{array}
  \]
\rule{0.9\textwidth}{.5pt}
\begin{mathpar}
  \inferdeft[CtxEmp]{ }{\ctxty{\emp}}
  \label{rule:adaptt-ctx-emp}
  \and
  \inferdeft[CtxExt]
  {\typing{\Gamma}{A}}
  {\ctxty{\Gamma \ext A}}
  \label{rule:adaptt-ctx-ext}
  \and
  \inferdeft[SubTy]
  {\typing{\Gamma}{\sigma}[\Delta]\\
    \typing{\Delta}{A}}
  {\typing{\Gamma}{\subs{A}{\sigma}}}
  \label{rule:adaptt-sub-ty}
  \and
  \inferdeft[SubTm]
  {\typing{\Gamma}{\sigma}[\Delta]\\
    \typing{\Delta}{t}[A]}
  {\typing{\Gamma}{\subs{t}{\sigma}}[\subs{A}{\sigma}]}
  \label{rule:adaptt-sub-tm}
  \and
  \inferdeft[SubEmp]
  {\ctxty{\Gamma}}
  {\typing{\Gamma}{\emp}[\emp]}
  \label{rule:adaptt-sub-emp}
  \and
  \inferdeft[SubExt]
  {\typing{\Gamma}{\sigma}[\Delta]\\
  \typing{\Gamma}{t}[\subs{A}{\sigma}]}
  {\typing{\Gamma}{\sigma \ext t}[\Delta \ext A]}
  \label{rule:adaptt-sub-ext}
  \and
  \inferdeft[Wk]
  {\typing{\Gamma}{A}}
  {\typing{\Gamma \ext A}{\wk}[\Gamma]}
  \label{rule:adaptt-wk}
  \and
  \inferdeft[VarZero]
  {\typing{\Gamma}{A}}
  {\typing{\Gamma\ext A}{\tmvz}[\subs{A}{\wk}]}
  \label{rule:adaptt-var-zero}
  \and
  % \inferdeft[SubId]
  % {\ctxty{\Gamma}}
  % {\typing{\Gamma}{\id}[\Gamma]}
  % \label{rule:adaptt-sub-id}
  % \and
  % \inferdeft[SubComp]
  % {\typing{\Gamma}{\sigma}[\Delta]\\
  % \typing{\Delta}{\tau}[\Xi]}
  % {\typing{\Gamma}{\tau \circ \sigma}[\Xi]}
  % \label{rule:adaptt-sub-comp}
  % \and
  % \inferdeft[SubEmpEq]
  % {\typing{\Gamma}{\sigma}[\emp]}
  % {\conv{\Gamma}{\sigma}{\emp}[\emp]}
  % \label{rule:adaptt-sub-emp-eq}
  % \and
  \inferdeft[SubExtWk]
  {\typing{\Gamma}{\sigma}[\Delta]\\
  \typing{\Gamma}{t}[\subs{A}{\sigma}]}
  {\conv{\Gamma}{\wk \circ (\sigma \ext t)}{\sigma}[\Delta]}
  \label{rule:adaptt-sub-ext-wk}
  \and
  \inferdeft[SubExtVar]
  {\typing{\Gamma}{\sigma}[\Delta]\\
    \typing{\Gamma}{t}[\subs{A}{\sigma}]}
  {\conv{\Gamma}{\subs{\tmvz}{\sigma \ext t}}{t}[\subs{A}{\sigma}]}
  \label{rule:adaptt-sub-ext-var} \and
  % \vspace*{-.5em} \\
  % \rule{0.9\textwidth}{.5pt}
  % \vspace*{-.5em} \\
  \inferdeft[Id]
  {\typing{\Gamma}{A}}
  {\typing{\Gamma}{\id}[\adso{A}{A}]}
  \label{rule:adaptt-id}
  \and
  \inferdeft[Comp]
  {\typing{\Gamma}{f}[\adso{A}{B}]\\
    \typing{\Gamma}{g}[\adso{B}{C}]}
  {\typing{\Gamma}{g \circ f}[\adso{A}{C}]}
  \label{rule:adaptt-comp}
  \and
  \inferdeft[SubAd]
  {\typing{\Delta}{f}[\adso{A}{B}]\\
    \typing{\Gamma}{\sigma}[\Delta]}
  {\typing{\Gamma}{\subs{f}{\sigma}}[\adso{\subs{A}{\sigma}}{\subs{B}{\sigma}}]}
  \label{rule:adaptt-sub-ad}
  \and
  \inferdeft[Adapt]
  {\typing{\Gamma}{f}[\adso{A}{B}]\\
  \typing{\Gamma}{a}[A]}
  {\typing{\Gamma}{\ad{a}{f}}[B]}
  \label{rule:adaptt-adapt}
  \and
  \inferdeft[Assoc]
  {\typing{\Gamma}{f}[\adso{A}{B}] \\
    \typing{\Gamma}{g}[\adso{B}{C}] \\
    \typing{\Gamma}{h}[\adso{C}{D}]}
  {\conv{\Gamma}{h \circ (g \circ f)}{(h \circ g) \circ f}[\adso{A}{D}]}
  \label{rule:adaptt-comp-assoc}
  \and
  \inferdeft[IdLeft]
  {\typing{\Gamma}{f}[\adso{A}{B}]}
  {\conv{\Gamma}{\id \circ f}{f}[\adso{A}{B}]}
  \label{rule:adaptt-id-left}
  \and
  \inferdeft[IdRight]
  {\typing{\Gamma}{f}[\adso{A}{B}]}
  {\conv{\Gamma}{f \circ \id}{f}[\adso{A}{B}]}
  \label{rule:adaptt-id-right}
  \and
  \inferdeft[AdaptId]
  {\typing{\Gamma}{a}[A]}
  {\conv{\Gamma}{\ad{a}{\id}}{a}[A]}
  \label{rule:adaptt-adapt-id}
  \and
  \inferdeft[AdaptComp]
  {\typing{\Gamma}{a}[A]\\
    \typing{\Gamma}{f}[\adso{A}{B}]\\
    \typing{\Gamma}{g}[\adso{B}{C}]}
  {\conv{\Gamma}{\ad{a}{g \circ f}}{\ad{\ad{a}{f}}{g}}[C]}
  \label{rule:adaptt-adapt-comp}
  \and
  % \inferdeft[SubExtEq]
  % {\typing{\Gamma}{\sigma}[\Delta \ext A]}
  % {\conv{\Gamma}{\sigma}{(\wk \circ \sigma) \ext \subs{\tmvz}{\sigma}}[\Delta \ext A]}
  % \and
  % \inferdeft[SubTyId]
  % {\typing{\Gamma}{A}}
  % {\conv{\Gamma}{\subs{A}{\id}}{A}}
  % \label{rule:adaptt-sub-ty-id}
  % \and
  % \inferdeft[SubTyComp]
  % {\typing{\Gamma}{A}\\
  %   \typing{\Gamma}{\sigma}[\Delta]\\
  %   \typing{\Delta}{\tau}[\Xi]}
  % {\conv{\Gamma}{\subs{A}{\tau \circ \sigma}}{\subs{\subs{A}{\sigma}}{\tau}}}
  % \label{rule:adaptt-sub-ty-comp}
  % \and
  % \inferdeft[SubTmId]
  % {\typing{\Gamma}{t}[A]}
  % {\conv{\Gamma}{\subs{t}{\id}}{t}[A]}
  % \label{rule:adaptt-sub-tm-id}
  % \and
  % \inferdeft[SubTmComp]
  % {\typing{\Gamma}{t}[A]\\
  %   \typing{\Gamma}{\sigma}[\Delta]\\
  %   \typing{\Delta}{\tau}[\Xi]}
  % {\conv{\Gamma}{\subs{t}{\tau \circ \sigma}}{\subs{\subs{t}{\sigma}}{\tau}}[\subs{A}{\tau \circ \sigma}]}
  % \label{rule:adaptt-sub-tm-comp}
  % \and
  \inferdeft[SubAdId]
  {\typing{\Gamma}{f}[\adso{A}{B}]}
  {\conv{\Gamma}{\subs{f}{\id}}{f}[\adso{A}{B}]}
  \label{rule:adaptt-sub-ad-id}
  \and
  \inferdeft[SubAdComp]
  {\typing{\Gamma}{f}[\adso{A}{B}]\\
    \typing{\Gamma}{\sigma}[\Delta]\\
    \typing{\Delta}{\tau}[\Xi]}
  {\conv{\Gamma}{\subs{f}{\tau \circ \sigma}}{\subs{\subs{f}{\sigma}}{\tau}}[\adso{\subs{A}{\tau \circ \sigma}}{\subs{B}{\tau \circ \sigma}}]}
  \label{rule:adaptt-sub-ad-comp}
  \and
  \inferdeft[AdaptSub]
  {\typing{\Gamma}{\sigma}[\Delta]\\
    \typing{\Delta}{t}[A]\\
    \typing{\Delta}{f}[\adso{A}{B}]}
  {\conv{\Gamma}{\subs{\ad{t}{f}}{\sigma}}{\ad{\subs{t}{\sigma}}{\subs{f}{\sigma}}}}
  \label{rule:adaptt-adapt-sub}
  \and
  \inferdeft[IdSub]
  {\typing{\Gamma}{\sigma}[\Delta]\\
    \typing{\Delta}{A}}
  {\conv{\Gamma}{\subs{\id}{\sigma}}{\id}[\adso{\subs{A}{\sigma}}{\subs{A}{\sigma}}]}
  \label{rule:adaptt-id-sub-ad}
  \and
  \inferdeft[CompSub]
  {\typing{\Gamma}{\sigma}[\Delta]\\
    \typing{\Delta}{f}[\adso{A}{B}]\\
    \typing{\Delta}{g}[\adso{B}{C}]}
  {\conv{\Gamma}{\subs{(g \circ f)}{\sigma}}{\subs{g}{\sigma} \circ \subs{f}{\sigma}}[\adso{\subs{A}{\sigma}}{\subs{C}{\sigma}}]}
  \label{rule:adaptt-comp-sub-ad}
\end{mathpar}
\end{small}

  \Description{The standard CwF rules, plus new rules for adapters: identity,
  composition, cast, and their equations.}
  \caption{Judgments and rules for \AdapTT (excerpt, see \cref{sec:full-rule-adaptt})}
  \label{fig:adaptt-def}
\end{figure}

\AdapTT{}, presented in \cref{fig:adaptt-def}, extends a type theory à la Martin-Löf
% with judgment forms for well-formed contexts, type, term, substitutions and their conversion counterpart.
%
with a judgment $\typing{\Gamma}{\Box}[\adso{A}{B}]$
for \emph{adapters} between two types.
This terminology, borrowed from \textcite{McBride2021}, emphasizes that
%these morphisms between types
adapters provide information to transform values from one type to another.
The rules at the top of \cref{fig:adaptt-def} govern the standard judgments,
omitting some standard rules given in \cref{sec:full-rule-adaptt}.
Overloading notations, \(\emp\) denote both the empty context (\nameref{rule:adaptt-ctx-emp}) and empty substitution (\nameref{rule:adaptt-sub-emp}), $\ext$ is used for extension of contexts (\nameref{rule:adaptt-ctx-ext}) and of substitutions (\nameref{rule:adaptt-sub-ext}), and action of substitutions an all objects is written \(\subs{\cdot}{\cdot}\) (\nameref{rule:adaptt-sub-ty}, \nameref{rule:adaptt-sub-tm}, \nameref{rule:adaptt-sub-ad}).
Variables use de Bruijn indices~\cite{deBruijn1972}, generated by the \(\tmvz\)'th variable and the weakening substitution \(\wk\).
For instance, the named judgment \(\typing{\Gamma, x : A, y : A}{x}[A]\) corresponds to \(\typing{\Gamma \ext A \ext \subs{A}{\wk}}{\subs{\tmvz{}}{\wk}}[\subs{\subs{A}{\wk}}{\wk}]\).

There is an identity adapter at any type (\nameref{rule:adaptt-id}),
and adapters compose (\nameref{rule:adaptt-comp}).
% composes two adapters with compatible target and source.
%
With equations \nameref{rule:adaptt-id-left}, \nameref{rule:adaptt-id-right} and \nameref{rule:adaptt-comp-assoc}, types in each context form a category, preserved by substitution (\nameref{rule:adaptt-sub-ad}, \nameref{rule:adaptt-adapt-sub},\nameref{rule:adaptt-id-sub-ad}, \nameref{rule:adaptt-comp-sub-ad}).
The cast \(\ad{a}{f}\) lets an adapter \(f \ty \adso{A}{B}\) act on a term \(a \ty A\)
to yield a term of type \(B\) (\nameref{rule:adaptt-adapt}).
Casts preserve identities (\nameref{rule:adaptt-adapt-id}) and composition (\nameref{rule:adaptt-adapt-comp}): these functor laws are crucial to ensure the coherence of
adapters' action.%
\footnote{
In the context of structural subtyping, \textcite{Subtyping2024}
show that coherent elaboration from (implicit) subsumptive subtyping to
(explicit) coercive subtyping requires exactly these functorial laws.}

% Another view is that this is a categorification of subtyping: in usual subtyping one has a judgment \(A \sub B\), so implicitly there is only one way to cast from \(A\) to \(B\), but by contrast there might \textit{a priori} be many adapters between two given types.\km{the proof-relevant subtyping metaphore is useful but I think it should go elsewhere}

% %
% Adapters $a : \GatAd\,A\,B$ act on terms $t : \GatTm A$ to obtain another term $\GatAdapt{t}{a} : \GatTm B$.
% %
% This action must respect identity adapters ($\GatAdaptId$) and composition of adapters ($\GatAdaptComp$), hence satisfying functoriality laws that yield a ($\cSet$-valued) functor on the category of types and adapters.

% \km{Add a glance toward what we aim to obtain syntactically on type formers ?}

\begin{example}[Function type in \AdapTT]
\label{ex:pi-adaptt}
\AdapTT{} provides the judgments to state functoriality of type formers when adding
them to the system.
% Once \AdapTT{} is extended with type formers, functoriality can be stated
% This core type theory defining \AdapTT{} is meant to be extended with type formers.
%
Illustrating this process on \(\P\)-types, for which we assume the usual \MLTT rules,
one can express the data required to build an adapter and how this adapter acts on terms of \(\P\) type,
which constitute the common core of \cref{fig:casts-comparison}.
% \begin{small}
\begin{mathpar}
  \inferdef
  { \typing{\Gamma}{a}[\adso{A'}{A}] \\\\
    \typing{\Gamma \ext A'}{b}[\adso{\subs{B}{\ad{\tmvz}{a}}}{B'}]}
  {\typing{\Gamma}{\trans{\P}{a \ext* b}}[\adso{\P A. B}{\P A'. B'}]}
  \and
  \inferdef
  { \typing{\Gamma}{f}[\P A.B] \\ \typing{\Gamma}{u}[A'] \\\\
  \typing{\Gamma}{a}[\adso{A'}{A}] \\
  \typing{\Gamma \ext A'}{b}[\adso{\subs{B}{\ad{\tmvz}{a}}}{B'}]
  }
  {\conv{\Gamma}{(\ad{f}{\trans{\P}{a \ext* b}})\;u}
    {\ad{(f\,(\ad{u}{a}))}{\subs{b}{u}}}[\subs{B'}{u}]}
\end{mathpar}
Each of the examples of \cref{fig:casts-comparison} can be represented by
making a different choice for what constitutes an adapter, which is made
possible by having a separate, dedicated sort of adapters.
\end{example}

For datatypes, structural adapters behave like generalized map functions, so for instance
\[\ad{(h \cons t)}{\List(f)} \convop (\ad{h}{f}) \cons (\ad{t}{\List(f)})\]
Our goal in \cref{sec:inductive-types} will be to derive these
functorial adapters for large classes of datatypes, together with the appropriate
computation rules such as the above.

% \end{small}

% \paragraph{Functorial laws for elaboration}
% Beyond the functorial structures propagating adapters that one can endow on a given type former, the functorial laws (preservation of identity and composition) themselves are important for elaboration considerations.
% \textcite{Subtyping2024} show that for subtyping, an elaboration procedure is possible from (implicit) subsumptive subtyping to (explicit) coercive subtyping exactly when type formers satisfy the functorial laws with respect to identity and composition.

\subsection[Models for AdapTT]{Models for \AdapTT}
\label{sec:adapters}

\AdapTT is naturally interpreted in a variation on natural models~\cite{Awodey2018},
% and categories with families~\cite{dybjerInternalTypeTheory1996},
which we introduce.
For a category \(C\), we write \(\psh{C} = C^{\op}{\to}\cSet\) for the presheaves over \(C\),
and \(\catpsh{C} = C^{\op}{\to}\Cat\) the \(2\)-category of \(\Cat\)-valued presheaves
(or pre-stacks) over \(C\).

\begin{definition}[\shepherd{\NMDO}]
  \label{def:CwFs}
  A \shepherd{\emph{natural model with discrete opfibration} (\NMDO)}
  consists of a category \(\Ctx\) with a terminal object \(\emp\),
  two \(\Cat\)-valued presheaves \(\Ty, \Tm \ty \Ctx^{\op} \to \Cat\)
  and a representable natural transformation \(p \ty \Tm \natt{} \Ty \)
  equipped with a discrete opfibration structure \shepherd{in the \(2\)-category
  \(\catpsh{\Ctx}\)}.
\end{definition}

By forgetting some of its structure, any \NMDO is also a natural model in
the standard sense.
% \km{What about splitting the definition \cref{def:CwFs} into 2/3: first the
%   categorical definition that should be much smaller; and second the SOGAT
%   definition of the syntax; finally a lemma stating that the two presentations
%   are interchangeable ?}
% \mlb{Agreed, but the local representability is annoying to express in terms of
%   CwF. Should we take the natural model version as the “categorical” definition?}
% \tb{Natural models and CwFs are literally the same, the context extension operation is exactly the representability of the map \(\Tm\to \Ty\), no?}\km{yes, but 1. I have very slight doubts on its generalization to pre-stack when we move to the 2-dim version, 2. we need to check that the example follow easily}
The representability of \(p\), a key insight of \textcite{Awodey2018}
(and, independently, \textcite{Fiore2012}),
characterizes context extension \(\ext\), weakening \(\wk\) and the variable \(\tmvz\),
as a pullback in \(\catpsh{\Ctx}\):
\[
  \begin{tikzcd}[row sep = scriptsize]
    \yon(\Gamma\ext A) \ar[r, "\tmvz"]\ar[d,"y\wk"'] \ar[dr,phantom,"\lrcorner"{very near start}] & \Tm \ar[d,"p"] \\
    \yon\Gamma \ar[r,"A"'] & \Ty
  \end{tikzcd}
\]
where \(\yon \ty \Ctx \to \catpsh{\Ctx}\) is the Yoneda embedding.%
\footnote{The fact that \(p\) is a discrete opfibration in the \(2\)-category \(\catpsh{C}\)
ensures that the strict pullback is a \(2\)-pullback~\cite{Street74},
so that all \(2\)-categorical constructions are well-behaved.}
Unfolding the definition of a discrete opfibration \shepherd{shows that \(p\) is
a discrete opfibration objectwise, together with a naturality condition. The
objectiwise structure gives,} for any context \(\Gamma \ty \Ctx\) and term \(t \ty \Tm(\Gamma)\), an adapter \(a \ty p_{\Gamma}\,t \to A'\) in \(\Ty(\Gamma)\) uniquely lifts through \(p_{\Gamma}\), inducing some \(\bar{a} \ty t \to t'\) in \(\Tm(\Gamma)\) that maps to \(a\) through \(p_{\Gamma}\).
The existence of the lifting interprets~\ruleref{rule:adaptt-adapt}, with \(t'\) above
being \(\ad{t}{a}\),
while the uniqueness validates~\ruleref{rule:adaptt-adapt-id}
and~\nameref{rule:adaptt-adapt-comp}. \shepherd{The
  naturality condition of \(p\) in the discrete opfibration structure in the
  \(2\)-category \(\catpsh{\Ctx}\) corresponds exactly to the validation of \ruleref{rule:adaptt-adapt-sub}.}
Collecting these observations, we obtain the following theorem.

% \km{Use this opportunity to explain that internal cats in presheaves are the same as Cat-valued presheaves}

% The key points in this definition are the two conditions on the natural transformation \(p\): it is representable and a split discrete opfibration. We now explain those two conditions.

% Here and in the whole paper, we systematically overload the \(\ext\) symbol to denote all “extensions” for objects that (in the syntax) behave like snoc\tb{snoc?} lists.
% %
% Similarly, we use \(\emp\) for all “empty” lists.
% \tb{Maybe here reiterate the point: In a \NMDO there is a category of types above each object, morphisms being subtyping witnesses, and a set of terms above each object and each type above that object, with adequate actions?}
% %

% %
% As the name of the components suggest, the functor $T = (\Ty, \Tm)$ in a \NMDO carries the structure to interpret the types $\GatTy$ and terms $\GatTm~A$ in \AdapTT.
% %
% The category structure on $\Ty\,\Gamma$ provides an interpretation of $\GatAd$ in context $\Gamma \ty \Ctx$ as well as the corresponding identities, compositions and categorical equations.
% %
% The functor $\Tm\,\Gamma$ on $\Ty\,\Gamma$ yields an interpretation of $\GatAdapt{\cdot}{\cdot}$ together with the functoriality laws.
% %
% Summing up

\begin{theorem}[Interpretation of \AdapTT]
  \label{thm:adaptt-cwfs}
  \AdapTT interprets soundly into any \NMDO.
\end{theorem}
\begin{proof}
  \AdapTT{} can be presented as a Second-Order Generalized Algebraic Theory (SOGAT) (see \cref{fig:sogat-cwfs}) that makes very explicit that it consists of a sort of types equipped with a category structure, and a functorial action on the dependent sort of terms.
  The presentation of \cref{fig:adaptt-def} has been obtained with an explicit SOGAT-to-GAT translation~\cite{KaposiX24}.
  Following \textcite{UemuraPhD}, a model of that SOGAT consists of a small category \(\Ctx\) with a terminal object, together with a presheaf of types \(\Ty \ty \Psh(\Ctx)\) equipped with an internal category structure, a presheaf \(\Tm \ty \Psh(\Ctx)\), and a natural transformation \(\Tm \natt \Ty\) equipped with the structure of an internal discrete opfibration.
  To finish the proof, observe that internal categories in \(\Psh(\Ctx)\) are equivalent to \(\Cat\)-valued presheaves \(\catpsh{\Ctx}\), so this data exactly amounts to that of \cref{def:CwFs}.
\end{proof}

\shepherd{
The proof of \cref{thm:adaptt-cwfs} also shows a form of completeness of \AdapTT
% with respect to \NMDO{}s,
since the notion of models of \AdapTT in its SOGAT presentation coincides with \NMDO{}s.
A notion of completeness including initiality of the syntax
would require a discussion of morphisms of models that could preserve all the
structure on the nose or only up to an isomorphism.
}

% \km{How formal do we want to get this theorem ? A more precise version would claim 1. that (the syntactic model of) \AdapTT yield a \NMDO 2. the existence part of the initiality of the syntactic model}

\NMDO{}s have many interesting examples. The admissible rules that we identify for some of these models (\cref{ex:subtyping,ex:ttobs-cwfs}) could more directly be used to \emph{define} type theories with structural type casts, using \AdapTT{} as a framework informing their design.
% \km{Filling in all proofs obligations in the following examples will be too much; should we defer that to appendices ?}
% \tb{Strong agree}

\newcommand\discr[0]{\operatorname{discr}}

\begin{example}[Natural models are discrete \NMDO{}s]
\label{ex:cwf-cwfs}
Since any set is trivially a category with only identity morphisms,
any natural model is trivially a \NMDO with only identity adapters.
Indeed, if \(\Ty(\Gamma)\) is a discrete category, \(p\) is automatically
a discrete opfibration.
% , \ie where the $\typing{\Gamma}{\star_{A,B}}[\adso{A}{B}] \iff \Gamma \vdash A \convop B$.
%
% Formally, it suffices to let $T^{\text{\NMDO}} \coloneq \discr \circ T^{\text{\CwF}}$ where the functor $\discr : \Famset \to \Famcat$ sends a pair $(X,F)$ of a set $X$ and a family of sets $F : X \to \cSet$ on $X$ to the pair of the discrete category on $X$ and $F$ seen as a functor $\discr X \to \cSet$.
% \km{Missing functoriality of $\discr$ (trivial) and argument for preservation of context extension}
\end{example}

\newcommand\Sect[1]{\mathrm{Sect}_{#1}}
\example[\(\cSet\) as a \NMDO]
\label{ex:set-cwfs}
% \km{After writing both this example and the full \NMDO example, I realize this one is a particular case of the latter construction. I think we may want to keep it still because it is ``concrete'' and simpler to grasp}\tb{I agree to keep it}
%
The standard natural model on \(\cSet\) extends to a non-discrete \NMDO.
Setting $\Ctx \coloneq \cSet$, contexts are interpreted as sets and substitutions as functions.
Types over a set $\Gamma$ are functors \(\Ty(\Gamma) \coloneq \discr \Gamma {\to} \cSet\) from the discrete category $\discr \Gamma$ to $\cSet$.
Terms over $\Gamma$ are functors \(\Tm(\Gamma) \coloneq \discr \Gamma {\to} \cSet_{\bullet}\) to the category of pointed set, and \(p\) is the post-composition by the functor \(\cSet_{\bullet} {\to} \cSet\) forgetting the point.
% Given such a type \(A \ty \discr \Gamma \to \cSet\), a term is a section of \(A\),
% \ie a dependent function \(\P (\gamma : \Gamma). A(\gamma)\).
% These assemble into a functor of sections $\Sect{\Gamma} : \Ty(\Gamma) \to \cSet$,
% and taking its Grothendieck construction, we define \(\Tm(\Gamma) \coloneq \int \Sect{\Gamma}\).
% defined on objects as $\Sect{\Gamma}(A : \Gamma \to \cSet) \coloneq (\gamma : \Gamma) \to A\,\gamma$.\km{Missing functorial actions of $\Sect{\Gamma}$ and of $T$}
 % \tb{the notation \(\Gamma\to\cSet\) is nonstandard. To me, it reads like an internal hom, but this is a coslice. The standard notation for the coslice is \(\Gamma\backslash\cSet\). I think this notation needs to be explained before this definition}\km{No, this is not a coslice, it's the category of functors from $\Gamma$ to $\cSet$}
% \km{explicit the notation for functor categories}
% so that a %
Context extension is inherited from the natural model:
the extension of a set $\Gamma$ with a family $A \ty \Gamma {\to} \cSet$ is given by the set $\Sigma(\gamma \ty \Gamma). (A\,\gamma)$.
Expliciting the morphisms of $\discr \Gamma {\to} \cSet$, an adapter $\typing{\Gamma}{f}[\adso{A}{B}]$ between $A, B \ty \Ty(\Gamma)$ is a natural transformation $A \natt B$,
thus a family of functions $f : (\gamma : \Gamma) {\to} A\, \gamma {\to} B\, \gamma$ since \(\Gamma\) is discrete.

\begin{example}[\(\Cat\) as a \NMDO]
  \label{ex:cat-cwfs}
The \NMDO structure on $\cSet$ extends to the $1$-category $\Cat$, substituting all instances of a discrete category \(\discr \Gamma\) by an arbitrary one \(\Gamma \ty \Cat\).
Context extension is given by the Grothendieck construction\footnote{The
  Grothendieck construction \(\int_{\Gamma} A\) we consider here for
  \(A : \Gamma {\to} \cSet\) yield an \emph{opfibration}
  \(\pi_\Gamma : \int_{\Gamma} A \to \Gamma\) and not a fibration. In
  this particular instance, the Grothendieck construction is is simply a
  category of elements.}~\cite[1.10.1]{JacobsCatlogTT} of a type
$A \ty \Gamma \to \cSet$ and the naturality condition on adapters
$\typing{\Gamma}{f}[\adso{A}{B}]$ between $A, B \ty \Gamma \to \cSet$ becomes
non-trivial.
%
% The discrete functor $\discr \ty \cSet \to \Cat$ induces a morphism of \NMDO: the notions of types, terms and extension over a discrete context coincide with that in $\cSet$.
% \tb{I think we need to make the adapters explicit somewhere here}
\end{example}

\begin{comment}
\begin{example}[\(\Cat\) as a \NMDO]
%
The \NMDO structure on $\cSet$ extends to the $1$-category $\Cat$.
%
A context is interpreted as a category and a substitution as a functor.
%
A type $A$ over a category $\Gamma$ is a functor $A \ty \Gamma \to \cSet$ and its set of terms is defined by the functorial sections $\Sect{\Gamma}(A) \coloneq (\gamma : \Gamma) \to A\,\gamma$.
%
Adapters $\typing{\Gamma}{f}[\adso{A}{B}]$ between $A, B \ty \Gamma \to \cSet$ still consist of natural transformations $A \natt B$.
%
Context extension is given by the Grothendieck construction\km{refnec} of a type $A \ty \Gamma \to \cSet$.
%
% The discrete functor $\discr \ty \cSet \to \Cat$ induces a morphism of \NMDO: the notions of types, terms and extension over a discrete context coincide with that in $\cSet$.
% \tb{I think we need to make the adapters explicit somewhere here}
\end{example}
\end{comment}

\begin{example}[Full \NMDO]
\label{ex:cwf-full}
In a \NMDO, an adapter \(a\) from \(A\) to \(B\) induces a term
\(\typing{\Gamma \ext A}{\ad{\vz}{\subs{a}{\wk}}}[\subs{B}{\wk}]\), and this
preserves the categorical structure given by the identity adapter and adapter
composition.
% \((\Ty(\Gamma),\Ad(\Gamma,\cdot,\cdot))\) embeds into
% \((\Ty(\Gamma),\Tm(\Gamma \ext \cdot,\subs{\cdot}{\wk}))\).
 A \emph{full} \NMDO is one where this embedding is an isomorphism,
\ie any open term induces an adapter.
Any natural model $C$ can be equipped with a full \NMDO structure by taking
\(\typing{\Gamma}{f}[\adso{A}{B}] \coloneq \typing{\Gamma \ext A}{f}[\subs{B}{\wk}]\) and $\ad{t}{f} \coloneq \subs{f}{\id_\Gamma \ext t}$.
The identity adapter is given by the variable \(\tmvz\), and composition by substitution.
%
% This is what \MLTTmap achieves in \textcite{Subtyping2024}.\footnote{Although \MLTTmap rather takes \(\Ad(\Gamma, A, B) \coloneq \Tm(\Gamma,A \to B)\), which
% is isomorphic by definition of the function type.}
\end{example}

\Cref{ex:cwf-cwfs,ex:cwf-full} relate to the two well-known ways to turn a natural model into a comprehension category \cite{Jacobs1993,Coraglia2024}, either discrete or full, by making either no or all terms into arrows in the category of types.
\Cref{ex:set-cwfs,ex:cat-cwfs} are instances of the latter construction.

% This gives us many interesting examples of \NMDO{}, for instance \cref{ex:set-cwfs}
% is obtained this way, and similarly there is a natural model of categories, where types are categories, substitutions are functors, types are functors from their contexts to \(\cSet\), and
% adapters are natural transformations. \mlb{refnec}
%
%
% The two examples correspond to the two ways of making this embedding uninteresting, by either having a trivial category as source, or an isomorphism.

\begin{example}[Subtyping]
\label{ex:subtyping}
  Coercive subtyping can be expressed with adapters:
  \MLTTcoe \cite{Subtyping2024} is an instance of \NMDO where an adapter \(\typing{\Gamma}{f}[\adso{A}{B}]\) is a witness that \(A\) is a subtype of \(B\).
  Since there is at most one subtyping witness between two types, this \NMDO validates the following “adapter irrelevance” rule, meaning that the category of types is a mere poset
  \[
    \inferdef[AdIrrel]{
      \typing{\Gamma}{f}[\adso{A}{B}] \\
      \typing{\Gamma}{f'}[\adso{A}{B}]
    }{
      \conv{\Gamma}{f}{f'}[\adso{A}{B}]
    }
    \label{rule:ad-unique}
  \]
  For an adapter $\typing{\Gamma}{f}[\adso{A}{B}]$, the cast \(\ad{a}{f}\)
  is given by the coercion operation \(\coe[A][B]{a}\).
  %
  % As explained by \textcite{Subtyping2024}, the functorial laws on the action of adapters are crucial to obtain a coherent elaboration from implicit (subsumptive) subtyping.

  % Categorically, this corresponds to \(\Ad\) being a poset rather than a
  % full-blown category.
  % As tracking this extra constraint brings more complication than
  % relief, and is too strong to capture other examples of interest,
  % However, adding this rule is a strong requirement on models.
  % For instance, in a set-theoretic model, it is natural to interpret adapters
  % as function between the relevant sets. But
  % interpreting \(\Ad(\emp,A,B)\) as the set of function
  % \(\llbracket B \rrbracket^{\llbracket A \rrbracket}\) will not satisfy \ruleref{rule:ad-unique},
  % since there are way too many functions, even though our syntax only “hits” one of them.
  % In a sense, this is similar to the fact that
  % \(\P x : A.B \convop \P x : A'.B'\) implies \(A \convop A'\)
  % and \(B \convop B'\), which is typically true in the syntax, but rarely in models.
  % Thus, it might be more
  % reasonable to \emph{show} that \ruleref{rule:ad-unique} is \emph{admissible} in
  % the syntax of systems where we want to think of adapters as subtyping,
  % rather than to impose it as part of the system.
\end{example}

\begin{example}[\TTobs as a \NMDO]
\label{ex:ttobs-cwfs}
Observational Type Theory (\TTobs)~\cite{Altenkirch2007} is a type theory where
the operation \(\castTTobs{}\ty \oId[\univ]{A}{B} \to A \to B\) associated
to the identity type \(\oId[\univ]{A}{B}\) computes structurally.
\posscite{Pujet2023} version of \TTobs is a \NMDO instance, by taking:
\begin{mathpar}
  \inferdef[EqAd]
  {\typing{\Gamma}{e}[\oId[\univ]{A}{B}]}
  {\typing{\Gamma}{\underline{e}}[\adso{A}{B}]}
  \label{rule:eq-ad}
  \and
  \inferdef[AdEq]
  {\typing{\Gamma}{e}[\oId[\univ]{A}{B}]\\
  \typing{\Gamma}{a}[A]}
  {\conv{\Gamma}{\ad{a}{\underline{e}}}{\castTTobs{}\,e\,a}[B]}
  \label{rule:ad-eq}
  \and
  \inferdef[EqReflId]
  {\typing{\Gamma}{A}}
  {\conv{\Gamma}{\underline{\refl}}{\id_A}[\adso{A}{A}]}
  \label{rule:cast-refl}
\end{mathpar}
By \ruleref{rule:eq-ad}, any proof of observational equality between types entails
an adapter with action on term given by \(\castTTobs{}\) (\nameref{rule:ad-eq}).
% ---\(\oId[A]{t}{u}\) is the observational equality of \(t\) and \(u\):
Casts along reflexivity proofs are identities by \ruleref{rule:cast-refl}, and
\textcite{Pujet2025} show that adding this equation preserves decidability of conversion.%
\footnote{
  Thanks to proof irrelevance, \ruleref{rule:cast-refl} implies
  the stronger rule where \(\refl\) is replaced by an arbitrary proof of
  \(\oId{A}{A}\), because such a proof is always convertible to \(\refl\).
}

It is tempting to add a similar rule for composition
(where \(\operatorname{trans}\) is transitivity of equality)
\begin{mathpar}
  \inferdef[EqCompTrans]
  {\typing{\Gamma}{e}[\oId[\univ]{A}{B}] \\
    \typing{\Gamma}{e'}[\oId[\univ]{B}{C}]}
  {\conv{\Gamma}{\underline{e} \circ \underline{e'}}{\underline{\operatorname{trans} e~e'}}[\adso{A}{C}]}
  \label{rule:obstt-eq-trans}
\end{mathpar}
However, \textcite[Appendix A]{Subtyping2024} show that, using
this rule, we can construct a model of pure λ-calculus in the system, by working
in a context with a hypothesis \(e \ty A = A \to A\).
To avoid this obstacle to decidability of conversion,
we can instead remove this equation and instead interpret composition freely: adapters are chains of
equalities \(\underline{e_1} \circ \ldots \circ \underline{e_n}\).
% Thus, just like \citeauthor{Pujet2023}, we do not add any extra equation for transitivity.
\end{example}

\begin{example}[CastCIC as a \NMDO]
  CastCIC \cite{LennonBertrand2022} has been proposed as a foundation for gradual dependent types.
  In CastCIC, \emph{any} two types \(A,B\) are related by an adapter
  \(\typing{\Gamma}{(A \Rightarrow B)}[\adso{A}{B}]\).
  Non-diagonal casts such as \(\ad{0}{\Nat \Rightarrow \Bool}\) reduce
  to runtime errors following the exceptional type theory of~\textcite{Pedrot2018}.
  \posscite{LennonBertrand2022} fire triangle,
  which shows that normalization cannot hold in a system satisfying the gradual guarantees, is another avatar of the obstruction exposed by \citeauthor{Subtyping2024} regarding the above rule for transitivity. %\km{Unsure about the precise relationship}
\end{example}

% \begin{example}[Dialectica] The dialectica model of dependent type theory~\cite{MossG18} where contexts are interpreted as polynomials (on a base \NMDO) can be equipped with a non-trivial non-full \NMDO structure.
% %
% The adapters $\Ad(\Gamma, (A, P_A), (B, P_B))$ are linear maps $(f, \phi)$ where $f : A \to B$ and $\phi : (a : A) \to P_B\,(f\,a) \to P_A\,a$ where each $\phi\,a$ are isomorphisms.\km{I didn't check the details}
% \end{example}

% \km{Removing for now because the notations for adapters are a headache}
% \begin{example}[Non-trivial and non-full \NMDO]
%  Any \NMDO $C$ give rise to a \NMDO $C^\cong$ whose adapters are adapter-based isomorphisms in $C$, meaning that $\Ad^{C^\cong}(\Gamma,A,B) \coloneq \Sigma(f : \Ad^C(\Gamma,A,B))(g : \Ad^C(\Gamma,B,A)) f \circ g = \id \times g \circ f = \id$
% \end{example}

\subsection{Comparison with Generalized Categories with Families (\texorpdfstring{\GCwF}{gCwF})}
\label{sec:other-models}

% \mlb{Title is pretty weak}\tb{What to you think of this title?}

\NMDO, our model of \AdapTT, is closely related to the generalized categories with families (\GCwF) of \textcite{Coraglia2024a}.
\GCwF, initially introduced under the name of “plain dependent type theories” \cite{Coraglia2021}, were used by \textcite{Coraglia2024} to establish a connection between two formalisms to model type theory: categories with families/natural models, and comprehension categories.
The same authors then remarked \cite{Coraglia2024a} that, despite not being their original purpose, \GCwF support some flavor of subtyping, without explicitly deriving a type cast operation.
\textcite{Najmaei2025} propose a sound syntax to work in comprehension
categories, and further the analysis of subtyping, deriving type casts for \(\Pi, \Sigma\) and \(\Id\) types.
Clarifying further the status of type casts, we establish a correspondence between \GCwF{}s and \NMDO.

Let us start with a review of \posscite{Coraglia2024a} notations and terminology.
Natural models can be reformulated as the following diagram~\cite[3.1.11]{UemuraPhD} where \(u\) and \(\dot{u}\) are discrete fibrations, and \(\Sigma\) is a fibration
morphism with right adjoint \(\Delta\) (which is \emph{not} a fibration morphism)
\[
  \begin{tikzcd}
    \dot{\mathcal{U}}
    \ar[rr, bend right=15, "\Sigma"']
    \ar[rr,phantom, "\top"]
    \ar[dr, "\dot{u}"']
    &
    &
      \mathcal{U}
      \ar[ll , bend right=15, "\Delta"']
      \ar[dl, "u"]
    \\
    & \mathcal{B}
  \end{tikzcd}
\]
The base category \(\mathcal{B}\) should be understood as that of contexts and substitutions, while \(\mathcal{U}\) is the category of types in contexts, and \(\dot{\mathcal{U}}\) that of typed terms in a context. The functors \(u\) and \(\dot{u}\) project the underlying context of a type and a term. The discrete fibration property of \(u\) and \(\dot{u}\) gives the action of substitutions on types and terms. \GCwF generalize the above, merely requiring \(u\) and \(\dot{u}\) to be fibration, without dicreteness.
% are just required to be fibrations, instead of discrete ones. \tb{Maybe an explanation of fibrations?}

Dropping the discreteness condition is relevant to model type casts. Indeed, considering an object \(\Gamma\) of \(\mathcal{B}\), and two objects \(A,B\) of \(\mathcal{U}\) in the fiber over \(\Gamma\) (\ie two types in a context \(\Gamma\)), it allows for non-trivial \emph{vertical morphisms} \(f : A \to B\) in \(\mathcal{U}\), those such that \(u(f) = \id_{\Gamma}\). These play precisely the role devoted to adapters in a \NMDO. However, it also relaxes the structure too much to interpret
\AdapTT{}, as it makes the action of morphisms of \(\mathcal{B}\) onto the fibers in \(\mathcal{U}\) pseudo-functorial instead of functorial. In particular, the equations \(\subs{A}{\sigma \circ \tau} \convop \subs{\subs{A}{\sigma}}{\tau}\) and \(\subs{A}{\id}\convop A\) and their counterparts on terms are not valid in a \GCwF. To account for this, we restrict our attention to
a subclass of \GCwF that we call \emph{split} \GCwF.

\begin{definition}
  A \emph{split} \GCwF equips \(u\) and \(\dot{u}\) as above with a functorial choice of lifts, preserved by \(\Sigma\) and \(\Delta\), such that the unit and counit are componentwise the chosen lift of their projection.
  % A \GCwF is split when both functors \(u\) and \(\dot{u}\) are split fibrations, the functors \(\Sigma\) and \(\Delta\) preserve chosen lifts, the unit and counit are component wise the chosen lift of their projection.
\end{definition}

To model coercive subtyping with \GCwF, \textcite{Coraglia2024a} define an additional judgment \(\Gamma \vdash a \ty_{f} B\) interpreted as the existence of a term \(a\) of type \(A\) in context \(\Gamma\) together with a vertical morphism \(f : A \to B\). But they do not derive a proper type cast corresponding to~\ruleref{rule:adaptt-adapt}: %Given the aforementioned pieces of data,
assuming this data,
we need to construct an element \(\ad{t}{a}\) of \(\dot{\mathcal{U}}\) such that \(\Sigma(\ad{t}{a}) = B\). The following proposition, proven in \cref{sec:gcwf-cwfs}, resolves this discrepancy.

\begin{proposition}\label{proposition:coercion-gcwf}
  Split \GCwF support a cast operation: the functor \(\Sigma\) induces a discrete opfibration between \(\Cat\)-valued presheaves, from the presheaf of fibers of \(\dot{u}\) to the presheaf of fibers of~\(u\).
\end{proposition}

\begin{theorem}
  \label{thm:cwfs-gcwf}
  Split \(\GCwF\)s are equivalent to \NMDO{}s.
\end{theorem}
\begin{proof}[Proof sketch]
  The proof closely parallels the correspondence between discrete \GCwF and natural models, while additionally accounting for adapters. We briefly spell out the correspondence here and refer to \cref{sec:gcwf-cwfs} for the complete proof.

  From a split \GCwF, we define a \NMDO with \(\mathcal{B}\) as the category of contexts.
  The categories of types and terms in a context \(\Gamma\) are respectively given by the fiber of \(u\) and \(\dot{u}\) over \(\Gamma\).
  The morphism of fibrations \(\Sigma\) induces the natural transformation \(p\ty\Tm \natt{}\Ty\).
  Its right adjoint \(\Delta\) then gives the context extension operation.
  Finally \cref{proposition:coercion-gcwf} shows that \(p\) is a discrete opfibration.% \mlb{This is the classical proof, maybe highlight what's new/different with adapters?}\tb{fixed?}

  Conversely, from a \NMDO, we define a split \GCwF by defining the base category \(\mathcal{B}=\Ctx\), the split fibrations \(u \ty \mathcal{U} \to \mathcal{B}\) and
  \(\dot{u} \ty \dot{\mathcal{U}} \to \mathcal{B}\) are obtained respectively as the Grothendieck construction of the functor \(\Ty \ty \Ctx^{\op} \to \Cat\), and of the functor \(\Tm \ty \Ctx^{\op} \to \Cat\).
  The natural transformation \(p: \Tm \natt{} \Ty\) then induces a morphism of fibrations \(\Sigma \ty \dot{\mathcal{U}} \to \mathcal{U}\). The right adjoint \(\Delta\) to \(\Sigma\) is given by the context extension operation and the action of adapters on terms.
\end{proof}

% \km{Could we use \NMDO to ``explain'' the relationship between MLTTmap and
%   MLTTcoe, e.g. building a binary logical relation between the two that exhibit
%   that morally $\coe[F\,A][F\,B]{m} \simeq \texttt{map}^F\,
%   \coe[A][B]{}\,m$ ?}

\section{Representing type formers}
\label{sec:type-vars}

\AdapTT, as introduced in the previous section, provides a judgmental structure
to capture type casts.
As it stands, however, the type theory does not give good tools to uniformly
describe functorial type formers.
Indeed, to be able to talk about functoriality, we need to have categories
between which a type former maps. We have just worked to make the codomain clear: it is the category
of types and adapters. However, the domain can be much more complex: as shown
by \cref{ex:pi-adaptt}, it can involve multiple types, dependency,
and variance information.

% For this, we need to extend the type theory.
%
% The issue is that \AdapTT cannot faithfully internalize a representation of type formers, and in particular the parametrizing data that a given type former take as input.
In the standard setting, the description of type formers' domain typically relies on
universes. In \Agda{}, for instance, the type former for lists would be declared by:
% \km{Consider not using Agda, that's the sole use of this syntax in the paper for now}
\newcommand{\agdaSet}[0]{\textcolor{BlueViolet}{\tt Set}}
\begin{center}
  \hspace{3em}
  \textcolor{ForestGreen}{\tt data} $\List$ ($X$ : \agdaSet{}) : \agdaSet{}\hfill\strut
\end{center}
% \begin{lstlisting}
% data List (A : Set) : Set
% \end{lstlisting}
The universe \agdaSet{} provides
a (partial) internalization of the judgment for well-formed types.
In \AdapTT, however, this will not do. As adapters are a type-level notion only,
and terms form a mere set,  %, that does not provide any categorical structure.
% This will not do for us, as our : in \AdapTT we only have a \emph{set} of terms for a given type, which is crucial to keep the theory tractable.\km{unsure about the tractability claim, we kind of throw that away with \AdapTTt}
universes in \AdapTT could internalize types, but not their categorical structure,
and are thus ill-suited to describe the category that is the domain of a type former.

To address this, we extend \AdapTT to a theory \AdapTTt with a primitive notion of type variables.
% and the accompanying structure, to faithfully capture type formers like \(\List\) above.
%
\AdapTTt will be used to satisfactorily describe the parameters of inductive types, such as $\List$ above, in \cref{sec:functorial-types}.
We start in \cref{sec:type-var-examples} by distilling how these type variables ought to work,
to explain the design of \AdapTTt.
Then, \cref{sec:adaptt2-def} formally describes the type theory per se,
while \cref{sec:back-adaptt} builds a model of it on top of a model of \AdapTT,
providing a formal construction of functorial type formers for \AdapTT out of
\AdapTTt's ones.

\subsection{What do we want of type variables?}
\label{sec:type-var-examples}

\shepherd{In this section, we motivate our choice of considering type variables
  and the form of the rules we provide for them by looking at examples and
  expected features. For the sake of presentation, we use an informal,
  named syntax with implicit weakenings. We switch
  back to a formal, name-free version in \cref{sec:adaptt2-def}. Moreover, none
  of the rules presented here are defining a theory, but rather they constitute
  desiderata for rules that must be derivable in our proposed theory. For this
  reason some are left unnamed, and many are presented in a simplified
  variant.}

\subsubsection{Lists: type variables and transformations.}
\label{sec:tyvar-simple}

As we see for \(\List\) above, the input of a type former
is specified by a context. In that case, this context
consists of a single type variable: \(\Gamma_{\List} \coloneq (X \ty \Ty)\).
This is sufficient to capture the \(\List\) type former, given by \ruleref{rule:list-univ}
below. Indeed, by the general \ruleref{rule:subs-ty},
\ruleref{rule:list-subs} is derivable. Moreover,
a substitution \(\typing{\Delta}{\sigma}[\Gamma_{\List}]\) consists
of a single type: \(\sigma \convop \emp \ext \subs{X}{\sigma}\).
We hence obtain the more standard
\ruleref{rule:list-ty}. Conversely, instantiating \ruleref{rule:list-subs}
with \(\id_{\Gamma_{\List}}\) or \ruleref{rule:list-ty} with \(\typing{(X : \Ty)}{X}\),
we recover \ruleref{rule:list-univ}.%
\footnote{
For the category theorist, this is a manifestation, on objects,
of the (2-categorical) Yoneda lemma,
which says that \(\Ty(\Gamma)\)
is isomorphic to the category of natural transformations
\(\Sub(\cdot,\Gamma) \to \Ty\).
}
As all three versions are equivalent, we favor the less verbose
style of \ruleref{rule:list-univ}.
\begin{mathpar}
\Gamma_{\List} \coloneq (X \ty \Ty)
\and
\inferdef[ListUniv]{
  % \Delta \vdash \sigma \ty \Gamma_{\List}
}{
  \typing{\Gamma_{\List}}{\List}
} \label{rule:list-univ} \and

\inferdef[ListSubs]{
  \typing{\Delta}{\sigma}[\Gamma_{\List}]
}{
  \typing{\Delta}{\subs{\List}{\sigma}}
} \label{rule:list-subs} \and

\inferdef[ListTy]{
  \typing{\Delta}{A}
}{
  \typing{\Delta}{\List(A)}
} \label{rule:list-ty}
\end{mathpar}

Type variables are given by a context extension operation \(\ext*\),
given by the rules below.
Any well-formed context can be extended with a type variable, which can then
be accessed to yield a type.
A substitution into an extension \(\Gamma \ext* (X \ty \Ty)\)
is built out of a substitution into \(\Gamma\) together with a
well-formed type instantiating the type variable.
\begin{mathpar}
  \inferdef{\ctxty{\Gamma}}{\ctxty{\Gamma \ext* (X \ty \Ty)}} \and
  \inferdef{\ctxty{\Gamma}}{\typing{\Gamma \ext* (X \ty \Ty)}{X}} \and
  \inferdef{\typing{\Gamma}{A}}{\typing{\Gamma \ext* (X \ty \Ty)}{A}} \and
  \inferdef{\typing{\Gamma}{\sigma}[\Delta] \\ \typing{\Gamma}{A}}{
    \typing{\Gamma}{\sigma \ext* A}[\Delta \ext* (X \ty \Ty)]
  }
\end{mathpar}

But there is more happening.
Indeed, consider two substitutions
\(\typing{\Delta}{\sigma, \tau}[\Gamma_{\List}]\),
% , corresponding to the following picture:
% \begin{center}
% \begin{tikzcd}[column sep = 10em]
%   \Delta
%     \arrow[r, bend left=20, "\sigma"] % \arrow[r,bend left,""{name=s,below}]
%     \arrow[r, bend right=20, "\tau"'] % \arrow[r,bend right,""{name=t,above}]
%     % \arrow[Rightarrow, from=s,to=t,"\mu"]
%     & \Gamma_{\List} % = (X \Ty)
% \end{tikzcd}
% \end{center}
amounting to two types \(\subs{X}{\sigma}, \subs{X}{\tau}\).
We have an interesting way to relate these: adapters!
Hence, in \AdapTTt, there is a natural
and interesting way to relate substitutions, which
we call \emph{transformations}, after~\textcite{Licata2011}.
Transformations are typed by a judgment
\(\typing{\Gamma}{\mu}[\transso[\Delta]{\sigma}{\tau}]\), where
\(\typing{\Gamma}{\sigma,\tau}[\Delta]\), and
collect adapters between their endpoints.
That is, to create a transformation relating
substitutions \(\sigma,\tau\) respectively extended by types \(A\) and
\(B\), we must provide a transformation \(\mu \ty \transso{\sigma}{\tau}\)
and an adapter \(f \ty \adso{A}{B}\).
\[
  \inferdef{
    \typing{\Gamma}{\sigma, \tau}[\Delta] \\
    \typing{\Gamma}{\mu}[\transso[\Delta]{\sigma}{\tau}] \\
    \typing{\Gamma}{A,B} \\
    \typing{\Gamma}{f}[\adso{A}{B}]
  }{
    \typing{\Gamma}{\mu \ext* f}[
      \transso[\Delta \ext* (X : \Ty)]{\sigma \ext* A}{\tau \ext* B}]
  }
\]
In categorical terms, \(\Ctx\) in \AdapTTt is a \emph{2-category},
whose objects are contexts, 1-morphisms substitutions,
and 2-morphisms the transformations just introduced.

\(\Ty\) also interacts with this structure: transformations act on types.
% \mlb{Slight lie, we don't explain here what happens to terms, this will be covered
% later.}
% In other words, assuming we have a type \(A \ty \Ty(\Gamma)\),
% two substitutions \(\mu, \tau \ty \Sub(\Delta,\Gamma)\)
% and a transformation \(\mu \ty \Trans(\sigma,\tau)\), how should \(A\) and \(\mu\) interact?
Indeed, the core idea of \AdapTTt is that a transformation \(\mu\) acting
on a type \(A\) induces an adapter:
\begin{mathpar}
  \inferdef[TransTy]{
    % \impl{\Gamma, \Delta \ty \Ctx} \\
    % \impl{\sigma, \tau \ty \Sub(\Gamma,\Delta)} \\\\
    \typing{\Gamma}{\mu}[\transso[\Delta]{\sigma}{\tau}] \\
    \typing{\Delta}{A}
  }{
    \typing{\Gamma}{\trans{A}{\mu}}[\adso{\subs{A}{\sigma}}{\subs{A}{\tau}}]
  }
  \label{rule:ty-trans}
\end{mathpar}
\AdapTTt{} internalizes the functoriality of types:
any type \(A\) in \(\Delta\), given a context \(\Gamma\),
gives rise to a functor from the category of substitutions from \(\Gamma\)
to \(\Delta\) to that of types in \(\Gamma\),
and \ruleref{rule:ty-trans} corresponds to the action on morphisms (transformations).
% This is in essence similar to \MLTTmap \cite{Subtyping2024},
% where each type comes with an associated
% map operation.\km{The ref looks a bit obscure here, would rather keep it out of the main/simple explanation}
% Our main innovation is that we not only have such an operation for a
% handful of carefully selected
% type formers,
This gives us a way to \emph{compute} the source category of \(A\)
as a functor, by analyzing the structure of its context \(\Delta\).
Thus, the more expressive the contexts,
the more interesting functorial type formers we can capture.

For instance, for lists, we derive \ruleref{rule:list-trans} below. Specializing it
as we did above, we obtain \ruleref{rule:list-ad},
corresponding to the usual subtyping rule for lists, or to a kind of map operation.
\begin{mathpar}
  \inferdef[TransList]{
    \typing{\Delta}{\sigma,\tau}[\Gamma_{\List}] \\
    \typing{\Delta}{\mu}[\transso[\Gamma_{\List}]{\sigma}{\tau}]
  }{
    \typing{\Delta}{\trans{\List}{\mu}}[\adso{\subs{\List}{\sigma}}{\subs{\List}{\tau}}]
  } \label{rule:list-trans} \and
  \inferdef[AdList]{
    \typing{\Delta}{A, B} \\
    \typing{\Delta}{f}[\adso{A}{B}]
  }{
    \typing{\Delta}{\List(f)}[\adso{\List(A)}{\List(B)}]
  } \label{rule:list-ad}
\end{mathpar}
As a more involved example, combining the formation rule for lists above and
that for sums below,
by a single application of \ruleref{rule:ty-trans},
we can derive functoriality for a combination of them.
\begin{mathpar}
\inferdef%[Sum]
{ }{
  \typing{(X \ty \Ty) \ext* (Y \ty \Ty)}{X+Y}
} \and
\inferrule*{
  \typing{\Delta}{A, A', B, B'} \\
  \typing{\Delta}{a}[\adso{A}{A'}] \\
  \typing{\Delta}{b}[\adso{B}{B'}]
}{
  \typing{\Delta}{\trans{(\List(X + \List(Y)))}{a \ext* b}}[
    \adso{\List(A + \List(B))}{\List(A' + \List(B'))}]
}
\end{mathpar}
The behavior of this complex adapter is obtained by
following the type's structure, \ie by combining
the functoriality of the \(\List\) and \(+\) types.
In general, in \AdapTTt, the functoriality of complex types
is derived compositionally from that of each type former.

\begin{comment}
Finally, just as we can compose substitutions
in \AdapTT, \AdapTTt has usual 2-categorical operations ---
composition of transformations, left and right whiskering of a
transformation by a substitution--- and the relevant equations.
All operations are componentwise:
composition of transformations derives from composition of adapters, left whiskering
from the action of a substitution on adapters (\nameref{rule:adaptt-sub-ad}),
and right whiskering from the action of transformations (\nameref{rule:ty-trans}).
\end{comment}

\subsubsection{Identity: transformations and terms}
\label{sec:tyvar-eq}

So far, we omitted to consider terms. However, just like types, terms should be acted upon
by transformations, and conversely we should have a way to extend transformations
to account for term variables. Here, the insight is that since we have only a set of
terms, the only way to relate them is by a definitional equality constraint, as follows:
\begin{mathpar}
  \inferdef[TransTm]{
      \typing{\Gamma}{\mu}[\transso[\Delta]{\sigma}{\tau}] \\
      \typing{\Delta}{A} \\
      \typing{\Gamma}{t}[\subs{A}{\sigma}] \\
      \typing{\Gamma}{u}[\subs{A}{\tau}] \\
      \conv{\Gamma}{\ad{t}{\trans{A}{\mu}}}{u}[\subs{A}{\tau}]}
      {\typing{\Gamma}{\mu \ext t}
        [\transso[\Delta \ext (x \ty A)]{(\sigma \ext t)}{(\tau \ext u)}]}
  \label{rule:trans-tm-intro}
  \and
  \inferdef[TransTyAdTm]{
    \typing{\Gamma}{\mu}[\transso[\Delta]{\sigma}{\tau}] \\
    \typing{\Delta}{A} \\
    \typing{\Delta}{t}[A]
  }{
    \conv{\Gamma}{\ad{\subs{t}{\sigma}}{\trans{A}{\mu}}}{\subs{t}{\tau}}[\subs{A}{\tau}]
  }
  \label{rule:tm-trans-intro}
\end{mathpar}
In particular, the equality constraint required in \ruleref{rule:trans-tm-intro}
can be exactly recovered by specializing \ruleref{rule:tm-trans-intro}
to \(t \defconvop \tmvz\), so the rules are in harmony.

An interesting example that uses term variables is the identity type, given by
% \[
% \inferdef[Id]{ }{
%   \typing{
%     \LaTeXunderbrace{(X \ty \Ty) \ext (x \ty X) \ext (y \ty X)}_{\Gamma_{\Id}}
%   }{
%     \subs{\Id}{X \ext x \ext y}
%   }
% }
% \]
\[\Gamma_{\Id} \coloneq (X \ty \Ty) \ext (x \ty X) \ext (y \ty X)\]
We can compute that a substitution \(\typing{\Delta}{\sigma}[\Gamma_{\Id}]\)
consists of a type \(\typing{\Delta}{\subs{X}{\sigma}}\)
and two of its inhabitants, while transformations between
these correspond to an adapter between the types preserving the terms.
In other words, we obtain the following rule:
\[
 \inferrule*{
    \typing{\Gamma}{A, B} \\
    \typing{\Gamma}{a, a'}[A] \\
    \typing{\Gamma}{f}[\adso{A}{B}]
  }{
    \typing{\Gamma}{\trans{\Id}{f \ext a \ext a'}}
    [\adso{\subs{\Id}{A \ext a \ext a'}}
    {\subs{\Id}{B \ext \ad{a}{f} \ext \ad{a'}{f}}}]
  }
\]
We recover the complex-looking typing rule of \MLTTmap \cite{Subtyping2024},
completely mechanically, simply from the description \(\Gamma_{\Id}\).
Furthermore, the computation rule we derive for it in \cref{sec:inductive-types} is
\[\ad{(\refl~a)}{\trans{\Id}{f \ext a \ext a}} \convop \refl~(\ad{a}{f})\]
which also agrees with the one of \MLTTmap.

\subsubsection{Functions and trees: contravariance}
\label{sec:tyvar-contra}

Type variables as just introduced are not
quite enough to adequately capture all interesting examples.
First, as seen in the case of \(\P\), we need a form of contravariance:
adapters for the domain need to go in the direction
opposite to that of the codomain.
To describe this, we annotate type variables in contexts
with a \emph{direction} \(\Dir \coloneq + \mid -\).%
% \footnote{We use “direction” instead of “variance” for this sort to avoid a conflict with
% a sort of variables \(\Var\).}
\begin{mathpar}
    \inferdef{\ctxty{\Gamma}}{\ctxty{\Gamma \ext* (X \ty \Ty_{d})}}[\(d : \Dir\)]
      \and
    \inferdef{\ctxty{\Gamma}}{\typing{\Gamma \ext* (X \ty \Ty_{+})}{X}}
\end{mathpar}
Type variables from \cref{sec:tyvar-simple} correspond to the \(+\) direction.
Crucially, we still only allow access to covariant variables:
since a type \(\typing{\Gamma}{A}\) is implicitly always covariant with
respect to its context, it does not make sense to directly access contravariant
variables.

To nonetheless be able to use contravariant variables,
and more generally describe contravariant
types, we introduce an operation on contexts called \emph{dualization},
written \(\dual{\cdot}{-}\). It reverses
variables' direction, \eg
\(\dual{((X \ty \Ty_{-}) \ext* (Y \ty \Ty_{+}))}{-} \convop (X \ty \Ty_{+})
\ext* (Y \ty \Ty_{-})\).
Combined with the rule for (covariant) variable,
this lets us \eg derive
\(\typing{\dual{((X \ty \Ty_{-}) \ext* (Y \ty \Ty_{+}))}{-}}{X}\).
A contravariant type in \(\Gamma\) is then simply a type in \(\dual{\Gamma}{-}\).
% Accordingly, we write \(\Ty_{-}(\Gamma)\) for \(\Ty(\dual{\Gamma}{-})\).
%
Equipped with dualization, we extend the rules for substitutions and transformations:
we can substitute a contravariant type for a contravariant type variable but,
more interestingly, the direction of adapters for a contravariant
type variable is \emph{reversed}. That is, to construct a transformation
\(\mu \ty \transso[\Delta \ext* (X : \Ty_{-})]{\sigma \ext* A}{\tau \ext* B}\), we must provide an
adapter \(f \ty \adso{B}{A}\).

\begin{mathpar}
  \inferdef{
    \typing{\Delta}{\sigma}[\Gamma] \\ \typing{\dual{\Delta}{-}}{A}}{
    \typing{\Delta}{\sigma \ext* A}[\Gamma \ext* (X \ty \Ty_{-})]
  } \and
  \inferdef{
    % \Gamma \vdash \sigma, \tau \ty \Gamma \\
    \typing{\Gamma}{\mu}[\transso[\Delta]{\sigma}{\tau}] \\
    \typing{\dual{\Gamma}{-}}{A, B} \\
    \typing{\dual{\Gamma}{-}}{f}[\adso{B}{A}]
  }{
    \typing{\Gamma}{\mu \ext* f}[\transso[\Delta \ext* (X : \Ty_{-})]{\sigma \ext* A}{\tau \ext* B}]
  }
\end{mathpar}
We also allow term variables of contravariant type, although as for type
variables we only allow access to those of covariant type, the other
must be accessed via dualization.
\[
\inferdef{
    \typing{\dual{\Gamma}{-}}{A}
  }{
    \ctxty{\Gamma \ext_{-} (x \ty A)}
  }
\]

This lets us construct the context
\(\Gamma_{\to} \coloneq (X \ty \Ty_{-}) \ext* (Y \ty \Ty_{+})\) of the non-dependent
function type, and derive the corresponding adapter typing rule.
\[
 \inferrule*{
    \typing{\dual{\Gamma}{-}}{A, A'} \\
    \typing{\Gamma}{B, B'} \\
    \typing{\dual{\Gamma}{-}}{a}[\adso{A'}{A}] \\
    \typing{\Gamma}{b}[\adso{B}{B'}] \\
  }{
    \typing{\Gamma}{\trans{\mathord{\to}}{a \ext* b}}[\adso{A \to B}{A' \to B'}]
  }
\]
We can give it the computation rule
\((\ad{f}{\trans{\mathord{\to}}{a \ext* b}})\ u \convop \ad{(f\ \ad{u}{a})}{b}\),
a simpler variant of \cref{ex:pi-adaptt}.

Another example is the type \(\Tree\) of \(Y\)-branching trees
with \(X\)-storing nodes, with context of parameters
\(\Gamma_{\Tree} \coloneq (X \ty \Ty_{+}) \ext* (Y \ty \Ty_{-})\) and
constructors and equations as follows (where \(f \ty \adso{X}{X'}\) and
\(g \ty \adso{Y}{Y'}\))
\begin{mathpar}
\small
  \inferrule{
    % \typing{\Gamma}{A} \\
    % \typing{\dual{\Gamma}{-}}{B}
  }{
    \typing{(X : \Ty_{+}) \ext* (Y : \Ty_{-})}{\const{leaf}}[\Tree]
  }
  \and
  \inferrule{
    % \typing{\Gamma}{a}[A] \\
    % \typing{\Gamma \ext x : B}{r}[\subs{\Tree}{A \ext* B}]
  }{
    \typing{(X : \Ty_{+}) \ext* (Y : \Ty_{-}) \ext (x : X) \ext (r : Y \to \Tree)}
    {\const{node}}[\subs{\Tree}{X \ext* Y}]
  }
  \\
  \ad{\subs{\const{leaf}}{X \ext* Y}}{\trans{\Tree}{f \ext* g}}
  \convop \subs{\const{leaf}}{X' \ext* Y'} \and
  \ad{\subs{\const{node}}{X \ext* Y \ext x \ext r}}{\trans{\Tree}{f \ext* g}}
  \convop \subs{\const{node}}{X' \ext* Y' \ext
    \ad{x}{f} \ext \ad{r}{\trans{Y \to \Tree}{f \ext* g}}}
\end{mathpar}
By η-expansion, we can further compute that
\(\ad{r}{\trans{Y \to \Tree}{f \ext* g}} \convop \l x.~
  \ad{\subs{r}{\ad{x}{g}}}{\trans{\Tree}{f \ext* g}}\).
Again this is an inductive type whose computation rules we can automatically derive.
As for functions, we see the need for the contravariant adapter \(g\)
which acts by “pre-composition” for \(\const{node}\).

\subsubsection{\(\Sig\): dependency}
\label{sec:tyvar-dep}

So far, all our examples have been non-dependent. But if we want to capture
dependent function or pair types, the type variables as introduced for now
are not sufficient. %, as they lack a way to represent dependency.
To gain the ability to represent dependency, we further generalize them
to bind a term-level variable.
For simplicity, we allow only a single binding for now,
but generalize to telescopes of types in \cref{sec:adaptt2-def}.
Our new rules for context extension and variables are now
\begin{mathpar}
  \inferdef{\typing{\Gamma}{A}}{
    \ctxty{\Gamma \ext* (X : \tyext{A})}}
  \and
  \inferdef{\typing{\Gamma}{A}}{
    \typing{(\Gamma \ext* (X : \tyext{A})) \ext (x : A)}{X}}
  \and
  \inferdef{
    \typing{\Gamma}{\sigma}[\Delta] \\ \typing{\Delta}{A} \\
    \typing{\Gamma \ext \subs{A}{\sigma}}{B}}
    {\typing{\Gamma}{\sigma \ext* B}[\Gamma \ext* (X : \tyext{A})]
  } \and
  \inferdef{
    % \Gamma \vdash \sigma, \tau \ty \Delta \\
    \typing{\Gamma}{\mu}[\transso[\Delta]{\sigma}{\tau}] \\
    \typing{\Delta}{A} \\
    \typing{\Gamma \ext (x : \subs{A}{\sigma})}{f}[\adso{B}{
        \subs{B'}{\id \ext \ad{x}{\trans{A}{\mu}}}}]
  }{
    \typing{\Gamma}{\mu \ext* f}[\transso[\Delta \ext* (X : \tyext{A})]
      {\sigma \ext* B}{\tau \ext* B'}]
  }
\end{mathpar}
Accessing the type variables gives a type with access to an extra binder,
and accordingly in a substitution that type variable must be replaced by one
allowed to use this extra binder. Finally, transformations must be extended by an
adapter between the dependent types, suitably lying over the base transformation.
For instance, the context for the dependent pair type is
\(\Gamma_{\Sig} \coloneq (X: \Ty_{+}) \ext* (Y: X.\Ty_{+})\) where
\(Y\) depends on a term of type \(X\).
Once \ruleref{rule:ty-trans} is unfolded, its adapter is typed as follows
\[
 \inferdef[AdPairEx]
 {
    \typing{\Delta}{A, A'} \\
      \typing{\Delta}{a}[\adso{A}{A'}] \\\\
    \typing{\Delta \ext (x : A)}{B} \\
    \typing{\Delta \ext (x : A')}{B'} \\
    \typing{\Delta \ext (x : A)}{b}[\adso{B}{\subs{B'}{\id \ext \ad{x}{a}}}]
  }{
    \typing{\Delta}{\trans{\Sig}{a \ext* b}}[\adso{\Sig (x : A). B}{\Sig (x : A'). B'}]
  }
\label{rule:ad-pair-dep-ex}
\]

\paragraph{\(\W\) and \(\P\): putting it all together}

As examples which combine both contravariance and dependency, we can look at
 the type \(\W\)
of well-founded trees, given by
\(\Gamma_{\W} \coloneq (X: \Ty_{+}) \ext* (Y: X.\Ty_{-})\),
and the one of dependent functions
\(\Gamma_{\P} \coloneq (X: \Ty_{-}) \ext* (Y: X.\Ty_{+})\),
for which we also need to allow the type of variables bound by a type variable
to be contravariant. We defer the general rules for variables for a moment,
but the derived typing rules for adapters in these examples are as follows:
\begin{mathpar}
\inferdef[AdWEx]{
    \typing{\Delta}{A, A'} \\
    \typing{\Delta}{a}[\adso{A}{A'}] \\\\
    \typing{\dual{(\Delta \ext (x : A))}{-}}{B} \\
    \typing{\dual{(\Delta \ext (x : A'))}{-}}{B'} \\
    \typing{\dual{(\Delta \ext (x : A))}{-}}{b}[\adso{\subs{B'}{\id \ext \ad{x}{a}}}{B}]
  }{
    \typing{\Delta}{\trans{\W}{a \ext* b}}[\adso{\W (x : A). B}{\W (x : A'). B'}]
  } \and
 \inferdef[AdFunEx]
 {
    \typing{\dual{\Delta}{-}}{A, A'} \\
      \typing{\dual{\Delta}{-}}{a}[\adso{A'}{A}] \\\\
    \typing{\Delta \ext (x : A)}{B} \\
    \typing{\Delta \ext (x : A')}{B'} \\
    \typing{\Delta \ext (x : A')}{b}[\adso{\subs{B}{\id \ext \ad{x}{a}}}{B'}]
  }{
    \typing{\Delta}{\trans{\P}{a \ext* b}}[\adso{\P (x : A). B}{\P (x : A'). B'}]
  }
\label{rule:ad-fun-dep-ex}
\end{mathpar}
As we can see, there are subtle but important discrepancies in the types
of the “dependent adapter” between \(\Sig\), \(\W\) and \(\P\), and
the judgmental structure offered by \AdapTTt is leveraged in correctly
handling these discrepancies.

\subsection{A 2-Dimensional Type Theory with Adapters}
\label{sec:adaptt2-def}

We are now finally in the position to give the official rules of \AdapTTt.
%
% \paragraph{Methodology}
The main judgments are collected in \cref{fig:adapttt-sort}.
Formally, we understand these
as describing a generalized algebraic theory (GAT)
\cite{Cartmell1986,Bezem2021}, whose sorts we also collect in
\cref{fig:adapttt-sort}. We use both interchangeably, \ie write
either \(\sigma \ty \Sub(\Gamma,\Delta)\) or \(\typing{\Gamma}{\sigma}[\Delta]\).
In premises, we often omit variables appearing in later objects, and whose
sort can be inferred.
The full rules are collected in \cref{sec:full-rules},
and have been mostly type checked in \Agda as
postulated symbols and equations, using rewrite rules
to emulate an extensional type theory as meta-theory.

\paragraph{Basic judgmental structure}

\begin{figure}
\begin{small}
\begin{mathpar}
  % \inferdef{ }{\Dir} \and
  \inferdef%[Ctx]
    { }{\Ctx} \and
  \inferdef%[Sub]
    {\Gamma, \Delta \ty \Ctx}{\Sub(\Gamma,\Delta)} \and
  \inferdef%[Trans]
    {\impl{\Gamma, \Delta \ty \Ctx} \\\\
    \sigma, \tau \ty \Sub(\Gamma,\Delta)}{
      \Trans(\impl{\Gamma},\impl{\Delta},\sigma,\tau)} \and
  \inferdef%[Ty]
  {\Gamma \ty \Ctx}{\Ty(\Gamma)} \and
  \inferdef%[Ad]
  {\impl{\Gamma \ty \Ctx} \\\\
    A, B \ty \Ty(\Gamma)}{
      \Ad(\impl{\Gamma},A,B)
    } \and
  \inferdef%[Tm]
  {\impl{\Gamma \ty \Ctx} \\\\ A \ty \Ty(\Gamma)}{\Tm(\Gamma,A)} \\
  \ctxty{\Box} \and
  \typing{\Gamma}{\Box}[\Delta] \and
  \typing{\Gamma}{\Box}[\transso[\Delta]{\sigma}{\tau}] \and
  \typing{\Gamma}{\Box} \and
  \typing{\Gamma}{\Box}[\adso{A}{B}] \and
  \typing{\Gamma}{\Box}[A]
  \vspace*{-.5em} \\
  \rule{0.9\textwidth}{.5pt}
  \vspace*{-.5em} \\
  % \inferdef%[Id1]
  %   {\ctxty{\Gamma}}{\typing{\Gamma}{\id_{\Gamma}}[\Gamma]} \and
  % \inferdef%[Comp]
  %   {%\Gamma, \Delta, \Xi \ty \Ctx \\
  %     \typing{\Delta}{\tau}[\Xi] \\
  %     \typing{\Gamma}{\sigma}[\Delta]
  %   }{\typing{\Gamma}{\tau \circ \sigma}[\Xi]} \and
  \inferdef[TransId]
  {
    \typing{\Gamma}{\sigma}[\Delta]
  }{\typing{\Gamma}{\id}[\transso[\Delta]{\sigma}{\sigma}]} \and
  % \inferdef%[Comp0]
  %   {%\Gamma, \Delta, \Xi \ty \Ctx \\
  %   \typing{\Gamma}{\mu}[\transso[\Delta]{\sigma}{\sigma'}] \\
  %   \typing{\Delta}{\nu}[\transso[\Xi]{\tau}{\tau'}] \\
  %   }{
  %     \typing{\Gamma}{\nu \circ_{0} \mu}
  %     [\transso[\Xi]{\tau \circ \sigma}{\tau'\circ\sigma'}]} \and
  \inferdef[TransComp]
    {%\Gamma, \Delta, \Xi \ty \Ctx \\
    \typing{\Gamma}{\mu}[\transso[\Delta]{\rho}{\sigma}] \\
    \typing{\Gamma}{\nu}[\transso[\Delta]{\sigma}{\tau}] \\
    }{
      \typing{\Gamma}{\nu \vcomp \mu}[\transso[\Delta]{\rho}{\tau}]} \\
  \inferdef[TransWhiskerLeft]
    {%\Gamma, \Delta, \Xi \ty \Ctx \\
    \typing{\Delta}{\tau}[\Xi] \\
    \typing{\Gamma}{\mu}[\transso[\Delta]{\sigma}{\sigma'}] \\
    }{
      \typing{\Gamma}{\tau \whisk \mu}
      [\transso[\Xi]{(\tau \circ \sigma)}{(\tau \circ \sigma')}]} \and
  \inferdef[TransWhiskerRight]
    {%\Gamma, \Delta, \Xi \ty \Ctx \\
    \typing{\Delta}{\nu}[\transso[\Xi]{\tau}{\tau'}] \\
    \typing{\Gamma}{\sigma}[\Delta] \\
    }{
      \typing{\Gamma}{\nu \whisk \sigma}
      [\transso[\Xi]{(\tau \circ \sigma)}{(\tau'\circ\sigma)}]} \\
  %
  % \inferdef%[IdAd]
  %   {%\Gamma \ty \Ctx \\
  %   \typing{\Gamma}{A} \\
  %   }{\typing{\Gamma}{\id_{A}}[\adso{A}{A}]} \and
  % \inferdef%[CompAd]
  %   {%\Gamma \ty \Ctx \\
  %     % \typing{\Gamma}{A,B,C}[\Ty] \\
  %   \typing{\Gamma}{g}[\adso{B}{C}] \\
  %   \typing{\Gamma}{f}[\adso{A}{B}]
  %   }{\typing{\Gamma}{g \circ f}[\adso{A}{C}]} \and
  % \inferdef%[AdTm]
  %   {%\Gamma \ty \Ctx \\
  %   % \typing{\Gamma}{A,B}[\Ty] \\
  %   \typing{\Gamma}{f}[\adso{A}{B}] \\
  %   \typing{\Gamma}{t}[A]
  %   }{\typing{\Gamma}{\ad{t}{f}}[B]} \\
  %
  % \inferdef%[TySub]
  % {%\Gamma, \Delta \ty \Ctx \\
  %   \typing{\Delta}{A}[\Ty] \\
  %   \typing{\Gamma}{\sigma}[\Delta] \\
  %   }{\typing{\Gamma}{\subs{A}{\sigma}}} \and
  % \inferdef%[AdSub]
  %   {%\Gamma, \Delta \ty \Ctx \\
  %   % \typing{\Delta}{A,B}[\Ty] \\
  %     \typing{\Delta}{f}[\adso{A}{B}]\\
  %     \typing{\Gamma}{\sigma}[\Delta]
  %   }{\typing{\Gamma}{\subs{f}{\sigma}}[\adso{\subs{A}{\sigma}}{\subs{B}{\sigma}}]} \and
  % \inferdef%[TmSub]
  % {%\Gamma, \Delta \ty \Ctx \\
  %    \typing{\Delta}{t}[A] \\
  %   \typing{\Gamma}{\sigma}[\Delta]
  %   % \typing{\Delta}{A,B}[\Ty] \\
  %   }{\typing{\Gamma}{\subs{t}{\sigma}}[\subs{A}{\sigma}]} \\
  %
  \inferdeft[TransTy]
    {%\Gamma, \Delta \ty \Ctx \\
    % \sigma,\tau \ty \Sub(\Gamma,\Delta) \\
      \typing{\Delta}{A} \\
      \typing{\Gamma}{\mu}[\transso[\Delta]{\sigma}{\tau}] \\
    }{\typing{\Gamma}{\trans{A}{\mu}}[\adso{\subs{A}{\sigma}}{\subs{A}{\tau}}]} \and
  \inferdeft[TransTm]
    {%\Gamma, \Delta \ty \Ctx \\
    % \sigma,\tau \ty \Sub(\Gamma,\Delta) \\
      \typing{\Delta}{t}[A] \\
      \typing{\Gamma}{\mu}[\transso[\Delta]{\sigma}{\tau}]
    }{\conv{\Gamma}{
        \ad{\subs{t}{\sigma}}{\trans{A}{\mu}}
      }{\subs{t}{\tau}
      }[\subs{A}{\tau}]
    }
  \label{rule:tm-trans} \and
  \inferdeft[TransAd]
  {%\Gamma, \Delta \ty \Ctx \\
  % \sigma,\tau \ty \Sub(\Gamma,\Delta) \\
    \typing{\Delta}{f}[\adso{A}{B}] \\
    \typing{\Gamma}{\mu}[\transso[\Delta]{\sigma}{\tau}]
  }{\conv{\Gamma}{
      \trans{B}{\mu} \whisk \subs{f}{\sigma}
    }{\subs{f}{\tau} \whisk \trans{A}{\mu}
    }[\adso{\subs{A}{\sigma}}{\subs{B}{\tau}}]
  } \\
  % \inferdef{ }{\ctxty{\emp}} \and
  % \inferdef{\ctxty{\Gamma}}{\typing{\Gamma}{\emp_{\impl{\Gamma}}}[\emp]} \and
  % \inferdef{ }{+ \ty \Dir} \and \inferdef{ }{- \ty \Dir} \and
  % \inferdef{d, d' \ty \Dir}{d\ d' \ty \Dir} \\
  % \inferdef{ }{+ + \convop + \convop - - \\ + - \convop - \convop - +} \\
  \inferdef[CtxDual]
  {
    % d \ty \Dir \\
    \ctxty{\Gamma}}{\ctxty{\dual{\Gamma}{-}}} \and
  \inferdef[SubDual]
  {
    % d \ty \Dir \\
    % \impl{\Gamma, \Delta \ty \Ctx} \\
    \typing{\Gamma}{\sigma}[\Delta]}{
    \typing{\dual{\Gamma}{-}}{\dual{\sigma}{-}}[\dual{\Delta}{-}]} \and
  \inferdef[TransDual]
  {
    % \impl{\Gamma, \Delta \ty \Ctx} \\
    % \impl{\sigma, \tau \ty \Sub(\Gamma,\Delta)} \\
    \typing{\Gamma}{\mu}[\transso[\Delta]{\sigma}{\tau}] \\
  }{
    \typing{\dual{\Gamma}{-}}{\dual{\mu}{-}}
      [\transso[\dual{\Delta}{-}]{\dual{\tau}{-}}{\dual{\sigma}{-}}]
  } \and
  %
  % \inferdef{\Gamma \ty \Ctx \\ d \ty \Dir}{
  %   \Ty_{d}(\Gamma) \defconvop \Ty(\Gamma^d)} \and
  % \inferdef{\Gamma \ty \Ctx \\ d \ty \Dir \\ A \ty \Ty_{d}(\Gamma)}{
  %   \Tm_{d}(\Gamma,A) \defconvop \Tm(\dual{\Gamma}{d},A)} \and
  % \inferdef{\Gamma \ty \Ctx \\ A,B \ty \Ty_{+}(\Gamma)}{
  %   \Ad_{+}(\Gamma,A,B) \defconvop \Ad(\Gamma,A,B)} \and
  % \inferdef{\Gamma \ty \Ctx \\ A,B \ty \Ty_{-}(\Gamma)}{
  %   \Ad_{-}(\Gamma,A,B) \defconvop \Ad(\dual{\Gamma}{-},B,A)} \and
  %
  % \inferdef%[EmpCtxDual]
  % { }{\dual{\emp}{-} \convop \emp \ty \Ctx} \and
  %   \inferdef%[EmpSubsDual]
  % {\Gamma \ty \Ctx}{
  %   \dual{(\emp_{\impl{\dual{\Gamma}{-}}})}{-} \convop \emp_{\Gamma} \ty \Sub(\Gamma,\emp)}
\end{mathpar}
\end{small}

\Description{Basic judgment forms of \AdapTTt
  (those of \AdapTT, plus a new one for transformations),
  rules for transformations (they form a category, and
  act on terms and types), and rules for duality, which acts on substitutions
  and transformations.}
\caption{Basic structure of \AdapTTt (excerpt)}
\label{fig:adapttt-sort}
\end{figure}

The core judgments of \AdapTTt, given on top of \cref{fig:adapttt-sort},
extend those of \AdapTT with
transformations, which relates two substitutions.
We also use the group of directions \(\Dir\) with elements \(+\) and \(-\),
isomorphic to \(\mathbb{Z}/2\mathbb{Z}\).

The base operations of \AdapTTt
are collected in the rest of \cref{fig:adapttt-sort}.
We omit the empty context \(\emp\)
and unique substitution \(\typing{\Gamma}{\emp_{\impl{\Gamma}}}[\emp]\), similar
to \AdapTT.
We can compose transformations,
% (written \(\vcomp\), as it is the composition along 1-morphisms),
and left and right whisker a
transformation by a substitution, which respectively generalize the
composition of transformations,
action of transformations on terms/types, and of substitution on adapters.
And indeed, the former compute like componentwise versions of the latter.
% are componentwise:
% composition of transformations derives from composition of adapters, left whiskering
% from the action of a substitution on adapters (\nameref{rule:adaptt-sub-ad}),
% and right whiskering from the action of transformations (\nameref{rule:ty-trans}).
Categorically, these operations and their equations amount to saying that
\(\Ctx\) is a 2-category with \(\Sub\) and \(\Trans\) respectively as 1- and 2-morphisms.

As in \AdapTT, \(\Ty(\Gamma)\) also forms a category (with \(\Ad\) as morphisms),
but the novelty is that these categories assemble in a \emph{2}-functor \(\Ty\).
That is, given \(\sigma \ty \Sub(\Gamma,\Delta)\), substitution
\(\subs{\cdot}{\sigma}\) is a functor between \(\Ty(\Delta)\) and \(\Ty(\Gamma)\)
(as in \AdapTT), but moreover \(\trans{\cdot}{\mu}\) is a natural transformation
between the functors corresponding to its endpoints.
Similarly, \(\Tm\) is a “dependent 2-functor”,
albeit a degenerate one:
the action of a transformation on a term is a definitional equality \nameref{rule:tm-trans}.

% \begin{figure}
% \begin{mathpar}
%   \inferdef%[EmpCtx]
%   { }{\emp \ty \Ctx} \and
%   \inferdef%[EmpSubs]
%   { }{\emp_{\impl{\Gamma}} \ty \Sub(\Gamma,\emp)} \and
%   \inferdef%[EmpSubsEq]
%   {\sigma \ty \Sub(\Gamma,\emp)}{\sigma \convop \emp \ty \Sub(\Gamma,\emp)} \and
%   \inferdef%[TransSubsEq]
%   {\mu \ty \Trans(\impl{\Gamma,\emp,}\emp,\emp)}
%   {\mu \convop \id_{\emp} \ty \Trans(\emp,\emp)} \\
%   % \inferdef[EmpCtxDual]{ }{\dual{\emp}{-} \convop \emp \ty \Ctx} \and
%   % \inferdef[EmpSubsDual]{\Gamma \ty \Ctx}{
%   %   \dual{\emp_{\impl{\dual{\Gamma}{-}}}}{-} \convop \emp \ty \Sub(\Gamma,\emp)}
% \end{mathpar}
% \caption{Empty context \mlb{Should we omit this?}}
% \label{fig:adapttt-empty}
% \end{figure}

Finally, we have the dualization operation \(\dual{\cdot}{-}\),
which lets us represent objects that depend contravariantly on the context.
Dualization also acts on substitutions and transformations, essentially as the identity,
although it formally reverses the direction of the latter.
The empty context and substitutions are mapped to themselves:
% \begin{mathpar}
  \(\dual{\emp}{-} \convop \emp\) and
  \(\dual{(\emp_{\impl{\dual{\Gamma}{-}}})}{-} \convop \emp_{\Gamma}\).
Given a direction \(d \ty \Dir\), we write \(\dual{(\cdot)}{d}\)
with the convention that \(\dual{(\cdot)}{-}\) is the one of \cref{fig:adapttt-sort}
and \(\dual{(\cdot)}{+}\) is the identity. As \(\dual{(\cdot)}{d d'}\) coincides with
\(\dual{(\dual{(\cdot)}{d})}{d'}\), this defines a group action of directions on contexts. Since directions are a meta-level set,
a rule involving direction variables should
be formally understood as a family of rules, one for each possible
concrete combination of directions.

% We extend this for all sorts, \eg write \(\Ty_{d}(\Gamma)\) for contravariant types.
% This is merely a shortcut, but does not mean
% that the sort of types in the generalized algebraic theory
% is indexed by direction. That way, we can specify rules that do not need to specifically
% accommodate contravariance only for covariant types, and derive the one for contravariant
% types by applying them to some \(\dual{\Gamma}{-}\).

\begin{figure}
\begin{small}

\begin{mathpar}
  \inferdeft[CtxExtTm]
  {
    \ctxty{\Gamma} \\
    \typing{\dual{\Gamma}{d}}{A}
  }{
    \ctxty{\Gamma \ext_{\impl{d}} A}
  }[\(d:\Dir\)]
   \and
  \inferdeft[SubExtTm]
  {
    % \impl{\Gamma, \Delta \ty \Ctx} \\
    \typing{\Gamma}{\sigma}[\Delta] \\
    \impl{\typing{\dual{\Delta}{d}}{A}} \\
    \typing{\dual{\Gamma}{d}}{t}[\subs{A}{\dual{\sigma}{d}}]
  }{\typing{\Gamma}{\sigma \ext_{\impl{d}} t}[\Delta \ext_{\impl{d}} A]
  }[\(d:\Dir\)] \and
  \inferdeft[TransTm+]{
    % \impl{\Gamma, \Delta \ty \Ctx} \\
    % \impl{\sigma, \tau \ty \Sub(\Gamma,\Delta)} \\
    \typing{\Gamma}{\mu}[\transso[\Delta]{\sigma}{\tau}] \\
    % \impl{A \ty \Ty(\Delta)} \\
    \typing{\Gamma}{t}[\subs{A}{\sigma}]}
  {
    \typing{\Gamma}{\mu \ext_{+} t}[\transso%[\Delta \ext_{+} A]
      {\sigma \ext_{+} t}{\tau \ext_{+} \had{t}{\trans{A}{\mu}}}]
  } \label{rule:trans-tm-plus} \and
  \inferdeft[TransTm-]{
    % \impl{\Gamma, \Delta \ty \Ctx} \\
    % \impl{\sigma, \tau \ty \Sub(\Gamma,\Delta)} \\
    \typing{\Gamma}{\mu}[\transso[\Delta]{\sigma}{\tau}] \\
    % \impl{A \ty \Ty_{-}(\Delta)} \\
    \typing{\dual{\Gamma}{-}}{t}[\subs{A}{\dual{\tau}{-}}]
  }{
    \typing{\Gamma}{\mu \ext_{-} t}[\transso%[\Delta \ext_{-} A]
      {\sigma \ext_{-} \had{t}{\trans{A}{\dual{\mu}{-}}}}{\tau \ext_{-} t}]
  } \label{rule:trans-tm-min} \and
  \inferdeft[WkTm]
  {
    \ctxty{\Gamma} \\
    \typing{\dual{\Gamma}{d}}{A}
  }{
    \typing{\Gamma \ext_{\impl{d}} A}{\wk_{A}}[\Gamma]
  }[\(d : \Dir\)]
  \and
  \inferdeft[VarZTm]{
    \impl{\ctxty{\Gamma}} \\ \impl{\typing{\Gamma}{A}}
  }{\typing{\Gamma \ext_{+} A}{\tmvz_{\impl{A}}}[\subs{A}{\wk}]}
  \label{rule:vartm-z} \and
  \inferdeft[TransTl+]{
    % \impl{\Gamma, \Delta \ty \Ctx} \\
    \typing{\Gamma}{\mu}[\transso[\Delta]{\sigma}{\tau}] \\
    % \impl{A \ty \Ty(\Delta)} \\
    \typing{\Gamma}{t}[\subs{A}{\sigma}]
  }{\conv{\Gamma}{\wk \circ (\mu \ext_{+} t)}{\mu}[\transso[\Delta]{\sigma}{\tau}]}
  \label{rule:trans-tl+} \and
    \inferdeft[TransTl-]{
    % \impl{\Gamma, \Delta \ty \Ctx} \\
    \typing{\Gamma}{\mu}[\transso[\Delta]{\sigma}{\tau}] \\
    % \impl{A \ty \Ty(\Delta)} \\
    \typing{\dual{\Gamma}{-}}{t}[\subs{A}{\dual{\tau}{-}}]
  }{\conv{\Gamma}{\wk \circ (\mu \ext_{-} t)}{\mu}[\transso[\Delta]{\sigma}{\tau}]}
  \label{rule:trans-tl-} \and
  \inferdeft[TransEta]{
    % \impl{\Gamma, \Delta \ty \Ctx} \\
    % \impl{A \ty \Ty(\Delta)} \\
    % \impl{\sigma, \tau \ty \Sub(\Gamma,\Delta \ext A)} \\
    \typing{\Gamma}{\mu}[\transso[\Delta\ext_+A]{\sigma}{\tau}] \\
  }{
    \conv{\Gamma}{\mu}{(\wk \circ \mu) \ext_+ \subs{\tmvz}{\sigma}}[\transso[\Delta\ext_+A]{\sigma}{\tau}]}
  \label{rule:trans-eta}
\end{mathpar}
\end{small}

\Description{Contexts can be extended by term variables, and substitutions and
  transformations can accordingly be extended by terms. For the latter, a term
  corresponds to a definitional constraint between the substitutions.}
\caption{Term variables in \AdapTTt (equations for \(\sigma \ext t\)
  and \(\dual{\cdot}{-}\) omitted)}
\label{fig:adapttt-tmvar}
\end{figure}

\paragraph{Term variables}
Term variables (\cref{fig:adapttt-tmvar})
behave broadly similarly to those of \AdapTT, with two differences.
% , although
% in the formalization\km{first mention ?} we have a separate sort of variables
% representing terms of the form \(\subs{\tmvz}{\wk^n}\),
% or, alternatively, de Bruijn variables, closer to what one would implement
% in practice. This poses no particular extra difficulties.
%
The first is the ability to bind variables of contravariant type, a
flexibility used to represent \eg \(\P\)-types, as explained in \cref{sec:tyvar-dep}.
Note, however, that we only allow accessing covariant variables in \ruleref{rule:vartm-z},
contravariant variables have to be accessed indirectly, as we can derive
\(\typing{\dual{(\Gamma \ext_{-} A)}{-}}{\tmvz_{\impl{A}}}[\subs{A}{\wk}]\)
% \(\typing{\dual{(\Gamma \ext_{-} A)}{-}}{\tmvz_{\impl{A}}}[\Tm(\subs{A}{\wk})]\)
from \ruleref{rule:vartm-z} and the equation \(\dual{(\Gamma \ext_{-} A)}{-} \convop \dual{\Gamma}{-}
\ext_{+} A\).

The second is the extension of transformations, which we touched upon in
\cref{sec:tyvar-eq}. Here we have two rules, one for each direction,
which need a different adjustment of the endpoints. Rules
\nameref{rule:trans-tm-plus} and \nameref{rule:trans-tm-min}
are more economical albeit less symmetric versions of \ruleref{rule:trans-tm-intro},
where we substitute \(u\) by \(\ad{t}{\trans{A}{\mu}}\) to which it must be equal.

\begin{comment}
  \km{Commenting out: first mention of local representability, does not help much with intuition (any reader that understand local representability has read Awodey and probably understand that point)}
For the categorically-minded reader, this entire structure ---for the \(+\) direction---
amounts to the familiar condition
of local representability, which captures variables in a natural model, but used in
\(\Cat\)-valued presheaves ---and with the strictest possible notion of pullback.
\mlb{Check this.}\mlb{Say something about \(-\)?}
\end{comment}

Dualization acts on the primitives of \cref{fig:adapttt-tmvar}, adjusting variance.
For instance, we have
\[\conv{\dual{\Gamma}{-}}{
  \dual{\left(\sigma \ext_{\impl{d}} t\right)}{-}}{
  \dual{\sigma}{-} \ext_{\impl{-d}} t}[
      \dual{\left(\Delta \ext_{\impl{d}} A\right)}{-}]\]
which is well-typed as
\(\dual{\left(\Delta \ext_{\impl{d}} A\right)}{-} \convop \dual{\Delta}{-} \ext_{-d} A\)
and \(\subs{A}{\dual{(\dual{\sigma}{-})}{-}} \convop \subs{A}{\sigma}\).

\paragraph{Functoriality of context extension}
From the primitives of \cref{fig:adapttt-tmvar}, we can derive an operation
\(\pext\), which categorically corresponds to the functoriality of context extension,
and its equations
\begin{mathpar}
\inferdef[SubAd]{
  % \impl{\Gamma,\Delta \ty \Ctx} \\
  \typing{\Gamma}{\sigma}[\Delta] \\
  \typing{\dual{\Gamma}{d}}{A} \\
  \typing{\dual{\Delta}{d}}{B} \\
  \typing{\dual{\Gamma}{d}}{a}[\adso{A}{\subs{B}{\dual{\sigma}{d}}}]
}{
  \defconv{\Gamma \ext_{d} A}{\sigma \pext_{d} a}
    {\dual{\left((\dual{\sigma}{d} \circ \wk) \ext \ad{\tmvz}{a[\wk]}\right)}{d}}
    [\Delta \ext_{d} B]
} \\
  (\id_{\Gamma} \pext_{d} \id_{A}) \convop \id_{\Gamma \ext A} \quad
  (\tau \pext_{d} b) \circ (\sigma \pext_{d} a) \convop
    (\tau \circ \sigma) \pext_{d} (\subs{b}{\sigma} \circ a) \quad
  (\tau \pext_{d} a) \circ (\sigma \ext_{d} t) \convop
  (\tau \circ \sigma) \ext_{d} \ad{t}{\subs{a}{\sigma}}
\end{mathpar}

\begin{figure}
\begin{small}
\begin{mathpar}
  \inferdef%[Tel]
    {\Gamma \ty \Ctx}{\Tel(\Gamma)} \and
  \inferdef%[TelAd]
    {\impl{\Gamma \ty \Ctx} \\
    \Theta, \Theta' \ty \Tel(\Gamma)}{
      \TelAd(\impl{\Gamma},\Theta,\Theta')
    } \and
  \inferdef%[Inst]
    {\impl{\Gamma \ty \Ctx} \\ \Theta \ty \Ty(\Gamma)}{\Inst(\Gamma,\Theta)} \\
  \hspace*{-2em}\typing{\Gamma}{\Box} \and
  \hspace*{1.25em}\typing{\Gamma}{\Box}[\adso{\Theta}{\Theta'}] \and
  \hspace*{3em}\typing{\Gamma}{\Box}[\Theta] \\
  \inferdef[CtxExtTel]
  {
    \ctxty{\Gamma} \\ \typing{\dual{\Gamma}{d}}{\Theta}
  }{\ctxty{\Gamma \ext_{d} \Theta}}[\(d : \Dir\)] \and
  \inferdef[TelEmp]
  {\ctxty{\Gamma}}{\typing{\Gamma}{\emp}} \and
  \inferdef[TelExtTy]
  {
    \typing{\Gamma} \\
    \typing{\Gamma}{\Theta} \\
    \typing{\Gamma \ext_{+} \Theta}{A}
  }{\typing{\Gamma}{\Theta \ext A}} \\
  \inferdeft[WkTel]
  {
    \ctxty{\Gamma} \\
    \typing{\dual{\Gamma}{d}}{\Theta}
  }{
    \typing{\Gamma \ext_{d} \Theta}{\wk_{\Theta}}[\Gamma]
  }[\(d : \Dir\)]
  \and
  \inferdeft[VarInst]
  {
    \ctxty{\Gamma} \\
    \typing{\Gamma}{\Theta}
  }{
    \typing{\Gamma \ext_{+} \Theta}{\vinst}[\subs{\Theta}{\wk_{\Theta}}]
  } \and
  \inferdeft[SubExtInst]
  {
    % \impl{\Gamma, \Delta \ty \Ctx} \\
    \typing{\Gamma}{\sigma}[\Delta] \\
    \impl{\typing{\dual{\Delta}{d}}{\Theta}} \\
    \typing{\dual{\Gamma}{d}}{\iota}[\subs{\Theta}{\dual{\sigma}{d}}]
  }{\typing{\Gamma}{\sigma \ext_{d} \iota}[\Delta \ext_{d} \Theta]}[\(d : \Dir\)]
  \\
  \inferdeft[TransExt+Inst]
  {
    % \impl{\Gamma, \Delta \ty \Ctx} \\
    % \impl{\sigma, \tau \ty \Sub(\Gamma,\Delta)} \\
    \typing{\Gamma}{\mu}[\transso[\Delta]{\sigma}{\tau}] \\
    \impl{\typing{\Delta}{\Theta}} \\
    \typing{\Gamma}{\iota}[\subs{\Theta}{\sigma}]
  }{
    \typing{\Gamma}{\mu \ext_{+} \iota}[\transso{\sigma \ext \iota}
      {\tau \ext \had{\iota}{\trans{\Theta}{\mu}}}]
  } \and
  \inferdeft[TransExt-Inst]
  {
    % \impl{\Gamma, \Delta \ty \Ctx} \\
    % \impl{\sigma, \tau \ty \Sub(\Gamma,\Delta)} \\
    \typing{\Gamma}{\mu}[\transso[\Delta]{\sigma}{\tau}] \\
    \impl{\typing{\dual{\Delta}{-}}{\Theta}} \\
    \typing{\dual{\Gamma}{-}}{\iota}[\subs{\Theta}{\dual{\tau}{-}}]
  }{
    \typing{\Gamma}{\mu \ext_{-} \iota}[\transso{\sigma \ext_{-}
      \had{\iota}{\trans{\Theta}{\dual{\mu}{-}}}}{\tau \ext \iota}]}
  %
  % \inferdef[SubsTelAd]{
  %     % \impl{\Gamma,\Delta \ty \Ctx} \\
  %     \sigma \ty \Sub(\Gamma,\Delta) \\
  %     \Theta \ty \Tel(\Gamma) \\
  %     \Theta' \ty \Tel(\Delta) \\
  %     \alpha \ty \TelAd(\Gamma,\Theta,\subs{\Theta'}{\sigma})
  %   }{\sigma \pext \alpha \defconvop (\sigma \circ \wk) \ext \ad{\vinst}{\alpha} \ty
  %     \Sub(\Gamma \ext \Theta, \Delta \ext \Theta')}
\end{mathpar}
\end{small}

\Description{Judgments for telescopes (well-formed telescope, telescope adapter,
  and instantiation), and basic rules for context extension of a telescope, empty
  telescope, telescope extension by a type, the corresponding weakening, variable
  instantiation, substitution extension (by an instantiation), and transformation
  extension.}
\caption{Rules for telescopes (excerpt) –
  Note that telescopes only contain term variables}
\label{fig:adapttt-tel}

\end{figure}

\paragraph{Telescopes} In \cref{sec:type-var-examples}, we saw the need for
dependent type variables.
Our basic examples only need 0 or 1 binder, but to allow for a uniform treatment of these
and more complex examples with more binders, we introduce telescopes,
whose main rules are collected in \cref{fig:adapttt-tel}. Telescopes
represent context extensions, although limited to term variables only.
Hence, they are lists of types (in a pre-existing context), and
informally behave as iterated \(\Sig\)-types.
Telescopes in context \(\Gamma\) are built out of the empty telescope \(\emp\) by successive telescope extension \(\Theta \ext A\), which extends \(\Theta\) with
a new term variable of type \(A\).
A context \(\Gamma\) can be extended by a telescope \(\Theta\) defined over it:
\begin{align*}
 \ctxconv{\Gamma \ext_{d} \emp}{&~\Gamma} &
 \ctxconv{\Gamma \ext_{d} (\Theta \ext A)}{(\Gamma \ext_{d} \Theta) \ext_{d} A}
\end{align*}
% Just like for types, we write \(\Tel_{d}(\Gamma) \defconvop \Tel(\dual{\Gamma}{d})\).
All types in a telescope have the same (covariant) direction. Just
like for types, we can represent a contravariant telescope \(\Theta\) in \(\Gamma\)
(where all types are contravariant) as an element of \(\Tel(\dual{\Gamma}{-})\).

Telescopes come with their own sort of adapters \(\TelAd\), which are lists of
componentwise adapters between two telescopes, and instantiations \(\Inst\),
which are lists of terms inhabiting the respective types of a telescope.
The behavior of telescopes, instantiations and telescope adapters
mirrors that of respectively types, terms and adapters,
so we reuse the notations for the judgments of the latter.
They come with a somewhat tedious although unsurprising calculus,
where all operations are componentwise.
We record only the most interesting ones: the
telescope weakening \(\wk_{\Theta}\), obtained by composing the individual
weakenings for the types in \(\Theta\), \ie % we have
\[\conv{\Gamma \ext_{d} (\Theta \ext A)}{
    \wk_{\Theta \ext A}}{\wk_{\Theta} \circ \wk_{A}}[\Gamma]\]
and the variable instantiation \(\vinst\), which consists
of a list of variables, in that
\[\conv{\Gamma \ext \Theta \ext A}
  {\vinst_{\Theta \ext A}}
  {\subs{\vinst_{\Theta}}{\wk_{A}} \ext \tmvz}
  [\subs{(\Theta \ext A)}{\wk}]\]
This instantiation plays a role very similar to that of \(\tmvz\).
Finally, just as for terms we can define the functorial action \(\pext\)
of context extension by a telescope:
\(\sigma \pext \alpha \defconvop (\sigma \circ \wk) \ext \ad{\vinst}{\subs{\alpha}{\wk}}\).

\begin{figure}
\begin{small}

\begin{mathpar}
    \inferdef[CtxTy]
    {\ctxty{\Gamma}\\
      \typing{\dual{\Gamma}{d'}}{\Theta}}
    {\ctxty{{\Gamma \ext*_{d'} (\tyext{\Theta}[d])}}}
    [\(d, d' : \Dir\)]
    \label{rule:ctx-ty}
    \and
    \inferdef[SubTy]
    {
      % \impl{\Gamma, \Delta \ty \Ctx} \\
      \typing{\Gamma}{\sigma}[\Delta] \\
      \typing{\dual{\Delta}{d'}}{\Theta} \\
      \typing{\dual{\left(\Gamma \ext_{d'} \subs{\Theta}
        {\dual{\sigma}{d'}}\right)}{d}}{A}
    }{\typing{\Gamma}{\sigma \ext*_{d} A}[\Delta \ext*_{d'} (\tyext{\Theta}[d])]}
    [\(d, d' : \Dir\)]
    \label{rule:subs-ty} \and
  \inferdef[TransAd++]{
    % \impl{\Gamma, \Delta \ty \Ctx} \\
    % \impl{\sigma, \tau \ty \Sub(\Gamma,\Delta)} \\
    \typing{\Gamma}{\mu}[\transso[\Delta]{\sigma}{\tau}] \\
    \typing{\Delta}{\Theta} \\
    \typing{\Gamma \ext_{+} \hsubs{\Theta}{\sigma}}{A} \\
    \typing{\Gamma \ext_{+} \hsubs{\Theta}{\tau}}{B} \\
    \typing{\Gamma \ext_{+} \hsubs{\Theta}{\sigma}}{f}[
      \adso{A}{\hsubs{B}{\id_{\Gamma} \pext_{+} \trans{\Theta}{\mu}}}]
  }{
    \typing{\Gamma}{\mu \ext*_{+} f}
      [\transso[\Delta \ext*_{+} (\tyext{\Theta}[+])]{\sigma \ext*_{+} A}{\tau \ext*_{+} B}]
  }
  \label{rule:trans-ty-plus}
  \and
  \inferdef[TransAd+-]{
    % \impl{\Gamma, \Delta \ty \Ctx} \\
    % \impl{\sigma, \tau \ty \Sub(\Gamma,\Delta)} \\
    \typing{\Gamma}{\mu}[\transso[\Delta]{\sigma}{\tau}] \\
    \typing{\Delta}{\Theta} \\
    \typing{\dual{\left(\Gamma \ext_{+} \hsubs{\Theta}{\sigma}\right)}{-}}{A} \\
    \typing{\dual{\left(\Gamma \ext_{+} \hsubs{\Theta}{\tau}\right)}{-}}{B} \\
    \typing{\dual{\left(\Gamma \ext_{+} \hsubs{\Theta}{\sigma}\right)}{-}}{f}[
      \adso{\hsubs{B}{\id_{\Gamma} \pext_{+} \trans{\Theta}{\mu}}}{A}]
  }{
    \typing{\Gamma}{\mu \ext*_{-} f}
      [\transso[\Delta \ext*_{+} (\tyext{\Theta}[-])]{\sigma \ext*_{-} A}{\tau \ext*_{-} B}]
  }
  \label{rule:trans-ty-min}
  % \and
  % %
  % \inferdef[TransAd-+]{
  %   % \impl{\Gamma, \Delta \ty \Ctx} \\
  %   % \impl{\sigma, \tau \ty \Sub(\Gamma,\Delta)} \\
  %   \typing{\Gamma}{\mu}[\transso[\Delta]{\sigma}{\tau}] \\
  %   \typing{\dual{\Delta}{-}}{\Theta} \\
  %   \impl{\typing{\Gamma \ext_{-} \hsubs{\Theta}{\dual{\sigma}{-}}}{A}} \\
  %   \impl{\typing{\Gamma \ext_{-} \hsubs{\Theta}{\dual{\tau}{-}}}{B}} \\
  %   \typing{\Gamma \ext_{\impl{-}} \hsubs{\Theta}{\dual{\tau}{-}}}{f}[
  %     \adso{\hsubs{A}{\id_{\Gamma} \pext_{-} \trans{\Theta}{\dual{\mu}{-}}}}{B}]
  % }{
  %   \typing{\Gamma}{\mu \ext*_{+} f}
  %     [\transso[\Delta \ext* (\tyext{\Theta}[+])]{\sigma \ext*_{+} A}{\tau \ext*_{+} B}]
  % }
  % \and
  % %
  % \inferdef[TransAd-{}-]{
  %   % \impl{\Gamma, \Delta \ty \Ctx} \\
  %   % \impl{\sigma, \tau \ty \Sub(\Gamma,\Delta)} \\
  %   \typing{\Gamma}{\mu}[\transso[\Delta]{\sigma}{\tau}] \\
  %   \typing{\dual{\Delta}{-}}{\Theta} \\
  %   \impl{\typing{\dual{\left(\Gamma \ext_{-} \hsubs{\Theta}{\dual{\sigma}{-}}\right)}{-}}{A}} \\
  %   \impl{\typing{\dual{\left(\Gamma \ext_{-} \hsubs{\Theta}{\dual{\tau}{-}}\right)}{-}}{B}} \\
  %   \typing{\dual{\left(\Gamma \ext_{\impl{-}} \hsubs{\Theta}{\dual{\tau}{-}}\right)}{-}}{f}[
  %     \adso{B}{\hsubs{A}{\dual{\left(\id_{\Gamma} \pext_{-} \trans{\Theta}{\dual{\mu}{-}}\right)}{d}}}]
  % }{
  %   \typing{\Gamma}{\mu \ext*_{-} f}
  %     [\transso[\Delta \ext* (\tyext{\Theta}[-])]{\sigma \ext*_{-} A}{\tau \ext*_{-} B}]
  % }
  \\
    \inferdef[WkTy]{
      \ctxty{\Gamma} \\
      \typing{\dual{\Gamma}{d'}}{\Theta}}{
      \typing{\Gamma \ext*_{d'} (\tyext{\Theta}[d])}{\wk}[\Gamma]
    }[\(d,d' : \Dir\)]
    \and
    \inferdef[VarZTy]{
      \ctxty{\Gamma} \\
      \typing{\dual{\Gamma}{d}}{\Theta}
    }{\typing{\Gamma \ext*_{d} (\tyext{\Theta}[+]) \ext_{d} \subs{\Theta}{\dual{\wk}{d}}}{\tyvz}
    }[\(d \ty \Dir\)]
    \label{rule:vz-ty}
  \end{mathpar}
\end{small}

\Description{Rules for type variables. Context can be extended by type variables
  indexed by a telescope and with a direction, and substitutions (resp. transformations)
  can be accordingly extended by a type family (resp. family of adapters).
  We have weakening and variable operations, too.
  }
\caption{Type variables}
\label{fig:adapttt-tyvar}
\end{figure}

\paragraph{Type variables}

Using telescopes and instantiations, we can turn to
type variables (\cref{fig:adapttt-tyvar}),
which combine contravariance (\cref{sec:tyvar-contra}), and
dependency (\cref{sec:tyvar-dep}).
The context \(\Gamma \ext*_{d'} (\tyext{\Theta}[d])\)
represents the extension of \(\Gamma\)
with a new type variable, dependent of the telescope \(\Theta\) defined over \(\dual{\Gamma}{d'}\),
and assumed to have direction \(d\).
\ruleref{rule:subs-ty} gives the corresponding substitution extension:
a type variable \({(\tyext{\Theta}[d])}\) must be replaced by a type with direction \(d\)
in a context extended by \(\Theta\).
% , the extra bound variables the type has access to.
Rules \nameref{rule:trans-ty-plus} and \nameref{rule:trans-ty-min}
give transformations' extension for covariant telescopes. The biggest difference is
the variance of the adapter: in the latter
rule, the direction of the adapter is reversed compared to the former.
% In particular, we see that the direction of the variable
% dictates the adapters' direction for the extension of a transformation:
% for a variable with direction \(-\), the adapter is contravariant.
\shepherd{Similar rules applying to telescopes with a contravariant dependency over
  the context are provided in \cref{sec:adapTTt-full-rules-type-variables}.}

It might not be evident why in these rules the direction of the adapter coincides
with that of the types substituted for the variable: we could imagine
contravariant adapters between covariant types or vice-versa.
However, this constraint arises from whiskering.
Indeed, consider the data in the diagram below, from which we construct
\((\emp \ext* A) \circ \mu \ty \Trans(\Gamma,\emp \ext* \subs{A}{\sigma},\emp \ext* \subs{A}{\tau})\).
\begin{center}
  \begin{tikzcd}[column sep = 8em]
    \Gamma
      \arrow[r, bend left=20, "\sigma", ""{name=s,below}]
      \arrow[r, bend right=20, "\tau"',""{name=t,above}]
      \arrow[Rightarrow, from=s,to=t,"\mu"]
      & \Delta
        \arrow[r,"\emp \ext* A"]
      & \emp \ext* (\tyext{\emp}[+])
  \end{tikzcd}
\end{center}
The obvious computation rule is to simplify it to \(\emp \ext* \trans{A}{\mu}\),
which is only valid if \(A\) is covariant.
% An entirely similar constraint appears for contravariant variables, too.

Moreover, Rules \nameref{rule:subs-ty},
\nameref{rule:trans-ty-plus} and \nameref{rule:trans-ty-min}
are in line with the action of dualization, as directions are just right for
the following equations to be type-correct:
\begin{mathpar}
  \conv{\Gamma}{\dual{(\sigma \ext*_{d} A)}{-}}{\dual{\sigma}{-} \ext*_{-d} A}
    [\Delta \ext*_{d'} (\tyext{\Theta}[d]) ] \and
  \conv{\Gamma}{\dual{(\mu \ext_{d} f)}{-}}{\dual{\mu}{-} \ext_{-d} f}
    [\transso{\tau \ext*_{-d} B}{\sigma \ext*_{-d} A}]
\end{mathpar}

Finally, as for term variables, we have a weakening operation, and a variable zero
\(\tyvz\). This variable lives in an extended context, corresponding to the
extra binders it has access to. These would typically be instantiated
by providing an instantiation, and indeed we can derive
\[
  \inferrule*[right=(\(d : \Dir\))]{
    \ctxty{\Gamma} \\
    \typing{\dual{\Gamma}{d}}{\Theta} \\
    \typing{\Gamma \ext*_{d} (\tyext{\Theta}[+])}{\Theta'} \\
    \typing{\Gamma \ext*_{d} (\tyext{\Theta}[+]) \ext \Theta'}{\iota}[\Theta]
  }{
    \typing{\Gamma \ext*_{d} (\tyext{\Theta}[+]) \ext \Theta'}{\subs{\tyvz}{\wk_{\Theta'} \ext \iota}}
  }
\]
That is, if we access a type variable and provide an instantiation of its binders,
then we obtain a type in the current context, \ie without extra extension.

% This concludes the judgmental structure of \AdapTTt.
% \mlb{Revisit examples from \cref{sec:type-var-examples}?}

% \km{Collect the different aspects of the theory explained in the previous section and give a formal definition}
\newcommand\Typ[0]{\Ty_{+}}
\newcommand\Tmp[0]{\Tm_{+}}
\newcommand\tyofp[0]{\mathrm{of}_{+}}
\newcommand\Tyn[0]{\Ty_{-}}
\newcommand\Tmn[0]{\Tm_{-}}
\newcommand\tyofn[0]{\mathrm{of}_{-}}
\newcommand\Tys[0]{\Ty_{\star}}
\newcommand\Tms[0]{\Tm_{\star}}
\newcommand\tyofs[0]{\mathrm{of}_{\star}}

\subsection[Back to AdapTT and natural models]{Back to \AdapTT and natural models}
\label{sec:back-adaptt}

A type former \(\typing{\Gamma}{F}\) in \AdapTTt{}
is automatically equipped with a functorial structure with respect to its category of parameters represented by the context $\Gamma$.
This structure can be transferred to any \NMDO,
\ie model of \AdapTT{}, which interprets the underlying type former \(F\).
To establish this correspondence, we first explain what it means for a \NMDO{} \(C\) to support a given type former in terms of presheaves over \(C\), and how the existence of a structural type cast derives from an internal category structure on these presheaves.
Such an internal category in presheaves over \(C\) can equivalently be seen
as a \(\Cat\)-valued presheaf over \(C\).
The main theorem of this section shows that, for any \NMDO \(C\),
these \(\Cat\)-valued presheaves \(\catpsh{C} \coloneq C^{\op} \to \Cat\)
on \(C\) support a model of \AdapTTt{}.
Applying this interpretation to \(\typing{\Gamma}{F}\) in \(\catpsh{C}\) thus
yields a structural type cast on the interpretation of the underlying type former in \(C\).

\begin{definition}[Type former in a natural model]
\label{def:type-former}
A type former in a natural model \(C\) is a pair \((D : \Psh(C), F : D \natt{} \Ty)\) of a presheaf \(D\) and a natural transformation \(F\) to the presheaf of types.
\end{definition}

For a context \(\Gamma \ty C\), the object \(D\ \Gamma\) represents the input parameters of the type former \((D,F)\), while \(F\) turn such data \(d \ty D\ \Gamma\) into a type \(F_{\Gamma}\ d \ty \Ty(\Gamma)\).

\begin{example}[\(\P\) type former]
    In a natural model, the \(\P\) type is given \cite[Prop. 2.4]{Awodey2018}
    by %a type former with
  \begin{align*}
    D^{\P}\ \Gamma &\coloneq (A \ty \Ty(\Gamma)) \times (B \ty \Ty(\Gamma \ext A))
    &F^{\P}_{\Gamma}\ (A, B) \coloneq \P A. B \ty \Ty\ \Gamma
  \end{align*}
\end{example}

\begin{definition}[Functorial structure on a type former]
\label{def:functorial-type}
A type former \((D, F : D \natt{} \Ty)\) in the natural model underlying
a \NMDO \(C\) has a functorial structure when \(D\) is equipped with an internal category structure and \(F\) is an internal functor from \(D\) to \(\Ty\),
equipped with its internal category structure of adapters.
\end{definition}

Given a functorial structure on a type former \((D,F)\) and a context \(\Gamma \ty C\), the morphisms between inputs \(d, d' \ty D\ \Gamma\) represents the data needed to build an adapter between \(F_{\Gamma}\ d\) and \(F_{\Gamma}\ d'\),
while the functorial action of \(F_{\Gamma}\) builds this adapter.
% Thus, such a type former is by definition functorial,
% with respect to whatever notion of morphisms we have chosen to use in \(D\).
% \(D\) being a presheaf means that substitutions act on it, and naturality says that this
% action commutes with the action of \(C\) on morphisms.
%
% \begin{example}[\(\P\)-types]
%   For \(\P\) types, \(D\) maps a context \(\Gamma\) to
%   pairs \((A \ty \Ty(\Gamma)) \times (B \ty \Ty(\Gamma \ext A))\), and an arrow
%   between \((A,B)\) and \((A',B')\) consists of pairs of adapters
%   \((a \ty \Ad(\Gamma,A',A)) \times (b \ty \Ad(\Gamma \ext A',\subs{B}{\ad{vz}{a}},B'))\).
%   These assemble into a \(\Cat\)-valued presheaf by having substitutions act componentwise.
%   The \(\P\) type formers acts
%   on pairs as above to construct the corresponding \(\P\) type, and also acts on
%   adapters to create a new adapter \(\P(a,b)\).
%   \mlb{Do we want to give the action of the adapter here? This is not part of
%   \cref{def:type-former} per se.}
%   % with action given by
%   % \[
%   %   (\ad{f}{\P(a,b)})\ u \convop \ad{(f\ (\ad{u}{a}))}{\subs{b}{\ad{u}{a}}}
%   % \]
%   % This is exactly the shared core of \cref{fig:casts-comparison}!
% \end{example}
%
% In this example, we could have taken
% \(D\) to be a discrete category, and we would have obtained a very boring functoriality
% of \(\P\). In general, we can always degenerate type former functoriality in this way,
% but of course we should have higher hopes.
%
Going back to the example of \(\P\)-types, we could equip \((D^{\P},F^{\P})\) with the functorial structure of~\cref{ex:pi-adaptt}.
But instead we can \emph{derive} this structure from an interpretation, via
\cref{thm:adapttt-over-adaptt}, of the judgment
\(\typing{\Ty_{-} \ext* (\tyvz.\Ty_{+})}{\P}[\Ty]\) in \AdapTTt{}, corresponding to an intrinsically functorial presentation of \(\P\)-types.

To properly establish the correspondence between \NMDO{}s models of \AdapTT and \AdapTTt, we need conditions on the size of the models.
Given a universe $\univ$, a category $C$ is $\univ$-small if its types of objects and homsets are in \(\univ\).
A \NMDO \(C\) is $\univ$-small if its underlying category is and the
$\Ty$ and $\Tm$ presheaves land into $\univ$-small categories.

\begin{theorem}
  \label{thm:adapttt-over-adaptt}
  Given a $\univ$-small \NMDO $C$, the 2-category $\catpsh{C}$ of $\Cat$-valued presheaf is a model of 2-dimensional \AdapTTt.
\end{theorem}

\begin{proof}[Proof sketch]
The construction follows ideas from $2$-level type theory~\cite{AnnenkovCKS23} where the outer layer is interpreted as presheaves over a model of the inner layer.
Here, the inner layer is an instance of \AdapTT while the outer layer is one of \AdapTTt.
We leverage an alternative description of  $\catpsh{C}$ as the \(2\)-category $\Cat(\psh{C})$ of categories, functors and natural transformations internal to presheaves over \(C\)~\cite[Ch. 7]{JacobsCatlogTT}.
To manipulate these structures internally to a presheaf category,  we use extensional \MLTT\footnote{Or intensional \MLTT extended with the principles of uniqueness of identity proof and function extensionality~\cite{Hofmann1997,WinterhalterST19}.} following \textcite[Ch. 4]{Hofmann1997b}.
Working within $\psh{C}$, we construct the 2-category $\univ$-small categories equipping it with all the components to interpret \AdapTTt, akin to \posscite{Licata2011} model of directed 2-dimensional type theory in \(\Cat\).
Categories, functor and natural transformation can be defined in a standard fashion in extensional \MLTT.
Relativizing these constructions to $\univ$, seen as an internal universe using Hofmann-Streicher lifting~\cite{Awodey24}, we use $\UCat$, the $2$-category of $\univ$-small categories, as the interpretation of \(\Ctx\).
$\UCat$ has a terminal object \(\emp\), the category with a single object and morphism.
A type \(A\) in a context $\Gamma \ty \UCat$ is interpreted as a functor $A \ty
\Gamma \to \UCat$, substitutions acts by precomposition and transformations by
whiskering.
A term \(t\) of type \(A\) in \(\Gamma\) is interpreted as a lax natural transformation \(\emp \natt{} A \)~\cite[definition 4.2.1]{JohnsonYauTwoDimCats}, where \(\emp \ty \Gamma \to \UCat \) is the constant functor to \(\emp\).
%
% Observe that whenever the type \(A\) takes value in discrete categories, terms are interpreted as plain natural transformations.
%
An adapter \(f\) between types \(A\) and \(B\) in \(\Gamma\) is interpreted as a
lax natural transformation \(f \ty A  \natt{} B \), and acts on a term \(t \ty
\emp \natt A\) by vertical composition \(f \circ t \ty \emp_{\Gamma} \natt B\).
In these two cases, substitutions interpreted as functors act by whiskering and transformations interpreted as natural transformations by horizontal composition which preserve lax natural transformations.
All these operations are functorial on the nose since they are interpreted by composition in a strict \(2\)-category.
Context extensions are interpreted by variants on the Grothendieck construction:
positive context extension \(\Gamma \ext_{+} A\) is interpreted by \(\int_{\Gamma}A\) while the negative context extension \(\Gamma \ext_{-} B\) is interpreted by \(\left ( \int_{\Gamma^{\op}}B \right )^{\op}\).
The interpretation of telescopes and instantiations reuses that of types and terms.
The empty telescope on \(\Gamma\) is interpreted as \( \emp_{\Gamma} \ty \Gamma \to \UCat \).
For a telescope $\Theta : \Gamma \to \UCat$ over \(\Gamma\) and a type $ A \ty \int_{\Gamma} \Theta \to \UCat$, the extension \(\Theta \ext A\) is interpreted as the functor \(\gamma \ty \Gamma \mapsto \int_{\Theta(\gamma)} A(\gamma, \cdot)\).
The final missing piece is the interpretation of type variables.
Observe that \(\UCat\) is cartesian closed as a $1$-category and take $\Rightarrow : \UCat^{\op} \times \UCat \to \UCat$ the bifunctor sending categories \(A,B\) to their functor category.
The assumption that the $\Cat$-valued presheaf \(\Ty\) is externally $\univ$-small implies that \(\Ty \ty \UCat\), as well as its opposite category \(\Ty^{\op} \ty \UCat\) where the direction of adapters is flipped.
Interpreting direction \(+\) as the identity functor \(\UCat \to \UCat\) and \(-\) as the opposite category functor, for any context \(\Gamma\) and telescope \(\Theta : \Gamma^d \to \UCat\), the interpretation of \(\Gamma \ext*_{d} (\Theta.\Ty_{d'})\) proceeds by first composing the pair of \(\Theta^{\op} : \Gamma^{-d} \to \UCat^{\op}\) and the constant functor with value \(\Ty^{d'}\) with the bifunctor \(\Rightarrow\) which yield \(\Theta^{\op}{\Rightarrow}\Ty^{d'} : \Gamma^{-d} \to \UCat\), and then taking its Grothendieck construction \((\int_{\Gamma^{-d}} \Theta^{\op}{\Rightarrow}\Ty^{d'} )^{-d}\).
\end{proof}

% We can additionally observe that the \NMDO $C$ embeds faithfully into this model of \AdapTTt through Yoneda.
\shepherd{
\paragraph{Expressivity of \AdapTTt{}'s type variables}

\AdapTTt{} does not offer many constructions on type variables: there are no $\Sigma$ or $\Pi$ formers that apply to type variables, even though the model sketched above would support such constructions.
This design is chosen to express the functoriality rules required for type formers in \AdapTT{}, only tracking the functorial dependencies between families of parameters.
The restriction of telescopes in family variables to term variables is compatible with a simple model where every type is discrete and enough to express many type formers we are interested in.
Extending these quantifications to family variables, which stand for non-discrete categories, would extend the expressivity of the framework but also require more intricate tracking on the directed aspect, as can be found in \cite{Licata2011} with restrictions on the combination of polarities.
}

\section{Functorial types}
\label{sec:functorial-types}

With the underlying judgmental structure sorted out, we can turn back to adding interesting
type formers to \AdapTTt. Normally,
to specify a type former one gives rules for
type formation, introduction, elimination, computation (β) and, possibly, uniqueness (η)
\cite[Rem. 1.5.1]{UniFoundationsProgram2013}.
We worked hard to make sure that the type formation rule amounts to giving a context, as showcased in \cref{sec:type-var-examples}.
This automatically implies a typing rule for the adapter ---by \ruleref{rule:trans-ty-plus}---, as well as the functoriality equations.
Thus, to extend a usual type former to integrate adapters, the main remaining
task is to specify the interaction of the adapter with
introduction and elimination forms.
% and should moreover ensure that the computation and functoriality
% rules for the adapters and the type's native rules cohabit well.

\subsection{Negative types: functions and pairs}
\label{sec:negative-types}

\begin{figure}
\begin{small}

\begin{mathpar}
  \inferdef[AdFun]{
    \typing{\Delta}{\sigma, \tau}[\Gamma_{\P}] \\
    \typing{\Delta}{\mu}[\transso{\sigma}{\tau}]
  }{
    \typing{\Delta}{\trans{\P}{\mu}}[\adso{\subs{\P}{\sigma}}{\subs{\P}{\tau}}]
  } \label{rule:fun-ad-dep}\and
  \inferdef[AdPair]{
    \typing{\Delta}{\sigma, \tau}[\Gamma_{\Sig}] \\
    \typing{\Delta}{\mu}[\transso{\sigma}{\tau}]
  }{
    \typing{\Delta}{\trans{\Sig}{\mu}}[\adso{\subs{\Sig}{\sigma}}{\subs{\Sig}{\tau}}]
  } \label{rule:pair-ad-dep} \and
  \inferdef[AdFunEq]{
    \typing{\Delta}{a}[\adso{A'}{A}] \\
    \typing{\Delta \ext A'}{b}[\adso{\subs{B}{\id \pext a}}{B'}] \\
    \typing{\Delta}{f}[\subs{\P}{A \ext* B}] \\
    \typing{\Delta}{u}[A']
  }{
    \conv{\Delta}
      {\ad{f}{\trans{\P}{a \ext* b}}~u}
      {\ad{(f~\ad{u}{a})}{\subs{b}{\id \ext \ad{u}{a}}}}
      [\subs{B'}{\id \ext u}]
  } \label{rule:ad-fun-eq} \and
  \inferdef[AdPairEq1]{
    % \Delta \vdash A, A' \ty \Ty_{+} \\
    % \Delta \ext A \vdash B \ty \Ty_{+} \\
    % \Delta \ext A' \vdash B' \ty \Ty_{+} \\
    \typing{\Delta}{a}[\adso{A}{A'}] \\
    \typing{\Delta \ext A}{b}[\adso{B}{\subs{B'}{\id \pext a}}] \\
    \typing{\Delta}{p}[\subs{\Sig}{A \ext* B}]
  }{
    \conv{\Delta}
      {\pi_{1}~(\ad{p}{\trans{\Sig}{a \ext* b}})}
      {\ad{(\pi_{1}~p)}{a}}
      [A'] \\
    \conv{\Delta}
      {\pi_{2}~(\ad{p}{\trans{\Sig}{a \ext* b}})}
      {\ad{(\pi_{2}~p)}{\subs{b}{\id \ext \pi_{1} p}}}
      [\subs{B'}{\id \ext \ad{(\pi_{1} p)}{a}}]
  }
  % \and
  % \inferdef[AdPairEq2]{
  %   % \Delta \vdash A, A' \ty \Ty_{+} \\
  %   % \Delta \ext A \vdash B \ty \Ty_{+} \\
  %   % \Delta \ext A' \vdash B' \ty \Ty_{+} \\
  %   \typing{\Delta}{a}[\adso{A}{A'}] \\
  %   \typing{\Delta \ext A}{b}[\adso{B}{\subs{B'}{\ad{\tmvz}{a}}}] \\
  %   \typing{\Delta}{p}[\subs{\Sig}{A \ext* B}]
  % }{
  %   \conv{\Delta}
  %     {\pi_{2}~(\ad{p}{\trans{\Sig}{a \ext* b}})}
  %     {\ad{(\pi_{2}~p)}{\subs{b}{\pi_{1} p}}}
  %     [\subs{B'}{\ad{(\pi_{1} p)}{a}}]
  % }
\end{mathpar}
\end{small}

\Description{Typing and computation rules for adapters of dependent function
and pair types. The first are derived from the given context, computation of
a function adapter amounts to pre/post composition, and computation for a
pair adapter applies the relevant adapter to each projection.}
\caption{Adapters for \(\P\) and \(\Sig\) types}
\label{fig:pi-sig-ad}
\end{figure}

For negative types (\(\P\) and \(\Sig\)), we can essentially follow \textcite{Subtyping2024}.
Their contexts, as given in \cref{sec:type-var-examples} %\km{The gap is really huge},
translate to the following
\begin{align*}
  & \Gamma_{\P} \defconvop \emp \ext*_{+} (\tyext{\emp}[-]) \ext*_{-} (\tyext{(\emp \ext \tyvz)}[+])
  & \Gamma_{\Sig} \defconvop \emp \ext*_{+} (\tyext{\emp}[+]) \ext*_{+} (\tyext{(\emp \ext \tyvz)}[+])
\end{align*}
The typing rules and equations for adapters are in \cref{fig:pi-sig-ad}.
If we compute concretely what a substitution/transformation targeting \(\Gamma_{\P}\)
and \(\Gamma_{\Sig}\) consists of, we exactly recover
rules \nameref{rule:ad-fun-dep-ex} and \nameref{rule:ad-pair-dep-ex} from \cref{sec:tyvar-dep}.
We can derive a contravariant version of the rules for free
by instantiating them in a dualized context, for instance \(\P A.B\) is
contravariant if \(B\) is and \(A\) is covariant.

Adapters compute according to the types' eliminator, following the idea
that negative types are characterized by their eliminators.
Extending \AdapTTt{} with \(\l\)-abstraction and application,
we recover equations on constructors, \eg
  \(\ad{(\l t)}{\trans{\P}{a \ext*b}} \convop \l \ad{(\subs{t}{\ad{\tmvz}{a}})}{b}\)
via η expansion.

Moreover, the functoriality laws up to casts are derivable from the computation
equations only. Indeed, we can obtain
  \(\ad{f}{\trans{\P}{(a_2 \circ a_1) \ext* (b_2 \circ b_1)}} \convop
    \ad{f}{\trans{\P}{a_1 \ext* b_1} \circ \trans{\P}{a_2 \ext* b_2}}\),
by first η-expanding both sides, then triggering \nameref{rule:ad-fun-eq} to
compute the \(\P\) adapters away, and conclude by functoriality of coercion,
\ie the fact that \(\ad{t}{a' \circ a} \convop \ad{\ad{t}{a}}{a'}\).
The situation for \(\trans{\P}{\id \ext* \id}\) and \(\Sig\) is similar.
% \mlb{Can we do better?}

% \km{The idea for $\Pi, \Sigma$ is already exposed in \cite{Subtyping2024}. Can
%   we give a more general statement thanks to the general presentation proposed
%   in this paper ?}

% \km{First level of presentation: Adapters for $\Pi, \Sigma$ have a standard and well-known action by coercion}

% \km{Second level of presentation: Any \NMDO equipped with an interpretation of
%   $\Pi, \Sigma$ on its terms can be extended to a \NMDO with functorial $\Pi,
%   \Sigma$ ?}

% \km{Third level of presentation: extend to other finitary types with
%   extensionality principles (records) using the signatures ?}

\subsection{Inductive types}
\label{sec:inductive-types}

To represent inductive types, we cannot rely on encodings through \eg \(\W\) types, as
these necessarily rely on term-level manipulations ---for instance \(A + B\) is encoded
as \(\Sig (b : \Bool). \operatorname{if} b \operatorname{then} A \operatorname{else} B\),
by using the large eliminator for booleans.
As we do not internalize adapters as a type, such an encoding would yield a degenerate notion of
adapters for the \(+\) type, which is unsatisfactory.

\begin{shepherdenv}
Instead, we provide a general scheme for parametrized and indexed inductive types,
capturing a class corresponding to that
originally described by \textcite{Coquand1990,PaulinMohring1993},
and still used as the backbone of most proof assistants.
% ---although in all these inductive types are nowadays somewhat more expressive.
We describe this class using a theory of signatures, inspired by
multiple existing ones \cite{KovacsPhD,Kaposi2020a,Escot2022}.
Compared to the former two \cite{KovacsPhD,Kaposi2020a}, we focus more closely
on parameters, as it is with respect to them
that (inductive) type formers are functorial, and use the machinery
of type variables and telescopes from
\cref{sec:adaptt2-def} in the signatures, rather than
relying on an external product type and universe.
Our descriptions are closer to the latter \cite{Escot2022},
although we frame them using GATs rather than of a universe of
description.
\end{shepherdenv}

\paragraph{Extended GAT} To lighten the rules, we rely on two extensions
to the standard setting of generalized algebraic theories.
The first is the list type ---at the meta-level of the GAT---,
which we denote with a vector notation. We still use \(\emp\) and \(\ext\) for
nil and cons, \([x ; y ; \dots]\) for an explicit list, and \(c \in cs\)
for witnesses that \(c\) appears in a list \(cs\), which are essentially de Bruijn indices.
Second, we allow ourselves a form of record types, expressed as follows:
\[
\inferdef{
    \ctxty{\Gamma_{par}} \\
    \typing{\Gamma_{par}}{\Theta_{ind}} \\
    \typing{\Gamma_{par}}{\Theta_{nr}} \\
    r \ty \mList{\RecDesc(\Gamma_{par},\Theta_{ind},\Theta_{nr})} \\
    \typing{\Gamma_{par} \ext \Theta_{nr}}{\iota}[\subs{\Theta_{ind}}{\wk}]
  }{
    \{\rfield{nrec} \coloneq \Theta_{nr};\rfield{rec} \coloneq r;\rfield{ind} \coloneq \iota\}
      \ty \ConDesc(\Gamma_{par},\Theta_{ind})
  }
\]
This rule adds a new “record sort” \(\ConDesc\), indexed by a context and a telescope,
with three fields \(\rfield{nrec}\), \(\rfield{rec}\) and \(\rfield{ind}\),
accessed by \(\proj{c}{\rfield{nrec}}\).
These extensions can be encoded in standard GATs at the cost of introducing
new sorts, operations and equations, which would only add distraction.

\begin{figure}
\begin{small}

\begin{mathpar}
  % \inferdef{
  %   \ctxty{\Gamma_{par}} \\
  %   \typing{\Gamma_{par}}{\Theta_{ind}} \\
  %   c \ty \mList{\ConDesc(\Gamma_{par},\Theta_{ind})}
  % }{\{\rfield{constrs} \coloneq c \} \ty \IndDesc(\Gamma_{par},\Theta_{ind})} \and
  \inferdef[DataDesc]{
    \ctxty{\Gamma_{par}} \\
    \typing{\Gamma_{par}}{\Theta_{ind}}
  }{
    \IndDesc(\Gamma_{par},\Theta_{ind}) \defconvop \mList{\ConDesc(\Gamma_{par},\Theta_{ind})}
  } \and
  \inferdef[ConDesc]{
    \ctxty{\Gamma_{par}} \\
    \typing{\Gamma_{par}}{\Theta_{ind}} \\
    \typing{\Gamma_{par}}{\Theta_{nr}} \\
    r \ty \mList{\RecDesc(\Gamma_{par},\Theta_{ind},\Theta_{nr})} \\
    \typing{\Gamma_{par} \ext \Theta_{nr}}{\iota}[\subs{\Theta_{ind}}{\wk}]
  }{
    \{\rfield{nrec} \coloneq \Theta_{nr};\rfield{rec} \coloneq r;\rfield{ind} \coloneq \iota\}
      \ty \ConDesc(\Gamma_{par},\Theta_{ind})
  } \\
  \inferdef[RecDesc]{
    \ctxty{\Gamma_{par}} \\
    \typing{\Gamma_{par}}{\Theta_{ind}} \\
    \typing{\Gamma_{par}}{\Theta_{nr}} \\
    \typing{\dual{(\Gamma_{par} \ext \Theta_{nr})}{-}}{\Theta_{ar}}\\
    \typing{(\Gamma \ext_{+} \Theta_{nr}) \ext_{-} \Theta_{ar}}{\iota}[\subs{\Theta_{ind}}{\wk}]
  }{
    \{\rfield{arit} \coloneq \Theta_{ar} ;\rfield{rind} \coloneq \iota \}
    \ty \RecDesc(\Gamma_{par},\Theta_{ind},\Theta_{nr})
  }
\end{mathpar}
% \km{An issue with using the same variable $\Theta$ for all telescopes is that the differentiating information is only in index, that makes the last 2 rules a bit hard to follow}
\end{small}

\Description{Formation rules for a well-formed inductive signature, relying
  on that for a well-formed constructor, itself relying on that for a well-formed
  recursive position description.}
\caption{Signatures for inductive types, constructors and recursive arguments}
\label{fig:signatures-desc}
\end{figure}

\paragraph{Signatures}
Signatures, presented in \cref{fig:signatures-desc}, are built out of three additional sorts: inductive descriptions \(\IndDesc\), constructors descriptions \(\ConDesc\) and recursive argument descriptions \(\RecDesc\).
The description of an inductive type \(I \ty \IndDesc\) is indexed by
a context $\Gamma_{par}$ of parameters and
a telescope $\Theta_{ind}$ of indices, and consists of a list
of constructor descriptions.
A constructor description \(c \ty \ConDesc\)
consists of a telescope \(\rfield{nrec}\) of non-recursive arguments,
a list \(\rfield{rec}\) of recursive argument description, and a final instantiation
$\rfield{ind}$ of indices, which can depend on the non-recursive arguments\footnote{Allowing indices to depend on recursive arguments would yield descriptors for inductive-inductive type former.
This is also what allows separating non-recursive arguments and recursive ones without losing generality.}.
Finally, the description of a recursive argument \(r \ty \RecDesc\) is given by a telescope of arities \(\rfield{arit}\) binding additional parameters for a recursive argument,
corresponding to a “branching” argument, and an instantiation \(\rfield{rind}\)
of the indices for the recursive occurrence.
% In other words, if the constructor of type \(Ind\) is instantiated with parameters
% \(p : \Gamma_{par}\) and non-recursive arguments
% \(n : \Theta_{nr}\), an \(r \ty \RecDesc(\Gamma_{par};\Theta_{ind};\Theta_{nr})\)
% represents an argument of type
% \(\P \proj{r}{arit}. Ind\,p\,(\subs{\proj{r}{rind}}{p \ext n})\).
%
Note that, crucially, the arity is a \emph{contravariant} telescope,
as an adapter acts on recursive arguments' parameters by pre-composition,
generalizing the examples of trees in \cref{sec:tyvar-contra,sec:tyvar-dep}.

\begin{example}[List]
\label{ex:sig-list}

  \(\List\) has one covariant type parameter and no indices, \ie
  \(\Gamma_{\List} \defconvop \emp \ext*_{+} (\tyext{\emp}[+]) \) is the context of
  parameters and \(\emp\) the telescope of indices.
  The description is given by
  \begin{align*}
    c_{nil} & \defconvop
      \{\rfield{nrec} \coloneq \emp;\rfield{rec} \coloneq \emp; \rfield{ind} \coloneq \emp\}
      \ty \ConDesc(\Gamma_{\List},\emp) \\
    c_{cons} &  \defconvop
      \{\rfield{nrec} \coloneq \emp \ext \tyvz ;
        \rfield{rec} \coloneq [\{
          \rfield{arit} \coloneq \emp;
          \rfield{rind} \coloneq \emp\}];
        \rfield{ind} \coloneq \emp\}
      \ty \ConDesc(\Gamma_{\List},\emp) \\
    \const{listDesc} & \defconvop
      [c_{nil} ; c_{cons}]
      \ty \IndDesc(\Gamma_{\List},\emp)
  \end{align*}
The constructor nil has no arguments, and cons has two arguments,
a non-recursive one typed by the unique type parameter, and a recursive one with
empty arity, \ie no branching.
\end{example}

We omit the type \(\Nat\), which is similar.
We use \(\const{0}\) and \(\const{S}\) for its two constructors.

\begin{example}[Vectors]
\label{ex:sig-vec}

  The type of vectors is similar to lists, except that there is now an index of type
  \(\Nat\), \ie we take \(\Theta_{\const{Vec}} \defconvop \emp \ext \Nat \ty \Tel(\Gamma_{\List})\). Constructors are given by
  \begin{align*}
    c_{nil} & \defconvop
      \{\rfield{nrec} \coloneq \emp;
        \rfield{rec} \coloneq \emp;
        \rfield{ind} \coloneq \emp \ext 0\}
      \ty \ConDesc(\Gamma_{\List},\Theta_{\const{Vec}}) \\
    c_{cons} &  \defconvop
      \{\rfield{nrec} \coloneq \emp \ext \tyvz \ext \Nat ;
        \rfield{rec} \coloneq [\{
          \rfield{arit} \coloneq \emp;
          \rfield{rind} \coloneq \emp \ext \tmvz \}];
        \rfield{ind} \coloneq \subs{\const{S}}{\tmvz}\} \\
      & \qquad \ty \ConDesc(\Gamma_{\List},\Theta_{\const{Vec}}) \\
    \const{vecDesc} & \defconvop [c_{nil} ; c_{cons}]
      \ty \IndDesc(\Gamma_{\List},\Theta_{\const{Vec}})
  \end{align*}
The index is instantiated in both constructors,
and in the recursive argument of \(c_{cons}\).
\end{example}

\begin{example}[Sum]
  \label{ex:sig-sum}
  The sum type has two non-dependent, covariant parameters, thus we take
  \(\Gamma_{+} \defconvop \emp \ext*_{+} (\tyext{\emp}[+]) \ext*_{+} (\tyext{\emp}[+])\),
  and no index. It is given by
  \begin{align*}
    c_{inl} & \defconvop
      \{\rfield{nrec} \coloneq \emp \ext \subs{\tyvz}{\wk};
        \rfield{rec} \coloneq \emp;
        \rfield{ind} \coloneq \emp\}
      \ty \ConDesc(\Gamma_{+},\emp) \\
    c_{inr} & \defconvop
      \{\rfield{nrec} \coloneq \emp \ext \tyvz ;
        \rfield{rec} \coloneq \emp;
        \rfield{ind} \coloneq \emp\}
      \ty \ConDesc(\Gamma_{+},\emp) \\
    \const{sumDesc} & \defconvop [c_{inl} ; c_{inr}]
      \ty \IndDesc(\Gamma_{+},\emp)
  \end{align*}
  We have two constructors with one non-recursive argument, of either
  the first or second parameter.
\end{example}

\begin{example}[Trees]
\label{ex:sig-w}

  The type \(\W\) of well-founded trees is given by the following:
  \begin{align*}
    \Gamma_{\W} & \defconvop \emp \ext*_{+} (\tyext{\emp}[+]) \ext*_{+} (\tyext{\emp \ext \tyvz}[-]) \ty \Ctx \\
    % \Theta_{\W} \defconvop \emp \ty \Tel(\Gamma_{\W}) \and
    c_{sup} & \defconvop \{
      \rfield{nrec} \coloneq \emp \ext \subs{\tyvz}{\wk};
      \rfield{rec} \coloneq \emp \ext \{
        \rfield{arit} \coloneq \tyvz;
        \rfield{rind} \coloneq \emp \};
      \rfield{ind} \coloneq \emp\}
      \ty \ConDesc(\Gamma_{\W},\emp) \\
    \const{WDesc} & \defconvop
      [c_{sup}] \ty \IndDesc(\Gamma_{\W},\emp)
  \end{align*}
This time we see a non-empty arity for its unique constructor, whose recursive
argument is branching. Note also that because the second variable of
\(\Gamma_{\W}\) appears in contravariant position (as “width” of the branching),
it is declared with direction \(-\).
\end{example}

\begin{example}[Identity]
\label{ex:sig-eq}

  The identity type has two parameters, a type and a term of that type,
  and one index, again a term of that type.%
  \footnote{This is the “based” identity type. Alternatively, we could consider
  both terms to be indices.}
  The single
  constructor for reflexivity has no arguments, and its index is the second
  parameter, thus
  \begin{align*}
    \Gamma_{\Id} & \defconvop \emp \ext*_{+} (\tyext{\emp}[+]) \ext \tyvz
    \hspace{5em}
    \Theta_{\Id} \defconvop \emp \ext \subs{\tyvz}{\wk} \\
    c_{refl} & \defconvop \{
      \rfield{nrec} \coloneq \emp ;
      \rfield{rec} \coloneq \emp ;
      \rfield{ind} \coloneq \tmvz \}
      \ty \ConDesc(\Gamma_{\Id},\Theta_{\Id}) \\
    \const{idDesc} & \defconvop
      [c_{refl}] \ty \IndDesc(\Gamma_{\Id},\Theta_{\Id})
  \end{align*}
\end{example}

\begin{figure}
\begin{small}

\begin{mathpar}
  \inferdef[IndTy]{
      % \impl{\Gamma \ty \Ctx(E)} \\ \impl{\Theta \ty \Tel(E,\Gamma)} \\
      \impl{I \ty \IndDesc(\Gamma_{par},\Theta_{ind})} \\
    }{
      \typing{\Gamma_{par}\ext \Theta_{ind}}{\ind(I)}
    } \label{rule:ind-ty} \and
  \inferdef[IndCstr]
  {
    I \ty \IndDesc(\Gamma_{par},\Theta_{ind}) \\
    \impl{c \ty \ConDesc(\impl{\Gamma_{par},\Theta_{ind}})} \\
    ic \ty c \in I \\
    % \ctxty{\Delta} \\
    % \typing{\Delta}{p}[\Gamma_{par}] \\
    % \typing{\Delta}{arg}[\conData(c,p,\subs{\ind(I)}{p \pext \id})]
  }{
    \typing{\Gamma_{par} \ext \subs{\conData(c)}{\id \ext \ind(I)}}
      {\constr(ic)}[\subs{\ind(I)}{\id \ext \subs{\proj{c}{ind}}{\wk}}]
  } \label{rule:ind-cstr} \and
  \inferdef[ConDataDef]{
    % \ctxty{\Gamma_{par}} \\
    % \typing{\Gamma_{par}}{\Theta_{ind}} \\
    c \ty \ConDesc(\Gamma_{par},\Theta_{ind}) \\
    % \ctxty{\Delta} \\
  }{
    \defconv{\Gamma_{par} \ext*_{+} (\tyext{\Theta_{ind}}[+])}
    {\conData(c)}
    {\left(\subs{\proj{c}{nrec}}{\wk} \ext \recDatas(\proj{c}{rec})\right)}
    % [\Tel]
  } \and
  \inferdef[RecDatas]{
    \ctxty{\Gamma_{par}} \\
    \typing{\Gamma_{par}}{\Theta_{ind}} \\
    \typing{\Gamma_{par}}{\Theta_{nr}} \\
    c \ty \mList{\RecDesc(\Gamma_{par},\Theta_{ind},\Theta_{nr})} \\
  }{
    \typing{\Gamma_{par} \ext*_{+} (\tyext{\Theta_{ind}}[+]) \ext \subs{\Theta_{nr}}{\wk}}
    {\recDatas(c)}% [\Tel]
  } \and
  \inferdef[RecDatasEmp]{
    \ctxty{\Gamma_{par}} \\
    \typing{\Gamma_{par}}{\Theta_{ind}} \\
    \typing{\Gamma_{par}}{\Theta_{nr}} \\
  }{
    \conv{\Gamma_{par} \ext*_{+} (\tyext{\Theta_{ind}}[+]) \ext \subs{\Theta_{nr}}{\wk}}
    {\recDatas(\emp)}{\emp}%[\Tel]
  } \and
  \inferdef[RecDatasExt]{
    % \ctxty{\Gamma_{par}} \\
    % \typing{\Gamma_{par}}{\Theta_{ind}} \\
    % \typing{\Gamma_{par}}{\Theta_{nr}} \\\\
    c \ty \mList{\RecDesc(\Gamma_{par},\Theta_{ind},\Theta_{nr})} \\
    r \ty \RecDesc(\Gamma_{par},\Theta_{ind},\Theta_{nr}) \\
  }{
    \conv{\Gamma_{par} \ext*_{+} (\tyext{\Theta_{ind}}[+]) \ext \subs{\Theta_{nr}}{\wk}}
    {\recDatas(c \ext r)}{\recDatas(c) \ext \subs{\recData(r)}{\wk}}%[\Tel]
  } \and
  \inferdef[RecDataDef]{
    % \ctxty{\Gamma_{par}} \\
    % \typing{\Gamma_{par}}{\Theta_{ind}} \\
    % \typing{\Gamma_{par}}{\Theta_{nr}} \\
    r \ty \RecDesc(\Gamma_{par},\Theta_{ind},\Theta_{nr}) \\
    % \ctxty{\Delta} \\
    % \typing{\Delta}{\sigma}[\Gamma_{par}] \\
    % \typing{\Delta \ext \subs{\Theta_{ind}}{p}}{A}
  }{
    \Gamma_{par} \ext*_{+} (\tyext{\Theta_{ind}}[+]) \ext \subs{\Theta_{nr}}{\wk} \vdash
      \recData(r) \defconvop \hspace{10em} \\
      \hspace{10em}
      \P \subs{\proj{r}{arit}}{\wk \pext \id}.
      \left(\hsubs{\tyvz}{(\wk \circ \wk) \ext (\subs{\proj{r}{rind}}{(\wk \circ \wk) \ext \subs{\vinst}{\wk} \ext \vinst}\right)})
     % \ty \Ty
  }
\end{mathpar}
\end{small}

\Description{Typing rules for inductive operations: inductive type, inductive
  constructor (relying on constructor data and recursive data).}
\caption{Inductive type and term constructors}
\label{fig:inductive-constr}
\end{figure}

\paragraph{Constructors}
Given a signature, we first construct the corresponding type and
term constructors, in \cref{fig:inductive-constr}.
The inductive type, \nameref{rule:ind-ty}, is easy:
we simply read off the parameters and indices from the description,
and declare the well-formedness of a type in the former extended by the latter.
As explained in \cref{sec:tyvar-simple}, this suffices to recover the type
in an arbitrary context by substitution.

For constructors, \ruleref{rule:ind-cstr} is a bit more complicated, and
relies on an auxiliary telescope \(\conData\), whose instantiations correspond
to the arguments of the constructor.
This telescope is computed with respect to an extra variable depending on
the telescope of indices, and the loop is tied
by instantiating this variable with the inductive itself.
In turn, \(\conData\) is computed by concatenating the telescope of non-recursive
arguments and that of recursive arguments. The latter are computed from their description by
\(\recData\), which, for each recursive argument, builds a \(\P\) with domain
the arity and codomain the extra type variable, properly instantiated.
% Categorically, \(\conData(c)\) decodes \(c \ty \ConDesc(\Gamma_{par},\Theta_{ind})\)
% to ????

\paragraph{Adapters}

\begin{figure}
\begin{small}

\begin{mathpar}
\inferdef[IndAdEqPar]{
    I \ty \IndDesc(\Gamma_{par},\emp) \\
    \impl{c \ty \ConDesc(\impl{\Gamma_{par},\emp})} \\
    ic \ty c \in I \\\\
    % \ctxty{\Delta} \\
    % \typing{\Delta}{p,p'}[\Gamma_{par}] \\
    \typing{\Delta}{\mu}[\transso[\Gamma_{par}]{p}{p'}] \\
    \typing{\Delta}{arg}[\subs{\conData}{p \ext* \subs{\ind(I)}{p}}]
}{
    \Delta \vdash
    \ad{\subs{\constr(ic)}{p \ext arg}}{\trans{\ind(I)}{\mu}}
        \convop \hspace{11.5em}\\
    \hspace{3em} \subs{\constr(ic)}{p' \ext
      \ad{arg}{
      \trans{\left(\subs{\conData(c)}{\id \ext* \ind(I)}\right)}{\mu}}}
      \ty \subs{\ind(I)}{p'}
} \label{rule:ind-ad-eq-par} \and
\inferdef[IndAdEq]{
    I \ty \IndDesc(\Gamma_{par},\Theta_{ind}) \\
    \impl{c \ty \ConDesc(\impl{\Gamma_{par},\Theta_{ind}})} \\
    ic \ty c \in I \\
    % \ctxty{\Delta} \\
    % \typing{\Delta}{p,p'}[\Gamma_{par}] \\
    \typing{\Delta}{\mu}[\transso[\Gamma_{par}]{p}{p'}] \\
    \typing{\Delta}{argn}[\subs{\proj{c}{nrec}}{p}] \\
    \typing{\Delta}{argr}[\subs{\recDatas(\proj{c}{rec})}
      {p \ext* \subs{\ind(I)}{p \pext \id} \ext argn}]
}{
    \Delta \vdash
    \ad{\subs{\constr(ic)}{p \ext argn \ext argr}}
      {\trans{\ind(I)}{\mu \pext \subs{\proj{c}{ind}}{p \ext argn}}}
        \convop \hspace{3em} \\
    \subs{\constr(ic)}{p' \ext
      \ad{(argn \ext argr)}{
      \trans{\left(\subs{\conData(c)}{\id \ext* \ind(I)}\right)}{\mu}}} \\
    \hspace{7em}
      \ty \subs{\ind(I)}{p' \ext \ad{\subs{\proj{c}{ind}}{p \ext argn}}{{\trans{\Theta_{ind}}{\mu}}}}
}
\label{rule:ind-ad-eq}
\end{mathpar}
\end{small}

\Description{Computation rule for inductive adapters on a constructor. A very
  generalised version of a map function, relying on the action of a transformation
  on a telescope.}
\caption{Inductive adapter equation, without and with indices}
\label{fig:inductive-ad}
\end{figure}

From the data of \cref{fig:inductive-constr}, we can already derive
an adapter \(\trans{\ind(I)}{\tilde{\mu}}\) given any
\(\typing{\Delta}{\tilde{\mu}}[\transso[\Gamma_{par} \ext \Theta_{ind}]{pi}{pi'}]\).
The missing piece is to give equality rules for
the action of this adapter on \(\constr\).
First, by extensionality for substitutions, such a \(\tilde{\mu}\)
must be of the form
\[\typing{\Delta}{\mu \ext \iota}[
    \transso[\Gamma_{par} \ext \Theta_{ind}]{p \ext \iota}{p' \ext \ad{\iota}{\trans{\Theta_{ind}}{\mu}}}]\]
for \(\typing{\Delta}{\mu}[\transso[\Gamma_{par}]{p}{p'}]\)
and \(\typing{\Delta}{\iota}[\subs{\Theta_{ind}}{p}]\). That is,
any “inductive adapter” consists of a transformation
between two parameter substitutions, and a constraint between
indices of the source and target types. Following this decomposition, we aim
to characterize a term of the form
\[\ad{\subs{\constr(ic)}{p \ext argn \ext argr}}{\trans{\ind(I)}
  {\mu \pext \subs{\proj{c}{ind}}{p \ext argn}}}\]
Indeed, as the term constructor is given in its universal context in
\cref{fig:inductive-constr}, we must also account for a substitution
\(p \ext argn \ext argr\).

In the end, we obtain \ruleref{rule:ind-ad-eq}
of \cref{fig:inductive-ad}. The core idea is more readable
without indices, in the simplified \ruleref{rule:ind-ad-eq-par}:
we rely on the fact that \(\conData\) is a telescope to derive
an action of the transformation \(\mu\) on it, which provides an
adapter by which we can adapt the arguments of the constructor.
The adapter we use,
  \(\trans{\left(\subs{\conData(c)}{\id \ext* \ind(I)}\right)}{\mu}\),
is obtained directly from the type of \(arg\). However, we can compute it
with the equation for the whiskering, and derive it is equal to
\(\trans{\conData(c)}{\left(\mu \ext* \trans{\ind(I)}{\mu} \right)}\).
This intuitively corresponds to the fact that the transformation acts on the
\(\conData\) by using \(\mu\) for the parameters, and recursively relies on
\(\trans{\ind(I)}{\mu}\) for recursive positions.
The final rule is more noisy, as it has to account for indices, but as
these are essentially constrained by a definitional equality, there is not much
more happening on their end.

\begin{example}[Computation of adapters]
We can now revisit \cref{ex:sig-list,ex:sig-vec,ex:sig-sum,ex:sig-w,ex:sig-eq} in \cref{fig:adapttt-adapters-from-ctx}.
We switch back to named syntax, and use named constants for
term and type constructors, in the obvious way:
\(\const{nil} \defconvop \constr(ic)\)
where \(ic : c_{nil} \in \const{listDesc}\),
and \(\List \defconvop \ind(\const{listDesc})\). We also conflate
adaptert \(f \ty \adso{A}{A'}\) and transformations
\(\emp \ext* f \ty \transso{\emp \ext* A}{\emp \ext* A}\), and similarly for
substitutions.
\end{example}
\begin{figure}
\begin{mathpar}
  \small
  \inferdef{ }{\typing{(X : \Ty_{+})}{\List}} \and
  \inferdef{ }{\typing{(X : \Ty_{+})}{\const{nil}}[\subs{\List}{X}]} \and
  \inferdef{ }{\typing{(X : \Ty_{+}) \ext (x : X) \ext (y : \subs{\List}{X})}
    {\const{cons}}[\subs{\List}{X}]} \and
  \inferdef{
    \typing{\Delta}{f}[\adso{A}{A'}] \\
    \typing{\Delta}{a}[A] \\
    \typing{\Delta}{l}[\subs{\List}{A}]
  }{
    \conv{\Delta}{\ad{\subs{\const{nil}}{A}}{\trans{\List}{f}}}
    {\subs{\const{nil}}{A'}}[\subs{\List}{A'}] \\
    \conv{\Delta}{\ad{\subs{\const{cons}}{A \ext a \ext l}}
      {\trans{\List}{f}}}
    {\subs{\const{cons}}{A' \ext \ad{a}{f} \ext \ad{l}{\trans{\List}{f}}}}
    [\subs{\List}{A'}]
  }
  \and
  \inferdef{ }{\typing{(X : \Ty_{+}) \ext (n : \Nat)}{\const{Vec}}} \and
  \inferdef{ }{\typing{(A : \Ty_{+})}{\const{nil}}[\subs{\const{Vec}}{A \ext \const{0}}]}
  \and
  \inferdef{ }{
    \typing{(A : \Ty_{+}) \ext (x : A) (y : \Nat) \ext (z : \subs{\const{Vec}}{A \ext y})}
    {\const{cons}}[\subs{\const{Vec}}{A \ext \subs{\const{S}}{y}}]
  } \and
  \inferdef{
    \typing{\Delta}{f}[\adso{A}{A'}] \\
    \typing{\Delta}{a}[A] \\
    \typing{\Delta}{n}{\Nat} \\
    \typing{\Delta}{v}[\subs{\const{Vec}}{A \ext n}]
  }{
    \conv{\Delta}{\ad{\subs{\const{nil}}{A}}{\trans{\const{Vec}}{f \ext \const{0}}}}
      {\subs{\const{nil}}{A'}}[\subs{\const{Vec}}{A \ext \const{0}}] \\
    \conv{\Delta}
      {\ad{\subs{\const{cons}}{A \ext a \ext n \ext v}}{
        \trans{\const{Vec}}{f \ext \subs{\const{S}}{n}}}}
      {\subs{\const{cons}}{A' \ext \ad{a}{f} \ext n \ext
        \ad{v}{\trans{\const{Vec}}{f \ext n}}}}
      [\subs{\const{Vec}}{A' \ext \subs{\const{S}}{n}}]
  }
  \and
  \inferdef{ }{\typing{(X : \Ty_{+}) \ext* (Y : \Ty_{+})}{+}} \and
  \inferdef{ }{\typing{(X : \Ty_{+}) \ext* (Y : \Ty_{+}) \ext (x : X)}{
      \const{inl}}[\subs{+}{X \ext* Y}]} \and
  \inferdef{ }{\typing{(X : \Ty_{+}) \ext* (Y : \Ty_{+}) \ext (y : Y)}
    {\const{inr}}[\subs{+}{X \ext* Y}]} \and
  \inferdef{
    \typing{\Delta}{f}[\adso{A}{A'}] \\
    \typing{\Delta}{g}[\adso{B}{B'}] \\
    \typing{\Delta}{a}[A] \\
    \typing{\Delta}{b}[B] \\
  }{
    \conv{\Delta}{\ad{\subs{\const{inl}}{A \ext* B \ext a}}{\trans{+}{f \ext* g}}}
      {\subs{\const{inl}}{A' \ext* B' \ext \ad{a}{f}}}[\subs{+}{A' \ext* B'}] \\
    \conv{\Delta}{\ad{\subs{\const{inr}}{A \ext* B \ext b}}{\trans{+}{f \ext* g}}}
      {\subs{\const{inr}}{A' \ext* B' \ext \ad{b}{g}}}[\subs{+}{A' \ext* B'}]
  }
  \and
  \inferdef{ }{\typing{(X : \Ty_{+}) \ext* (Y : X.\Ty_{-})}{\W}} \and
  \inferdef{ }{\typing{(X : \Ty_{-}) \ext* (Y : X.\Ty_{+}) \ext (x : X)
    \ext (z : \subs{Y}{\id \ext x} \to \subs{\W}{X \ext* Y})}{
      \const{sup}}[\subs{\W}{X \ext* Y}]} \and
  \inferdef{
    \typing{\Delta}{f}[\adso{A}{A'}] \\
    \typing{\Delta \ext (x : A)}{g}[\adso{\subs{B'}{\id \ext \ad{x}{f}}}{B}] \\
    \typing{\Delta}{a}[A] \\
    \typing{\Delta}{s}[\subs{B}{\id \ext a} \to \subs{\W}{A \ext B}]
  }{
    \conv{\Delta}{\ad{\subs{\const{sup}}{A \ext* B \ext a \ext s}}{\trans{\W}{f \ext* g}}}
      {\subs{\const{sup}}{A' \ext* B' \ext \ad{a}{f} \ext \ad{s}{\trans{\to}{\subs{g}{\id \ext a} \ext* \trans{\W}{f \ext* g}}}}}
      [\subs{\W}{A' \ext* B'}]
  }
  \and
  \inferdef{ }{\typing{(X : \Ty_{+}) \ext (x : X) \ext (y : X)}{\Id}} \and
  \inferdef{ }{\typing{(X : \Ty_{+}) \ext (x : X)}{
      \const{refl}}[\subs{\Id}{X \ext x \ext x}]} \and
  \inferdef{
    \typing{\Delta}{f}[\adso{A}{A'}] \\
    \typing{\Delta}{a}[A]
  }{
    \conv{\Delta}
      {\ad{\subs{\const{refl}}{A \ext a}}{\trans{\Id}{f \ext a \ext a}}}
      {\subs{\const{refl}}{A' \ext \ad{a}{f}}}
      [\subs{\Id}{A' \ext \ad{a}{f} \ext \ad{a}{f}}]
  }
\end{mathpar}

  \Description{Explicit computation of the computation rules for various
    inductive adapters. They closely resemble what one would naturally write
    as the definition of a "map" operation for these various types.}
  \caption{Example computation of adapters for inductive types}
  \label{fig:adapttt-adapters-from-ctx}
\end{figure}

\begin{nonexample}[Mixed variance]
\label{ex:mixed-variance}

Types in \AdapTTt being fully covariant (or contravariant)
slightly limits our expressivity.
% is a slight limitation as to the inductive types we can cover.
%
Consider the (simplified version of the) type of indexed
trees, \(\IW\), given by parameters
\(\Gamma_{\IW} \defconvop (I : \Ty_+)\,(X : \Ty_+)\,(Y : \Ty_-)\, (d : X \to Y \to I)\),
indices \(\Theta_{\IW} \defconvop I\), and constructor
\(\const{sup} \ty \P (i : I)\,(x : X)\,(\P (y : Y).\IW I\,X\,Y\,(d\, x\, y)).\IW X\,Y\,i \).
However, as \(Y \to I \ty \Ty_{+}\) but \(X \ty \Ty_{+}\),
the type \(X \to Y \to I\) of \(d\) is actually ill-formed. Making
\(X\) contravariant instead is not a solution, as it would break the quantification
over \(x : X\) ---and be inconsistent with plain \(\W\).

This is not an oversight. Indeed,
if one tries to describe an ad-hoc adapter between \(\IW\,I\,X\,Y\,d\)
and \(\IW\,I'\,X'\,Y'\,d'\), one certainly needs
adapters \(a_{I} \ty \Ad(I,I')\), \(a_{A} \ty \Ad(A,A')\) and \(a_{B} \ty \Ad(B',B)\).
But when trying to write the constraint between \(d\) and \(d'\), one ends up needing
\[\conv{(x : X) \ext (y : Y)}
  {\ad{\left(d\,\ad{x}{a_{A}}\,y\right)}{a_{I}}}{d'\,x\,\ad{y}{a_{B}}}[I']\]
The adapters \(a_{I}\), \(a_{A}\) and \(a_{B}\)
act \emph{simultaneously on both sides}, which goes beyond
Rules \nameref{rule:trans-tm-intro} and \nameref{rule:tm-trans-intro}.
Describing such mixed actions seems possible using spans,
% however the corresponding categorical structure and syntax
but this would be considerably more involved than our
current setup, so for now we refrain from exploring it.
This issue also explain why our telescopes have a uniform variance.
\end{nonexample}

\begin{comment}
For example, given a signature for the type $\List$ of lists, with a context of parameters consisting of a single type variable $\Gamma_{par} \coloneq (x \ty \Ty)$ and an empty telescope of indices, we can say that, given two
substitutions $\sigma,\tau : \Sub(\Delta,\Gamma_{par})$, and a transformation $\mu:  \Trans(\Delta,\Gamma_{par},\sigma,\tau)$, there exists an adapter between $\ind(\List)[\sigma]$ and $\ind(\List)[\tau]$.
Thanks to eta-rules for substitutions and transformations, we know that the data of $\sigma$ (resp. $\tau$) is a mapping of the type variable $x$ to types $A,B \ty \Ty(\Delta)$, and the data of $\mu$ is that of an adapter $ad \ty \Ty(A,B)$.
What we get from this is in effect an adapter between $\List A$ and $\List B$: our type former is indeed functorial, and acts covariantly over its parameter.
\end{comment}

\begin{comment}
\subsection{Examples \& Non-Examples}

\paragraph{Invariance}

Since we only maintain covariance or contravariance information, we cannot
express types that mix the two variance information such as $A \vdash A \to A$.
%

% \begin{example}[In subtyping]
%   \label{ex:func-subty}
%   As adapters are irrelevant, the action of coercion should never “look”
%   at the witness, only at the endpoints. \mlb{Explain the example of \(\P\)}
% \end{example}
\end{comment}

\paragraph{Recursors and fusion law} As it is mainly orthogonal to our
functoriality concerns and very tedious, we do not describe the derivation of
a recursor for our datatype descriptions here.

There is, however, one interaction worth investigating, that was dubbed \emph{fusion law}
\cite{Subtyping2024}. In \cref{sec:negative-types}, we described the action of
adapters only on destructors, but, thanks to η laws, this
entailed a derivable equality for adapters on constructors. For
inductive types, we only have an equation for adapters on constructors,
which entails no equation for the recursor in the absence of an η law.
The fusion law is such an equation, saying that applying a recursor on a
term \(\ad{t}{\indAd(I,\mu)}\) amounts to applying to \(t\) a
recursor where \(\mu\) is pushed in the branches.
Categorically, this says that the family (over \(\Gamma_{par}\))
of initial algebras \(\ind(I)\) is natural, in that
computing the section of a displayed algebra after the action of \(\mu\) coincides with
the section of a different displayed algebra, directly pre-composed with \(\mu\).
% We leave the exploration of this fusion law for future work.

% \subsection{Description for \NMDO}

% \km{Can we describe a construction of the type of descriptors on presheaves over a \NMDO ?}
\paragraph{Functorial type formers from descriptions}
To wrap up this section, we revisit \cref{thm:adapttt-over-adaptt}
with inductive type descriptions at hand.
% we now have all the pieces required to describe what it means for a \NMDO to support a functorial inductive type presented by a description in \AdapTTt{}.
%
Unfolding the construction of the model of \AdapTT{} over a \NMDO{}, the constant introduced by a description \(I \ty \IndDesc(\Gamma_{par},\Theta_{ind})\) in \AdapTTt{}
lets us compute the rules \(C\) must satisfy to support
the functorial inductive type former presented by \(I\).

To illustrate this, consider the description of
\(\W\)-types from \cref{ex:sig-w}.
The interpretation of \(\Gamma_{\W} \coloneq (X : \tyext{\emp}[+]) \ext*_{+} (Y \ty \tyext{\emp \ext \tyvz}[-])\) via \cref{thm:adapttt-over-adaptt}
yields a \(\Cat\)-valued presheaf that can be evaluated at an object \(\Gamma \) of \(C\): its objects are pairs \((\typing{\Gamma}{A}, \typing{\Gamma \ext (x : A)}{B})\)
of a type in \(\Gamma\) and a family over it,
while its morphisms are pairs of adapters \((\typing{\Gamma}{f}[\adso{A}{A'}], \typing{\Gamma \ext ( x : A )}{g}[\adso{\subs{B'}{\id \ext \ad{x}{f}}}{B}])\).
Requiring functorial \(\W\) types to exist then corresponds to the admissibility of the two following rules, whose premises have been computed from \(\Gamma_{\W}\).
\begin{mathpar}
  \small
  \inferdef{
    \typing{\Gamma}{A}\\
    \typing{\Gamma, x : A}{B}}
  {\W A\,B}
  \and
  \inferdef{
    \typing{\Gamma}{f}[\adso{A}{A'}]\\
    \typing{\Gamma \ext ( x : A )}{g}[\adso{\subs{B'}{\id \ext \ad{x}{f}}}{B}]}
  {\typing{\Gamma}{\trans{\W}{f \ext* g}}[\adso{\W\,A\,B}{\W\,A'\,B'}]}
\end{mathpar}
We can also compute the action of this adapter on the constructor.
\begin{mathpar}
  \inferdef{
    \typing{\Delta}{f}[\adso{A}{A'}] \\
    \typing{\Delta \ext (x : A)}{g}[\adso{\subs{B'}{\id \ext \ad{x}{f}}}{B}] \\
    \typing{\Delta}{a}[A] \\
    \typing{\Gamma \ext (b : \subs{B}{\id \ext a})}{k}[\W\,A\,B]}
  {\conv{\Delta}{\ad{\const{sup}\,a\,(\l b. k)}{\trans{\W}{f \ext* g}}}
      {\const{sup}\,(\ad{a}{f})\,(\l\, b'. \ad{(k\,(\ad{b'}{g}))}{\trans{\W}{f \ext* g}})}[\W\,A'\, B']
  }
\end{mathpar}

\begin{comment}
For completeness, we also state the corresponding rules for the unique constructor of \(\W\), computed from the constructor description \(c_{sup}\)
\begin{mathpar}
  \small
  \inferdef{
    \typing{\Gamma}{a}[A]\\
    \typing{\Gamma \ext (b : \subs{B}{\id \ext a})}{k}[\W\,A\,B]}
  {\typing{\Gamma}{\const{sup}\,a\,(\l\, b. k)}[\W\,A\,B]}
\end{mathpar}
\end{comment}

In general, considering descriptions in \AdapTTt which are more constrained
and informative thanks to variance information, we can derive
everything needed for a functorial type former
in the sense of \cref{def:type-former,def:functorial-type}:
the source category of the type former as functor, the way to construct
a type and adapter from this information, and the equations satisfied by
said adapter.

% \subsection[From AdapTT2 to AdapTT]{From \AdapTTt to \AdapTT}

% \mlb{Explain how we can use \AdapTTt to derive the extra structure needed in
% \AdapTT/a \NMDO to support a given type formers "properly".}

\section{Related work}
\label{sec:related-work}

\paragraph{Type casting}
Type casting is ubiquitous in dependent type theory, and has been
studied in many forms. The most pervasive one is provided by propositional
equality, which itself comes in many flavours: inductive \cite{MartinLoef1984},
univalent \cite{UniFoundationsProgram2013}, cubical \cite{Cohen2015},
observational \cite{Altenkirch2007,Pujet2022}… Observational equality is the
most directly structural, and it is hence the most natural to relate to \AdapTT. However,
since we do not force adapters to be unique,
other forms of equality could a priori work with \AdapTT,
similarly to \cref{ex:ttobs-cwfs}. It might be particularly interesting
to explore how \AdapTT could interact with the ongoing work of higher observational
type theory \cite{Altenkirch2024}.

Another common approach to type casting is subtyping. Here the main
divide is between subsumptive subtyping, where the same underlying term is given
multiple types, and coercive subtyping, with an explicit operation.%
\footnote{The subsumptive/coercive terminology is due to \textcite{Luo1999}.}
Subsumptive subtyping is closer to what users expect ---not having to
insert coercions--- but has a non-algebraic character which makes it
impossible to capture with GATs, and causes theoretical issues \cite{Lungu2018},
although it can be described with a more refined fibrational view
\cite{Mellies2015}. However, \textcite{Subtyping2024} show that focusing
on coercive subtyping, which is the one \AdapTT allows (see \cref{ex:subtyping}),
is not really an issue, as a flexible enough coercive system can emulate
a subsumptive one. For this, though, functoriality equations are paramount.

\begin{shepherdenv}
\paragraph{Comprehension categories}

\textcite{Coraglia2024,Coraglia2024a} analyze comprehension categories \cite{Jacobs1993} as models of type theory in terms of natural models and remark they carry structure akin to (proof-relevant) subtyping.
\textcite{Najmaei2025} further this comparison, chiefly motivated by semantics, and explore models generated by weak algebraic factorization system with a natural induced notion of morphism between types.
We arrive at a similar understanding from the opposite perspective: even though our initial motivation lie in the syntax for casts/subtyping, we also converge to the notion of comprehension categories.
This coincidence of two different research lines with rather different motivations is remarkable, and shows how natural \NMDO{}s are.

The emphasis on models in these works comes with an important difference regarding
strictness: \textcite{Najmaei2025} focus on a syntax faithfully representing comprehension categories, where fibrations are not necessarily split.
Hence, many correspondences are only up to isomorphism, \eg \(\subs{A}{\id}\) and \(A\) are isomorphic rather than convertible.
As we aim primarily for a usable syntax, we were happy to strictify these equations --
as done in \citeauthor{Najmaei2025}'s CCTT${}_{\mathrm{split}}$ --,
trading splitness for a lighter syntax.
\citeauthor{Coraglia2024a} similarly mention \enquote{[they] have written the action of reindexing as if the fibrations involved were split [which] allows [them] to simplify notation in the rules}.

The treatment of functorial types presents a second difference.
\citeauthor{Najmaei2025} focus on a handful of types (\(\P\),\(\Sig\) and \(\Id\)), for which they give dedicated rules in their equivalent CCTT of our \AdapTT.
This presentation does not provide a general framework to capture the functoriality of types, as granted by \AdapTTt.
On the other hand, they are careful to separate the structure for a given type former on a model from the extra subtyping structure for that type former.
We take the different view, embodied by \AdapTTt, that \emph{every} type (former) should be functorial --- possibly with respect to a discrete source category.
\end{shepherdenv}

\paragraph{Representing inductive types}
\textcite{Chapman2010} pioneered the description of inductive types through a type of descriptions.
\textcite{Escot2022} leveraged similar descriptions for metaprogramming in~\Agda{}, introducing telescopes as we do.
Our notion of inductive descriptions remain external as a judgment, heavily inspired by the theories of signatures of \textcite{KovacsPhD,Kaposi2020a}.
The very structured descriptions provided by these approaches are crucial for us,
and we could not use encodings via (indexed) containers \cite{Abbott2005,Altenkirch2015},
which fundamentally rely on a detour via universes and large elimination.
% \mlb{More? Epigram?}

\paragraph{Functoriality and parametricity} Beyond the works already
mentioned featuring structural coercions
\cite{Subtyping2024,LennonBertrand2022,Altenkirch2007,Pujet2022,Pujet2025},
another line with strong connections to ours is that on univalent parametricity
\cite{Tabareau2018}. Although relying on (binary) parametricity rather than
functoriality, the end result is a form of transport that follows the structure
of type formers. The subsequent work on \textsc{Trocq} \cite{Cohen2024}
extends this approach to a rich hierarchy of
possible relations between types along which one can (partially) transport,
from the basic existence of a function to full-blown equivalence,
the only one covered by univalent parametricity.
This careful study could inform the extension of \AdapTTt
with subtler forms of variance, while our work could inform how
to extend \textsc{Trocq} to handle inductive types.
\textcite{Arkor2022} also models type variables in System F using 2-categories.

\paragraph{Directed and Multimodal Type Theories}
\Cref{thm:adapttt-over-adaptt} is directly inspired by the work of \textcite{Licata2011} on a \(2\)-dimensional directed type theory corresponding to \(\Cat\).
Further work on \(2\)-dimensional directed type theory have focused on the internalization of morphisms as a type~\cite{North19,NeumannA24} and on bicategorical models~\cite{AhrensNW23}.
\textcite{Garner09} presents a \(2\)-dimensional variant of MLTT with \emph{undirected} \(2\)-cells as a step toward higher-models of identity types, recently extended to all type formers of \MLTT with weak computational rules by \textcite{Spadetto25}.
In the setting of homotopy type theory, \textcite{riehl2017type} propose a framework for synthetic \((\infty,1)\)-categories where every function between types behaving as categories (Segal types) are automatically functorial, with a cubical and constructive variant investigated by \textcite{WeaverL20}.
\textcite{GratzerWB25} design another variant based on simplicial type theory relying on multi-modal dependent type theory~\cite{GratzerKNB21} which we could take inspiration from for extending \AdapTTt{}.

%
% His construction of the syntactic $(2,1)$-comprehension categories features $2$-cells between substitutions in the context category as telescopes of identity proofs, a groupoidal version of our transformations.\km{Got reminded the existence of this paper by Matteo Spadetto who is currently writing something similar for Axiomatic Type Theory; also \cite{licataCanonicity2DimensionalType} but entry should be corrected}

\paragraph{Presheaf Models of Dependent Type Theory}
Using \posscite[Ch. 4]{Hofmann1997b} model of MLTT in presheaves is a recurrent theme in our work.
\textcite{UemuraPhD} proposes to use presheaf categories %(or rather the corresponding fibered version of categories of discrete fibrations)
as a semantics for Second-Order Generalized Algebraic Theories (SOGAT), that we employ (\cref{fig:sogat-cwfs}) to give a concise description of \AdapTT{}.
The articulation of \AdapTT{} and \AdapTTt{} is reminiscent of $2$-level type theory~\cite{AnnenkovCKS23} and the internal construction of the $2$-category of categories mirrors the axiomatic construction of cubical models in~\textcite{OrtonP18}.
\textcite{Weber2007} introduces the notion of (elementary) \(2\)-topos to abstract over the typical case of a \(2\)-category of internal categories, suitable for a general semantics of \AdapTTt{}.
%

% \km{TODO: Related work by Andras Kovacs on using 2LTT for metaprogramming !!!}

\section{Perspectives}
\label{sec:future-work}

\shepherd{
We have provided a foundational framework to understand type casting operations
in dependent type theory. This framework has close kinship to existing
classes of models, particularly comprehension
categories~\cite{Jacobs1993,Coraglia2024a,Najmaei2025}, although it is closer to
established syntaxes and type systems, making it appealing for potential
implementations. Our main contribution is the systematic study of functorial
type formers in this framework, providing a general explanation of the
structural aspects of type casting. This relies on the introduction
of variable for dependent types, giving rise to an elegant
2-categorical structure which let us easily derive functoriality for
complex composite types. The work culminates in the derivation of functoriality
for a general class of indexed inductive types, from their description
in a suitable theory of signatures.

We hope our work will inform the design of novel coercion
systems in proof assistants and dependently typed
programming languages, and pave the way for further study of topics
which are both critical for practical usability and theoretically
rather poorly understood, despite longstanding interest.
Many avenues are still open for further theoretical and practical investigations.
}

\paragraph{Variations on variance}
\AdapTTt only features co- and contravariant types and type variables, leaving the field open to extensions that we sketch here.
% but
% there is a richer nuance of variance that one might want to consider.
\shepherd{
First, allowing mixed variance in types and telescopes would support~\cref{ex:mixed-variance}.
From the analysis there, handling mixed variance, that is, types whose action on
a transformation generates some more complex data than a single adapter in one
or the other direction, seem to require working with spans.

Second, an equivariant direction characterizing types with only trivial adapters would be useful to describe types from \AdapTT which cannot be directly typed in \AdapTTt.
For instance, the identity function's type \(A \to A\) needs \(A\) to be both co- and contravariant, which could be obtained for an equivariant \(A\).
We conjecture this to be the sole obstruction to embed \AdapTT{} into \AdapTTt{}.
}
% with such an equivariant direction one
% could easily embed \AdapTT in an “equivariant fragment” of \AdapTTt.
% \mlb{Semantics?}
% Similarly to contravariance, would also
% have an operation on contexts which makes all variables equivariant, and corresponds semantically
% \mlb{Complete.}
% The identity's type, then would be well-formed for any equivariant type \(A\).

% \km{Pretty convinced this issue is not specific to universes}
% Such an equivariance direction would also be useful to handle universes in \AdapTTt.
% We did not consider universes in \AdapTTt as they have no interesting functorial aspect.
% Moreover, as we have no term-level order, universes are necessarily discrete,
% and would thus decode to equivariant types.
% This is in line with typing the polymorphic identity's type
% \(\P x : \univ.\El(x) \to \El(x)\).

Third, strict positivity of inductive types can itself be expressed using variance~\cite{Escot2023}.
Doing so with the infrastructure of \AdapTTt{} would simplify the description of inductive types, and enable nested types, \ie inductive types where a recursive occurrence appears as
parameter of another inductive type.
Moreover, \citeauthor{Escot2023} report that their \Agda implementation suffers from
interactions of variance with structural subtyping, which would
could alleviated using \AdapTTt.

\paragraph{Generating adapters}
Beyond the examples of \TTobs, \MLTTcoe and \CastCIC, many more instances of ground adapters would make sense as extensions \AdapTT{}: user-defined coercions à la \Rocq or \Lean, cumulativity as an adapter
\(\Ad(\univ[i],\univ[j])\) between universes whenever $i \leq j$, record coercions as a building block for algebraic hierarchies~\cite{Pollack2000}, refinement between subset types as employed in PVS~\cite{Shankar2000}, \Fstar~\cite{Swamy2016} and liquid type systems~\cite{Rondon2008}.
%
% Analyzing these instances of type cast through the lens of \AdapTT{} could shed light on the potential obstructions to structural type casting.
%
\AdapTT gives a good lens to (re)analyze these instances of type casts,
and the framework provided by \AdapTTt could greatly help %and strengthen their structural lifting,
in obtaining enhanced, structural versions,
well-behaved thanks to definitional functoriality.

\paragraph{Metatheory \& Implementation}
We did not attempt to prove normalization or decidability of type-checking for our languages, though we conjecture these properties should hold.
Indeed, very few such proofs cover a general class of inductive types, limiting the interest of the challenge.
% The main reason is that there are very few, if any, such proofs that cover
%
However, despite its rich equational theory, conversion in \AdapTTt is intentional enough to be decidable.
At a high level, apart from the unproblematic β and η rules, deciding the additional functor laws should follow the strategy of \MLTTmap{}~\cite{Subtyping2024}
and reduce to functoriality of adapters at inductive types that can be dealt
by using ideas from \posscite{Allais2013} ν-equations.

An important insight by \textcite{McBride2021}, which does not appear in
our algebraic approach to typing,
is the close relationship between adapters and bidirectional typing:
adapters naturally mediate
% where one has “one type too much”, \ie
the phase change between inference and checking
% , as a way to mediate between the inferred and checking types.
% for a term \(t\) and the one against which \(t\) is checked.
We expect that these aspects would surface in a future implementation of \AdapTT{} and \AdapTTt{}.
% As for normalization, these aspects
% would likely surface in an implementation of our ideas, which is an
% intriguing perspective.

\section*{Data Availability Statement}

The \Agda formalisation used to guide the rules' design by
type-checking them is freely available on Zenodo \cite{Formalisation}.

\begin{acks}
We would like to acknowledge and thank Théo Laurent who shared his insights from preliminary investigations on this topic.
We thank Conor McBride and Fredrik Nordvall Forsberg for hosting a research visit that fueled our interest in adapters, Rafael Bocquet, Thiago Felicissimo, Josselin Poiret, András Kovács, Benedikt Ahrens, Niyousha Najmaei, Paige North and Niels van der Weide for very valuable discussions, and the anonymous reviewers for their insightful comments.
%km: Rafael Bocquet, Josselin Poiret, András Kovacs ; also discussions with Niyousha Najmaei, Niels van der Weide (?)
% mlb/tb/aa: people in Cambridge? Lingyuan? --@meven: how much did you discuss this with Ioannis ? What about people from TYPES? Fernando ?
% Nathanael Arkor

This work was supported in part by a
\grantsponsor{erc}{European Research Council (ERC)}{https://erc.europa.eu/homepage}
Consolidator Grant for the project “TypeFoundry”, funded under the European Union’s Horizon 2020 Framework Programme (grant agreement no. \grantnum{erc}{101002277}).
\end{acks}

\printbibliography

\appendix

\section{Second order presentation of \AdapTT{}}
\label{sec:full-rule-adaptt}

\begin{figure}
\begin{align*}
  &\GatTy \ty \GatSort
  &&\GatTm~:~\GatTy \to \GatRepSort \\
  &\GatAd~:~\GatTy\to\GatTy \to \GatSort
  &&\GatAdapt{\cdot}{\cdot}~:~\{A\,B: \GatTy\}\to \GatAd\,A\,B \to \GatTm\,A\to\GatTm\,B\\
  &\GatId~:~\{A: \GatTy\} \to \GatAd\,A\,A
  &&\GatComp~:~\{A\,B\,C: \GatTy\} \to \GatAd\,B\,C \to \GatAd\,A\,B \to \GatAd\,A\,C\\
  &\GatIdL~:~\{A\,B : \GatTy\}(a : \GatAd\,A\,B) \to \GatId\GatComp a \equiv a
  &&\GatIdR~:~\{A\,B : \GatTy\}(a : \GatAd\,A\,B) \to a\GatComp \GatId \equiv a\\[.1em]
  &\mathrlap{\GatAssoc~:~\{A\,B\,C\,D : \GatTy\}(a : \GatAd\,A\,B)(b : \GatAd\,B\,C)(c : \GatAd\,C\,D)\to c\GatComp(b \GatComp a) \equiv (c \GatComp b)\GatComp a}\\
  &\GatAdaptId~:~\{A : \GatTy\}(t : \GatTm\,A) \to \GatAdapt{t}{\GatId} \equiv t
  &&\\
  &\mathrlap{\GatAdaptComp~:~\{A\,B\,C : \GatTy\}(a : \GatAd\,A\,B)(b : \GatAd\,B\,C)(t : \GatTm\,A) \to \GatAdapt{t}{b \GatComp a} \equiv \GatAdapt{\GatAdapt{t}{a}}{b}}
\end{align*}
  \caption{SOGAT definition of \AdapTT}
  \label{fig:sogat-cwfs}
\end{figure}

\Cref{fig:sogat-cwfs} presents a definition of \AdapTT{} using a second-order generalized algebraic theory (SOGAT)~\cite{UemuraPhD}.
In this presentation, only the part of the context that is required to specify a judgment is explicit: informally, all the sorts, operations and equations can be understood as parametrized by an implicit additional context.
The presentation of \cref{fig:adaptt-def} is derived from this presentation by the
translation of a SOGAT into a generalized algebraic theory (GAT) \cite{KaposiX24}.

The presentation of a bare category with families, which we extend, features
a sort $\GatTy$ classifying well-formed types, and a dependent sort
$\GatTm\,A$ classify well-formed terms of type $A \ty \GatTy$.
The sort $\GatTm$ of terms is flagged as a representable sort:
the construction of the corresponding (first-order) generalized algebraic theory uses this information to construct explicit contexts populated with variables for representable sorts, here only $\GatTm$. This is enough to derive all structural rules of a CwF, including
Rules \nameref{rule:adaptt-var-zero}, \nameref{rule:adaptt-wk},
\nameref{rule:adaptt-sub-ext} and their equations.

To this basic structure, we add $\GatAd\,A\,B$, the sort
of adapters between the types $A$ and $B$.
The operation $\GatId$ provides the identity adapter, while $\GatComp$ composes two adapters with compatible target and source.
Together with the equations $\GatIdL$, $\GatIdR$ and $\GatAssoc$, these operations equip $\GatTy$ with the structure of a category.
Again, because this is a SOGAT, all these sorts operations are automatically stable under
substitution in the derived GAT, \ie Rules \nameref{rule:adaptt-sub-ad}, \nameref{rule:adaptt-sub-ad-id},
and \nameref{rule:adaptt-sub-ad-comp}.

Adapters $a : \GatAd\,A\,B$ act on terms $t : \GatTm A$ to obtain another term $\GatAdapt{t}{a} : \GatTm B$.
This action must respect identity adapters ($\GatAdaptId$) and composition of adapters ($\GatAdaptComp$), hence satisfying functoriality laws that make \(\Ty\) a
($\cSet$-valued) functor on the category of types and adapters, thus an internal category
to presheaves on \(\Ctx\) after the translation.

% \begin{figure}
%   \[
%     \begin{array}{cl}
%       \ctxty{\Box} & \text{well-formed context} \\
%       \typing{\Gamma}{\Box}[\Delta] & \text{well-formed substitution in } \ctxty{\Gamma}, \ctxty{\Delta} \\
%       \typing{\Gamma}{\Box} & \text{well-formed type in } \ctxty{\Gamma}\\
%       \typing{\Gamma}{\Box}[\adso{A}{B}] & \text{well-formed adapter in } \typing{\Gamma}{A}, \typing{\Gamma}{B} \\
%       \typing{\Gamma}{\Box}[A] & \text{well-formed term in }\typing{\Gamma}{A} \\
%       \ctxconv{\Gamma}{\Delta} & \text{context conversion assuming } \ctxty{\Gamma}, \ctxty{\Delta} \\
%       \conv{\Gamma}{\sigma}{\tau}[\Delta] & \text{substitution conversion assuming} \typing{\Gamma}{\sigma}[\Delta], \typing{\Gamma}{\sigma}[\Delta] \\
%       \conv{\Gamma}{A}{B} & \text{type conversion assuming } \typing{\Gamma}{A}, \typing{\Gamma}{B} \\
%       \conv{\Gamma}{f}{g}[\adso{A}{B}] & \text{adapter conversion assuming } \typing{\Gamma}{f}[\adso{A}{B}], \typing{\Gamma}{g}[\adso{A}{B}] \\
%       \conv{\Gamma}{t}{u}[A] & \text{term conversion assuming }\typing{\Gamma}{t}[A], \typing{\Gamma}{u}[A]
%     \end{array}
%   \]
%   \caption{Judgments for \AdapTT}
%   \label{fig:appendix-adaptt-judgments}
% \end{figure}

\section{Correspondence between \texorpdfstring{\GCwF{}s}{GCwFs} and \texorpdfstring{\NMDO{}s}{NatModDO}}
\label{sec:gcwf-cwfs}

In this section, we show the correspondence between split \GCwF{}s and \NMDO{}s,
corresponding to \cref{sec:other-models} and in particular \cref{thm:cwfs-gcwf}.

\subsection{Categorical refresher}

\paragraph{2-categorical Yoneda}
We start by recalling the categorical material we need. First, recall the (2-categorical) Yoneda lemma~\cite{kelly1982}: given \(A\) in \(\catpsh{C}\), for every object \(c\) of \(C\), there is an isomorphism of categories \(A(c) \simeq \yon c \to A\) between \(A(c)\) and the category of functors \(\yon c \to A\).

\paragraph{Discrete opfibrations}
To assess the task at hand, we expand on~\cref{def:CwFs}, in particular regarding the condition that \(p\) is a discrete opfibration. We recall that a functor \(F \ty C \to D\) is a discrete opfibration if for every object \(a\) in \(C\) and any morphism \(f' \ty F(a) \to b'\) in \(D\), there exists a unique morphism \(f \ty a \to b\) in \(C\) such that \(F(f) = f'\), called the lift of \(f'\) at \(a\). Given any \(2\)-category \(C\), we say that a morphism \(f \ty a \to b\) in \(C\) is a discrete opfibration if for every \(c\) in \(C\), the functor \(C(c,f) \ty C(c,a) \to C(c,b)\) is a discrete opfibration of categories, and for any \(g \ty c\to d\), the following diagram is a morphism of discrete opfibrations,
\ie the lift of the image under the lower horizontal map is the image of the lift under the upper horizontal map:
\[
  \begin{tikzcd}
    C(d,a) \ar[r,"{C(g,a)}"]\ar[d,"{C(d,f)}"'] & C(c,a)\ar[d,"{C(c,f)}"]  \\
    C(d,b) \ar[r,"{C(g,b)}"'] & C(c,b)
  \end{tikzcd}
\]
On the morphism \(p \ty \Tm \to \Ty\) in the \(2\)-category \(\catpsh{\Ctx}\), the first part of this definition when evaluated on a representable presheaf \(\yon \Gamma\) gives that \(p_{\Gamma}\) is a discrete opfibration, which entails the existence of the type casting operation \(\ad{t}{a}\) and satisfaction of Rules \nameref{rule:adaptt-adapt-id} and \nameref{rule:adaptt-adapt-comp}. The second condition, when evaluated on \(\yon \sigma\) for some \(\sigma\ty \Gamma\to\Delta\) in \(\Ctx\) gives exactly the validation of Rule~\nameref{rule:adaptt-adapt-sub}.

\paragraph{Grothendieck construction}
We also make extensive use of the Grothendieck constructions and its properties, as well as the special case of the category of elements, and their link with fibrations and discrete fibrations. Given a functor \(F \ty C^{\op}\to \Cat\), we define the Grothendieck construction \(\int F\) as the category whose objects are pairs \((c,x)\) where \(c\) is an object of \(C\) and \(x\) is an object of \(F(c)\), and whose morphisms \((c,x) \to (d,y)\) are pairs \((f,\phi)\) where \(f \ty c \to d\) is a morphism in \(C\) and \(\phi \ty x \to F(f)(y)\) is a morphism in \(F(c)\). The first projection \(\int F \to C\) induces a split fibration. In fact, the Grothendieck construction induces an equivalence of \(2\)-categories between the \(2\)-category of functors \(C^{\op} \to \Cat\) and the \(2\)-category of split fibrations over \(C\)~\cite{JacobsCatlogTT}. The reciprocal functor is obtained by taking a split fibration \(F\) over \(C\) to the functor \(F^{-1}\ty C^{\op} \to \Cat\) sending every object to its fiber, the splitness axiom giving the functorial action.

\subsection[Proofs for Section 2.3]{Proofs for \cref{sec:other-models}}

We consider a split \GCwF given by the following data
\[
 \begin{tikzcd}
    \dot{\mathcal{U}}
    \ar[rr, bend right=15, "\Sigma"']
    \ar[rr,phantom, "\top"]
    \ar[dr, "\dot{u}"']
    &
    &
      \mathcal{U}
      \ar[ll , bend right=15, "\Delta"']
      \ar[dl, "u"]
    \\
    & \mathcal{B}
 \end{tikzcd}
\]
and we denote respectively \(\eta \ty \id \Rightarrow \Delta\Sigma\) and \(\epsilon \ty \Sigma\Delta \Rightarrow \id\) the unit and the counit of the adjunction. Given a morphism \(\sigma \ty \Gamma \to \Gamma'\) in the category \(\mathcal{B}\), and objects \(t\) and \(A\) in the fiber over \(\Gamma'\) of \(u\) and \(u'\) respectively, we denote respectively \(\bar{\sigma} \ty t[\sigma]\to t\) and \(\bar{\sigma} \ty A[\sigma]\to A\) the chosen cartesian lifts of \(\sigma\) at \(t\) and \(A\). Since \(\dot{u}\) and \(u\) are split this action respects composition and identities.

In order to provide the type cast operation of \cref{proposition:coercion-gcwf}, it will be convenient to consider a particular family of morphism in the split \GCwF. Given an object \(t\) of \(\dot{\mathcal{U}}\) and a vertical morphism \(a : \Sigma(t) \to B\) in \(\mathcal{U}\), we define
\[
  \tau_{t,a} = \dot{u}(\Delta(a)\circ \eta_{t})
\]
The special role of these morphisms can be understood in the light of \cref{thm:cwfs-gcwf}, where under the stated equivalence, \(t \ty \Tm(\Gamma)\) is a term of type \(A \ty \Ty(\Gamma)\) and \(a \ty A \to B\) is an adapter. Then the morphism \(\tau_{t,a}\) corresponds to the substitution \(\id_{\Gamma} \ext \ad{t}{a} \ty \Gamma \to \Gamma\ext B\). We thus use these morphisms, which are easy to construct in the \GCwF formalism, in order to recover the type cast operations. We first show the following technical result
\begin{lemma}\label{lemma:gcwf}
  With the same notations as above, the following equation holds
  \[
    \epsilon_{B} \circ \Sigma(\overline{\tau_{t,a}}) = \id_{B}
  \]
\end{lemma}
\begin{proof}
  First note that using the naturality of \(\epsilon\), the triangle equation
  for adjunctions and the verticality of \(a\) we have:
  \begin{align*}
    u(\epsilon_{B} \circ \Sigma(\overline{\tau_{t,a}}))
    &= u(\epsilon_{B}) \circ \tau_{t,a} \\
    &= u(\epsilon_{B}) \circ u(\Sigma (\Delta(a) \circ\eta_{t})) \\
    &= u(\epsilon_{B} \circ \Sigma\Delta(a) \circ \Sigma\eta_{t}) \\
    &= u(a \circ \epsilon_{\Sigma(t)} \circ \Sigma\eta_{t}) \\
    &= u(a) \\
    &= \id_{u(B)}
  \end{align*}
  Moreover, we note that the maps \(\epsilon_{B}\) and \(\overline{\tau_{t,a}}\) are cartesian, and \(\Sigma\) preserves cartesian maps, hence \(\epsilon_{B}\circ\Sigma(\overline{\tau_{t,a}})\) is cartesian. Being a cartesian map whose image is the identity, it is itself an identity by splitness of the fibration \(u\).
\end{proof}

We now prove a more precise version of \cref{proposition:coercion-gcwf}.
\begin{proposition}\label{proposition:coercion-gcwf-precise}
  in a split \GCwF, the functor \(\Sigma\) induces a discrete opfibration of \(\Cat\)-valued presheaves from \(\dot{u}^{-1}\) to \(u^{-1}\). Explicitly, considering an object \(t\) of \(\dot{\mathcal{U}}\) and a vertical morphism \(a : \Sigma(t) \to B\) in \(\mathcal{U}\), there exists a unique object \(\ad{t}{a}\) in \(\dot{\mathcal{U}}\) equipped with a vertical morphism \(\tilde{a} : t \to \ad{t}{a}\) such that
  \begin{align*}
    \Sigma(\ad{t}{a}) &= B
    &
      \Sigma(\tilde{a}) &= a
  \end{align*}
  This operation moreover satisfies the equation \(\ad{t}{a}[\sigma] = \ad{t[\sigma]}{a[\sigma]}\).
\end{proposition}
\begin{proof}[Proof of \cref{proposition:coercion-gcwf-precise}]
  Considering an object \(t\) of \(\dot{\mathcal{U}}\) and a vertical morphism \(a : \Sigma(t) \to B\) in \(\mathcal{U}\), we define \(\ad{t}{a} = \Delta(B)[\tau_{t,a}]\) with the vertical morphism \(\tilde{a} : t \to \ad{t}{a}\) given by the lifting property justified by the equation \(\dot{u}(\ad{t}{a}) = \dot{u}(t)\), depicted below in the category \(\dot{\mathcal{U}}\):
  \[
    \begin{tikzcd}[column sep = large]
      \ad{t}{a} \ar[r,"\overline{\tau_{t,a}}"] & \Delta(B) \\
      t \ar[u, dashed, "\tilde{a}"] \ar[ru, "\Delta(a)\circ\eta_{t}"']
    \end{tikzcd}
  \]

  It suffices to show that these satisfy the required equations. For the first equation, we note that since \(\epsilon_{B}\) is cartesian, we have \(\Sigma\Delta(B) = B[u(\epsilon_{B})]\). Thus, using \cref{lemma:gcwf}, we have
  \begin{align*}
    \Sigma(\ad{t}{a}) &= \Sigma\Delta(B)[\Sigma\overline{\tau_{t,a}}] =
                        B[\epsilon_{B}\circ \Sigma\overline{\tau_{t,a}}] = B
  \end{align*}

  The second equation is given by the following commutative diagram, where the left square commutes by definition of \(\tilde{a}\), the right square commutes by naturality, the top region commutes by the triangle equation, and the lower region commutes by \cref{lemma:gcwf}:
  \[
    \begin{tikzcd}[column sep = huge]
      \Sigma t
      \ar[r,"\Sigma(\eta_{t})"]
      \ar[d,"\Sigma\tilde{a}"']
      \ar[rr,bend left,equal]
      & \Sigma\Delta\Sigma t
        \ar[r,"\epsilon_{\Sigma t}"]
        \ar[d,"\Sigma\Delta a"]
      & \Sigma t
      \ar[d, "a"]
      \\
      B
      \ar[r, "\Sigma(\bar{\tau_{a}})"']
      \ar[rr,bend right,equal]
      & B[u(\epsilon_{B})] \ar[r,"\epsilon_{B}"']
      & B
    \end{tikzcd}
  \]

  We next prove uniqueness. Consider an object \(t'\) and a vertical morphism \(b \ty t \to t'\) such that \(\Sigma t' = B\) and \(\Sigma b = a\), we show that necessarily \(b = \tilde{a}\). For this we consider the following naturality square
  \[
    \begin{tikzcd}
      t \ar[r,"\eta_{t}"]\ar[d,"b"'] & \Delta\Sigma t\ar[d,"\Delta\Sigma b"] \\
      t' \ar[r,"\eta_{t'}"'] & \Delta\Sigma t'
    \end{tikzcd}
  \]
  Taking the image by \(\dot{u}\) of this naturality square yields by
  verticality of the morphism \(b\),
  \begin{align*}
    \dot{u}(\eta_{t'}) &= \dot{u}(\tau_{a})
  \end{align*}
  Since \(\eta_{t'}\) is cartesian it is its own chosen lift by assumption, we conclude that \(\eta_{t'} = \overline{\tau_{t,a}}\), which proves that \(t' = \ad{t}{a}\). Moreover, the morphism \(b\) is a vertical morphism making the following diagram commute
  \[
    \begin{tikzcd}
      \ad{t}{a} \ar[r,"\overline{\tau_{t,a}}"]& \Delta(B) \\
      t \ar[u,"b"] \ar[ur,"\tau_{t,a}"']
    \end{tikzcd}
  \]
  By uniqueness of such a morphism, we conclude that \(b = \tilde{a}\).

  Finally, we consider a morphism \(\sigma : \Gamma' \to \Gamma\) and show the equation \(\ad{t}{a}[\sigma] = \ad{t[\sigma]}{a[\sigma]}\). This is given by the following computation, where the second line is obtained by the fact that \(\Delta\) preserves chosen lifts, the third line is given by definition of \(a[\sigma]\), and the fourth line is given by naturality of \(\eta\) and by the fact that \(\Sigma\) preserves chosen lifts:
  \begin{align*}
    \ad{t[\sigma]}{a[\sigma]} &= \Delta(B[\sigma])[\tau_{t[\sigma],a[\sigma]}] \\
                              &= \Delta B [\Delta\sigma][\Delta(a[\sigma])\circ\eta_{t[\sigma]}] \\
                              &= \Delta B [\Delta(a)\circ\Delta(\sigma)\circ \eta_{t[\sigma]}] \\
                              &= \Delta B [\tau_{t,a}][\sigma] \\
                              &= \ad{t}{a}[\sigma] \tag*{\qedhere}
  \end{align*}
\end{proof}

\begin{theorem}
  The category of split \GCwF{}s and the category of \NMDO{}s are equivalent.
\end{theorem}
\begin{proof}
  Starting with a split \GCwF, we define a \NMDO by considering the presheaves of \(\Ty\) and \(\Tm\) to be given respectively by \(u^{-1}\) and \(\dot{u}^{-1}\). The morphism of fibrations \(\Sigma\) then induces a natural transformation \(p\ty \Tm\natt{}\Ty\). By \cref{proposition:coercion-gcwf-precise}, \(p\) is a discrete opfibration. Moreover, the existence of \(\Delta\) implies that \(p\) is representable. Indeed, given an object \(A\) in \(\Ty(\Gamma)\), we define \(\Gamma\ext A = \dot{u}(\Delta A)\), together with \(\wk = \dot{u}(\epsilon_{A})\) and \(\tmvz\) to be the element of \(\dot{\mathcal{U}}\) corresponding to \(\Delta(A)\). Since \(\epsilon_{A}\) is its own lift, \(\Sigma(\tmvz)=A[\wk]\) and so \(\tmvz\) is an object of \(\Tm(\Gamma\ext A,A[\wk])\). Given a morphism \(\sigma \ty \Gamma \to \Gamma'\) in \(\mathcal{B}\), and objects \(A \ty \Ty(\Gamma')\) and \(t \ty \Tm(\Gamma,A[\sigma])\), we define the morphism \(\sigma\ext t = \dot{u}(\Delta(\bar{\sigma})\circ \eta_{t})\) where \(\bar{\sigma} \ty A[\sigma] \to A\) is the cartesian lift of \(\sigma\) at \(A\) through \(u\). In order to check that these satisfy the adequate universal property, we verify the three following conditions:
  \begin{enumerate}
  \item Given \(\sigma \ty \Gamma \to \Gamma'\) with \(A \ty \Ty(\Gamma')\) and \(t \ty\Tm(\Gamma,A[\sigma])\), we have \(\wk\circ(\sigma\triangleright t) = \sigma\). This is proven by the following computation using the naturality of \(\epsilon\) and one triangle identity for adjunctions
    \begin{align*}
      \wk\circ(\sigma\triangleright t)
      &= u(\epsilon_{\Sigma\Delta\Sigma t})\circ\dot{u}(\Delta(\bar{\sigma})\circ\eta_{t}) \\
      &= u(\epsilon_{\Sigma\Delta\Sigma t} \circ\Sigma\Delta(\bar{\sigma})\circ\Sigma\eta_{t})\\
      &= u(\bar{\sigma} \circ \epsilon_{\Sigma t}\circ \Sigma\eta_{t}) \\
      &= u(\bar{\sigma})\\
      &= \sigma
    \end{align*}
  \item Given \(\sigma \ty \Gamma \to \Gamma'\) with \(A \ty \Ty(\Gamma')\) and \(t \ty\Tm(\Gamma,A[\sigma])\), we have \(\tmvz[\sigma\ext t] = t\). This is given by the fact that since \(\Delta\) preserves chosen lifts and \(\eta_{t}\) is the chosen lift of its projection, by splitness of \(\dot{u}\), the following morphism is the chosen lift of its projection
    \[
      \Delta(\bar{\sigma})\circ\eta_{t} \ty t \to \tmvz
    \]
  \item For \(\sigma \ty \Gamma \to \Gamma'\ext A\), we have \(\sigma = (\wk\circ\sigma)\ext \tmvz[\sigma]\). To show this, we denote \(\bar{\sigma} \ty \Delta(A)[\sigma] \to \Delta(A) \) the cartesian lift of \(\sigma\) at \(\Delta(A)\). Using the fact that \(\epsilon_{A}\) is the chosen lift of its projection, as well as the naturality of \(\eta\) and the other triangle identity for adjunction, we obtained the desired identity by the following computation:
    \begin{align*}
      (\wk\circ\sigma)\ext\tmvz[\sigma]
      &= \dot{u}(\Delta(\epsilon_{A}\circ\Sigma\bar{\sigma}) \circ\eta_{\Delta(A)[\sigma]}) \\
      &= \dot{u}(\Delta\epsilon_{A} \circ\eta_{\Delta(A)} \circ \bar{\sigma}) \\
      &= \dot{u}(\bar{\sigma}) \\
      &= \bar{\sigma}
    \end{align*}
  \end{enumerate}

  Conversely, starting with a \NMDO \(\Ctx\), we define a split \GCwF by letting \(\mathcal{B}\) be \(\Ctx\), the split fibration \(u\ty\mathcal{U}\to \mathcal{B}\) to be the Grothendieck construction of the functor \(\Ty \ty \Ctx \to \Cat\) and the split fibration \(\dot{u}\ty\dot{\mathcal{U}}\to\Cat\) to be the Grothendieck construction of the functor \(\Tm \ty \Ctx \to \Cat\). The natural transformation \(p \ty \Tm\natt{}\Ty\) then induces a morphism of split fibrations \(\Sigma\), which fits into the following picture:
  \[
    \begin{tikzcd}
      \int \Tm \ar[rr,"\int p"]\ar[rd,"\dot{u}"'] & & \int\Ty\ar[ld,"u"] \\
      & \mathcal{B}
    \end{tikzcd}
  \]
  where both \(u\) and \(\dot{u}\) are split opfibrations and \(\Sigma\) is a morphism of fibration that preserves chosen lifts. To complete the \GCwF structure, it suffices to check that \(\int_{\Gamma}\tyop_{\Gamma}\) has a right adjoint that preserves chosen lifts, with unit and counit being componentwise the chosen lift of their projection. We explicitly define the right adjoint
  \[
    \begin{tikzcd}[row sep = tiny]
      \int \Ty \ar[r, shorten >= 7pt]
      & \int\Tm \\
      (\Gamma,A)  \ar[r, mapsto, shorten <= 8pt ]
      & (\Gamma\ext A, \tmvz) \\
      (\Gamma,A) \ar[d,"{(\sigma,a)}"',""{name=1}]
      &  (\Gamma\ext A,\tmvz)
        \ar[d, "{((\sigma\circ\wk)\ext\ad{\tmvz}{a[\wk]} {,} a[\wk])}",""'{name=2}] \\[2em]
      (\Delta,B)
      & (\Delta\ext B,\tmvz)
        \ar[from=1,to=2,mapsto, shorten <= 14pt, shorten >= 20pt]
    \end{tikzcd}
  \]
  Note that \(\Delta\) sends \((\sigma,\id)\) onto a pair whose second component is an identity, and hence it preserves chosen lifts. Moreover, we have natural transformations
  \begin{align*}
    \eta_{\Gamma,t} & \ty  (\Gamma,t) \to  (\Gamma\ext p_{\Gamma}(t),\tmvz)
    & \epsilon_{\Gamma,A} & \ty (\Gamma\ext A, A[\wk]) \to (\Gamma,A) \\
    \eta_{\Gamma,t} & =  (\id \ext t, \id)
    &
      \epsilon_{\Gamma,A} & = (\wk,\id)
  \end{align*}
  Since the second component is always the identity, both these natural transformations are componentwise the chosen lift of their projection. In order to check the triangle identities it thus suffices to check their that they hold on the first components. The first triangle identity is given, for \(t \ty \Tm(\Gamma)\) by the equation
  \[
     \wk_{\Gamma,p_{\Gamma}(t)} \circ (\id_{\Gamma}\ext t) = \id_{\Gamma}
  \]
  The second equation is given, for \(A\ty\Ty(\Gamma)\), by the following computation:
  \begin{align*}
    ((\wk_{\Gamma,A} \circ \wk_{\Gamma\ext A,A[\wk]}) & \ext \tmvz_{\Gamma\ext A,A[\wk]}) \circ (\id_{\Gamma,A}\ext \tmvz_{\Gamma,A}) \\
    &= (\wk_{\Gamma,A} \circ \wk_{\Gamma\ext A,A[\wk]} \circ (\id_{\Gamma,A}\ext \tmvz_{\Gamma,A})) \ext \tmvz_{\Gamma\ext A,A[\wk]}[\id_{\Gamma,A}\ext \tmvz_{\Gamma,A}] \\
    &= \wk_{\Gamma,A}\ext \tmvz_{\Gamma,A} \\
    &= \id_{\Gamma\ext A}
  \end{align*}

  These two functors define an equivalence of \(2\)-categories, induced by the equivalence between split fibrations over \(\Ctx\) and functors \(\Ctx^{\op}\to \Cat\) defined by the Grothendieck construction and the fiber.
\end{proof}

\begin{shepherdenv}
\section{Alternative presentation of \AdapTT{}}
In this section, we present an alternate description of the theory
\AdapTT{}, that was suggested by an anonymous reviewer. This version is a
bit more verbose but gives insight regarding the required equality on term
constructors. We present here the rules together with a description of their
intended semantics, keeping in mind that it should have the same models as
\AdapTT{}, namely, \CwFs. This presentation is obtained by removing all the
equations that we impose on terms, and instead introducing an additional
judgment:
  \[
    \inferrule{
      \typing{\Gamma}{a}[A] \\
      \typing{\Gamma}{b}[B] \\
      \typing{\Gamma}{f}[\adso{A}{B}] \\
    }{\typing{\Gamma}{e}[\convover{f}{a}{b}]}
  \]
  The reader may think of this judgment as denoting
  \(\conv{\Gamma}{\ad{a}{f}}{b}[B]\), as this is what we will force it to be.
  We write this judgment with explicit witnesses, even thought
  the discreteness condition we impose for the opfibration enforces that
  there is (definitionally) at most one inhabitant. Semantically, in the \CwFs
  structure, \(\typing{\Gamma}{e}[\convover{f}{a}{b}]\)
  describes morphisms \(e\) in \(\Tm(\Gamma)\) whose image
  by \(p\) is the morphism \(f\) in the category \(\Ty(\Gamma)\).

  \subsection{Category structure}
  The following rules axiomatise the category structure of
  \(\Tm(\Gamma)\), together with the functoriality of the map
  \(p_{\Gamma} : \Tm(\Gamma) \to \Ty(\Gamma)\).
  \begin{mathpar}
    \inferdef{
      \typing{\Gamma}{a}[A]
    }
    {\typing{\Gamma}{\id}[\convover{id}{A}{A}]} \and
    \inferdef{
      \typing{\Gamma}{e}[\convover{f}{a}{b}]\\
      \typing{\Gamma}{e'}[\convover{g}{b}{c}]\\
    }
    {\typing{\Gamma}{e'\circ e}[\convover{g\circ f}{a}{c}]} \and
    \inferdef{
      \typing{\Gamma}{e}[\convover{f}{a}{b}]
    }
    {\conv{\Gamma}{e \circ \id}{e}[\convover{f}{a}{b}]} \and
        \inferdef{
      \typing{\Gamma}{e}[\convover{f}{a}{b}]
    }
    {\conv{\Gamma}{\id \circ e}{e}[\convover{f}{a}{b}]} \and
        \inferdef{
      \typing{\Gamma}{e}[\convover{f}{a}{b}]\\
      \typing{\Gamma}{e'}[\convover{g}{b}{c}]\\
      \typing{\Gamma}{e''}[\convover{h}{d}{c}]
    }
    {\conv{\Gamma}{e'' \circ (e'\circ e)}{(e''\circ e') \circ
        e}[\convover{h\circ g\circ f}{a}{d}]}
  \end{mathpar}

  \subsection{Action of substitutions}
  The following rules make the assignment \(\Gamma \mapsto \Tm(\Gamma)\)
  functorial in \(\Gamma\), at the level on morphisms, and ensure that the map
  \(p\) is natural.
  \begin{mathpar}
    \inferdef{
      \typing{\Gamma}{e}[\convover{f}{a}{b}]\\
      \typing{\Delta}{\sigma}[\Gamma]
    }
    {
      \typing{\Gamma}{\subs{e}{\sigma}}[\convover{\subs{f}{\sigma}}{\subs{a}{\sigma}}{\subs{b}{\sigma}}]
    }\and
    \inferdef{
      \typing{\Gamma}{e}[\convover{f}{a}{b}]
    }
    {
      \conv{\Gamma}{\subs{e}{\id}}{e}[\convover{f}{a}{b}]
    }\and
    \inferdef{
      \typing{\Gamma}{e}[\convover{f}{a}{b}]\\
      \typing{\Delta}{\sigma}[\Gamma] \\
      \typing{\Xi}{\tau}[\Delta]
    }
    {
      \conv{\Gamma}{\subs{e}{\sigma \circ \tau}}{\subs{\subs{e}{\sigma}}{\tau}}[\convover{\subs{\subs{f}{\sigma}}{\tau}}{\subs{\subs{a}{\sigma}}{\tau}}{\subs{\subs{b}{\sigma}}{\tau}}]
    }
  \end{mathpar}

  \subsection{Discrete opfibration structure}
  Finally, we add rules to axiomatize the discrete opfibration structure of the
  morphism \(p\) as follows:
  \begin{mathpar}
    \inferrule{
      \typing{\Gamma}{a}[A] \\
      \typing{\Gamma}{f}[\adso{A}{B}]
    }{
      \typing{\Gamma}{\eqover{f}{a}}[\convover{f}{a}{\ad{a}{f}}]
    } \and
    \inferrule{
      \typing{\Gamma}{e}[\convover{f}{a}{b}]
    }{
      \conv{\Gamma}{\ad{a}{f}}{b}
    } \and
    \inferrule{
      \typing{\Gamma}{e}[\convover{f}{a}{b}]
    }{
      \conv{\Gamma}{e}{\eqover{f}{a}}[\convover{f}{a}{b}]
    }
  \end{mathpar}
  From these rules, it follows that there is at most one witness for the
  judgment $\convover{f}{a}{b}$ up to conversion: if
  $\typing{\Gamma}{e}[\convover{f}{a}{b}]$ and
  $\typing{\Gamma}{e'}[\convover{f}{a}{b}]$
  then $\conv{\Gamma}{e}{e'}[\convover{f}{a}{b}]$.
  In particular, all the categorical equations introduced before are derivable.

  This completes the alternate presentation of \AdapTT{}. One can check that
  \ruleref{rule:adaptt-adapt-id} and \ruleref{rule:adaptt-adapt-comp} are
  derivable in this presentation. Conversely, the alternate presentation of
  \AdapTT{} given here is derivable from the one given in the main text.

  \subsection{Functoriality of term constructors}
  This presentation allows us to rewrite the equality that we require on
  term constructors in a different, perhaps more natural or systematic way.
  We illustrate this with the example of lists, for which we
  have required that adapters behave like generalized map functions as follows
  \[
    \ad{(h\cons t)}{\List(f)}\convop (\ad{h}{f})\cons (\ad{t}{\List(f)})
  \]
  The above constraint can be rephrased in the new formulation as a requirement
  for the term constructor to act on morphisms in \(\Tm(\Gamma)\), as follows
  \begin{mathpar}
    \inferdef{
      \typing{\Gamma}{e^h}[\convover{f}{h}{h'}] \\
      \typing{\Gamma}{e^t}[\convover{\List(f)}{t}{t'}]
    }
    {
      \typing{\Gamma}{e^h\cons e^t}[\convover{\List(f)}{h \cons t}{h' \cons t'}]
    }
  \end{mathpar}
We did not explore it further, but we believe
there should be an alternate presentation
of \AdapTTt{} that follows this style.
\end{shepherdenv}

\section{Full rules for \texorpdfstring{\AdapTTt{}}{AdapTT2}}
\label{sec:full-rules}

\subsection{All sorts and corresponding judgment forms}

\begin{mathparpagebreakable}
    % \inferdef{ }{\Dir} \and
  \inferdef%[Ctx]
    { }{\Ctx} \and
  \inferdef%[Sub]
    {\Gamma, \Delta \ty \Ctx}{\Sub(\Gamma,\Delta)} \and
  \inferdef%[Trans]
    {\impl{\Gamma, \Delta \ty \Ctx} \\\\
    \sigma, \tau \ty \Sub(\Gamma,\Delta)}{
      \Trans(\impl{\Gamma},\impl{\Delta},\sigma,\tau)} \and
  \inferdef%[Ty]
  {\Gamma \ty \Ctx}{\Ty(\Gamma)} \and
  \inferdef%[Ad]
  {\impl{\Gamma \ty \Ctx} \\\\
    A, B \ty \Ty(\Gamma)}{
      \Ad(\impl{\Gamma},A,B)
    } \and
  \inferdef%[Tm]
  {\impl{\Gamma \ty \Ctx} \\\\ A \ty \Ty(\Gamma)}{\Tm(\Gamma,A)} \\
  \ctxty{\Box} \and
  \typing{\Gamma}{\Box}[\Delta] \and
  \typing{\Gamma}{\Box}[\transso[\Delta]{\sigma}{\tau}] \and
  \typing{\Gamma}{\Box} \and
  \typing{\Gamma}{\Box}[\adso{A}{B}] \and
  \typing{\Gamma}{\Box}[A] \\
  \inferdef%[Tel]
    {\Gamma \ty \Ctx}{\Tel(\Gamma)} \and
  \inferdef%[TelAd]
    {\impl{\Gamma \ty \Ctx} \\
    \Theta, \Theta' \ty \Tel(\Gamma)}{
      \TelAd(\impl{\Gamma},\Theta,\Theta')
    } \and
  \inferdef%[Inst]
    {\impl{\Gamma \ty \Ctx} \\ \Theta \ty \Ty(\Gamma)}{\Inst(\Gamma,\Theta)} \\
  \hspace*{-2em}\typing{\Gamma}{\Box} \and
  \hspace*{1.25em}\typing{\Gamma}{\Box}[\adso{\Theta}{\Theta'}] \and
  \hspace*{3em}\typing{\Gamma}{\Box}[\Theta] \\
\end{mathparpagebreakable}

\subsection{Basic structure of \texorpdfstring{\AdapTTt{}}{AdapTT2}}

\begin{mathparpagebreakable}
    \inferdef[SubId]
    {\ctxty{\Gamma}}{\typing{\Gamma}{\id_{\Gamma}}[\Gamma]} \and
  \inferdef[SubComp]
    {%\Gamma, \Delta, \Xi \ty \Ctx \\
      \typing{\Delta}{\tau}[\Xi] \\
      \typing{\Gamma}{\sigma}[\Delta]
    }{\typing{\Gamma}{\tau \circ \sigma}[\Xi]} \and
  \inferdef[SubRightUnitality]
  {%\Gamma, \Delta \ty \Ctx \\
    \typing{\Gamma}{\sigma}[\Delta] \\
    }{\conv{\Gamma}{\sigma \circ \id_\Gamma}{\sigma}[\Delta]} \and
  \inferdef[SubLeftUnitality]
  {%\Gamma, \Delta \ty \Ctx \\
    \typing{\Gamma}{\sigma}[\Delta] \\
    }{\conv{\Gamma}{ \id_\Delta \circ \sigma }{\sigma}[\Delta]} \and
  \inferdef[SubAssoc]
  {%\Gamma, \Delta \ty \Ctx \\
    \typing{\Phi}{\sigma}[\Delta] \\
    \typing{\Xi}{\delta}[\Phi] \\
    \typing{\Gamma}{\tau}[\Xi] \\
    }{\conv{\Gamma}{\sigma \circ (\delta \circ \tau)}{(\sigma \circ \delta) \circ \tau}[\Delta]} \and
  \inferdef[AdId]
    {%\Gamma \ty \Ctx \\
    \typing{\Gamma}{A} \\
    }{\typing{\Gamma}{\id_{A}}[\adso{A}{A}]} \and
  \inferdef[AdComp]
    {%\Gamma \ty \Ctx \\
      % \typing{\Gamma}{A,B,C}[\Ty] \\
    \typing{\Gamma}{g}[\adso{B}{C}] \\
    \typing{\Gamma}{f}[\adso{A}{B}]
    }{\typing{\Gamma}{g \circ f}[\adso{A}{C}]} \and
  \inferdef[TransId]
  {
    \typing{\Gamma}{\sigma}[\Delta]
  }{\typing{\Gamma}{\id_{\sigma}}[\transso[\Delta]{\sigma}{\sigma}]} \and
  % \inferdef%[Comp0]
  %   {%\Gamma, \Delta, \Xi \ty \Ctx \\
  %   \typing{\Gamma}{\mu}[\transso[\Delta]{\sigma}{\sigma'}] \\
  %   \typing{\Delta}{\nu}[\transso[\Xi]{\tau}{\tau'}] \\
  %   }{
  %     \typing{\Gamma}{\nu \circ_{0} \mu}
  %     [\transso[\Xi]{\tau \circ \sigma}{\tau'\circ\sigma'}]} \and
  \inferdef[TransComp]
    {%\Gamma, \Delta, \Xi \ty \Ctx \\
    \typing{\Gamma}{\mu}[\transso[\Delta]{\rho}{\sigma}] \\
    \typing{\Gamma}{\nu}[\transso[\Delta]{\sigma}{\tau}] \\
    }{
      \typing{\Gamma}{\nu \vcomp \mu}[\transso[\Delta]{\rho}{\tau}]} \and
  \inferdef[TransWhiskerLeft]
    {%\Gamma, \Delta, \Xi \ty \Ctx \\
    \typing{\Delta}{\tau}[\Xi] \\
    \typing{\Gamma}{\mu}[\transso[\Delta]{\sigma}{\sigma'}] \\
    }{
      \typing{\Gamma}{\tau \whisk \mu}
      [\transso[\Xi]{(\tau \circ \sigma)}{(\tau \circ \sigma')}]} \and
  \inferdef[TransWhiskerRight]
    {%\Gamma, \Delta, \Xi \ty \Ctx \\
    \typing{\Delta}{\nu}[\transso[\Xi]{\tau}{\tau'}] \\
    \typing{\Gamma}{\sigma}[\Delta] \\
    }{
      \typing{\Gamma}{\nu \whisk \sigma}
      [\transso[\Xi]{(\tau \circ \sigma)}{(\tau'\circ\sigma)}]} \and
  \inferdef[TransLeftUnitality]
    {%\Gamma, \Delta, \Xi \ty \Ctx \\
    \typing{\Gamma}{\mu}[\transso[\Delta]{\sigma}{\sigma'}] \\
    }{
      \conv{\Gamma}{\mu \vcomp \id_{\sigma} }{\mu}
      [\transso[\Delta]{\sigma}{\sigma'}]} \and
  \inferdef[TransRightUnitality]
    {%\Gamma, \Delta, \Xi \ty \Ctx \\
    \typing{\Gamma}{\mu}[\transso[\Delta]{\sigma}{\sigma'}] \\
    }{
      \conv{\Gamma}{\id_{\sigma'} \vcomp \mu }{\mu}
      [\transso[\Delta]{\sigma}{\sigma'}]} \and
  \inferdef[TransAssoc]
    {%\Gamma, \Delta, \Xi \ty \Ctx \\
    \typing{\Gamma}{\mu}[\transso[\Delta]{\sigma}{\tau}] \\
    \typing{\Gamma}{\nu}[\transso[\Delta]{\tau}{\rho}] \\
    \typing{\Gamma}{\xi}[\transso[\Delta]{\rho}{\psi}] \\
    }{
      \conv{\Gamma}{\xi \vcomp (\nu \vcomp \mu) }{(\xi \vcomp \nu) \vcomp \mu}
      [\transso[\Delta]{\sigma}{\psi}]} \and
  \inferdef[TransWhiskerLeftRightUnitality]
    {%\Gamma, \Delta, \Xi \ty \Ctx \\
    \typing{\Gamma}{\sigma}[\Delta] \\
    \typing{\Delta}{\tau}[\Xi] \\
    }{
      \conv{\Gamma}{\tau \whisk \id_\sigma}{\id_{\tau \circ \sigma}}
      [\transso[\Xi]{\tau \circ \sigma}{\tau \circ \sigma}]} \and
  \inferdef[TransWhiskerRightLeftUnitality]
    {%\Gamma, \Delta, \Xi \ty \Ctx \\
    \typing{\Gamma}{\sigma}[\Delta] \\
    \typing{\Delta}{\tau}[\Xi] \\
    }{
      \conv{\Gamma}{\id_\tau \whisk \sigma}{\id_{\tau\circ\sigma}}
      [\transso[\Xi]{\tau \circ \sigma}{\tau\circ\sigma}]} \and

  \inferdef[TransWhiskerLeftLeftUnitality]
    {%\Gamma, \Delta, \Xi \ty \Ctx \\
    \typing{\Gamma}{\mu}[\transso[\Delta]{\sigma}{\sigma'}] \\
    }{
      \conv{\Gamma}{\id_\Delta \whisk \mu}{\mu}
      [\transso[\Delta]{\sigma}{\sigma'}]} \and
  \inferdef[TransWhiskerRightRightUnitality]
    {%\Gamma, \Delta, \Xi \ty \Ctx \\
    \typing{\Gamma}{\mu}[\transso[\Delta]{\sigma}{\sigma'}] \\
    }{
      \conv{\Gamma}{\mu \whisk \id_\Gamma}{\mu}
      [\transso[\Delta]{\sigma}{\sigma}]}
  \and
  \inferdef[TransWhiskerLeftAssoc]
  {
    \typing{\Gamma}{\mu}[\transso[\Delta]{\sigma}{\sigma'}] \\
    \typing{\Delta}{\tau}[\Xi] \\
    \typing{\Xi}{\tau'}[\Psi] \\
  }{
      \conv{\Gamma}{ (\tau' \circ \tau) \whisk \mu}{\tau' \whisk (\tau \whisk \mu)}
      [\transso[\Psi]{(\tau'\circ\tau\circ\sigma)}{(\tau'\circ\tau\circ\sigma')}]}
  \and
  \inferdef[TransWhiskerRightAssoc]
  {
    \typing{\Gamma}{\sigma}[\Delta] \\
    \typing{\Delta}{\sigma'}[\Xi] \\
    \typing{\Xi}{\mu}[\transso[\Psi]{\tau}{\tau'}] \\
  }{
      \conv{\Gamma}{ \mu \whisk (\sigma' \circ \sigma) }{(\mu \whisk \sigma') \whisk \sigma}
      [\transso[\Psi]{(\tau\circ\sigma'\circ\sigma)}{(\tau'\circ\sigma'\circ\sigma')}]}
  \and
  \inferdef[TransWhiskerLeftDistr]
  {
    \typing{\Delta}{\sigma}[\Xi] \\
    \typing{\Gamma}{\mu}[\transso[\Delta]{\tau}{\tau'}] \\
    \typing{\Gamma}{\nu}[\transso[\Delta]{\tau'}{\tau''}] \\
  }{
      \conv{\Gamma}{\sigma \whisk (\mu \vcomp \nu)}{(\sigma \whisk \mu) \circ (\sigma \whisk \nu)}
      [\transso[\Xi]{(\delta \circ \tau)}{(\delta\circ\tau'')}]}
  \and
  \inferdef[TransWhiskerRightDistr]
  {
    \typing{\Gamma}{\sigma}[\Delta] \\
    \typing{\Delta}{\mu}[\transso[\Xi]{\tau}{\tau'}] \\
    \typing{\Delta}{\nu}[\transso[\Xi]{\tau'}{\tau''}] \\
  }{
      \conv{\Gamma}{ (\mu \vcomp \nu) \whisk \sigma }{ (\mu \whisk \sigma ) \circ (\nu \whisk \sigma )}
      [\transso[\Xi]{(\delta \circ \tau)}{(\delta\circ\tau'')}]} \and
  \inferdef[AdRightUnitality]
    {%\Gamma \ty \Ctx \\
    \typing{\Gamma}{a}[\adso{A}{B}]\\
    }{\conv{\Gamma}{a \circ \id_A}{a}[\adso{A}{B}]} \and
  \inferdef[AdLeftUnitality]
    {%\Gamma \ty \Ctx \\
    \typing{\Gamma}{a}[\adso{A}{B}]\\
    }{\conv{\Gamma}{\id_B \circ a}{a}[\adso{A}{B}]} \and
  \inferdef[AdAssoc]
  {%\Gamma, \Delta \ty \Ctx \\
    \typing{\Gamma}{a}[\adso{A}{B}] \\
    \typing{\Gamma}{b}[\adso{B}{C}] \\
    \typing{\Gamma}{c}[\adso{C}{D}] \\
    }{\conv{\Gamma}{c \circ (b \circ a)}{(c \circ b) \circ a}[\adso{A}{D}]} \and
  \inferdef[AdTm]
    {%\Gamma \ty \Ctx \\
    % \typing{\Gamma}{A,B}[\Ty] \\
    \typing{\Gamma}{f}[\adso{A}{B}] \\
    \typing{\Gamma}{t}[A]
    }{\typing{\Gamma}{\ad{t}{f}}[B]} \\

  \inferdef[SubTy]
  {%\Gamma, \Delta \ty \Ctx \\
    \typing{\Delta}{A} \\
    \typing{\Gamma}{\sigma}[\Delta] \\
    }{\typing{\Gamma}{\subs{A}{\sigma}}} \and

  \inferdef[SubTyId]
    {\typing{\Gamma}{A}}
    {\conv{\Gamma}{\subs{A}{\id_\Gamma}}{A}}
  \inferdef[SubTyComp]
    {\typing{\Gamma}{A}\\
      \typing{\Xi}{\sigma}[\Gamma]\\
      \typing{\Delta}{\tau}[\Xi]}
    {\conv{\Delta}{\subs{A}{\sigma \circ \tau}}{\subs{\subs{A}{\sigma}}{\tau}}}

  \inferdef[SubAd]
    {%\Gamma, \Delta \ty \Ctx \\
    % \typing{\Delta}{A,B}[\Ty] \\
      \typing{\Delta}{f}[\adso{A}{B}]\\
      \typing{\Gamma}{\sigma}[\Delta]
    }{\typing{\Gamma}{\subs{f}{\sigma}}[\adso{\subs{A}{\sigma}}{\subs{B}{\sigma}}]} \and

  \inferdef[SubAdId]
    {\typing{\Gamma}{a}[\adso{A}{B}]}
    {\conv{\Gamma}{\subs{a}{\id_\Gamma}}{a}[\adso{A}{B}]}
  \and
  \inferdef[SubAdComp]
    {\typing{\Gamma}{a}[\adso{A}{B}]\\
      \typing{\Xi}{\sigma}[\Gamma]\\
      \typing{\Delta}{\tau}[\Xi]}
    {\conv{\Delta}{\subs{a}{\sigma \circ \tau}}{\subs{\subs{a}{\sigma}}{\tau}}
      [\adso{\subs{A}{\sigma \circ \tau}}{\subs{B}{\sigma \circ \tau}}]}
  \and
  \inferdef[SubAdOnId]
    {
      \typing{\Delta}{A} \\
      \typing{\Gamma}{\sigma}[\Delta]}
    {
      \conv{\Gamma}
        {\subs{\id_A}{\sigma}}
        {\id_{\subs{A}{\sigma}}}
        [\adso{\subs{A}{\sigma}}{\subs{A}{\sigma}}]
      }
  \and
  \inferdef[SubAdOnComp]
    { \typing{\Delta}{a}[\adso{A}{B}]\\
      \typing{\Delta}{b}[\adso{B}{C}]\\
      \typing{\Gamma}{\sigma}[\Delta]}
    {\conv{\Gamma}{\subs{(b \circ a)}{\sigma}}{(\subs{b}{\sigma}) \circ (\subs{a}{\sigma})}
      [\adso{\subs{A}{\sigma}}{\subs{C}{\sigma}}]}

  \inferdef[SubTm]
  {%\Gamma, \Delta \ty \Ctx \\
     \typing{\Delta}{t}[A] \\
    \typing{\Gamma}{\sigma}[\Delta]
    % \typing{\Delta}{A,B}[\Ty] \\
    }{\typing{\Gamma}{\subs{t}{\sigma}}[\subs{A}{\sigma}]} \\

  \inferdef[SubTmId]
    {\typing{\Gamma}{t}[A]}
    {\conv{\Gamma}{\subs{t}{\id_\Gamma}}{t}[A]}
  \inferdef[SubTmComp]
    {\typing{\Gamma}{t}[A]\\
      \typing{\Xi}{\sigma}[\Gamma]\\
      \typing{\Delta}{\tau}[\Xi]}
    {\conv{\Delta}{\subs{t}{\sigma \circ \tau}}{\subs{\subs{t}{\sigma}}{\tau}}
    [\subs{A}{\sigma \circ \tau}]}

  \inferdef[AdTmId]
  {
    \typing{\Gamma}{t}[A] \\
  }{\conv{\Gamma}{\ad{t}{\id_A}}{t}[A]} \\
  \inferdef[AdTmComp]
  {
    \typing{\Gamma}{t}[A] \\
    \typing{\Gamma}{a}[\adso{B}{C}]\\
    \typing{\Gamma}{b}[\adso{A}{B}]\\
  }{\conv{\Gamma}{\ad{t}{a \circ b}}{\ad{\ad{t}{b}}{a}}[C]} \\

  \inferdef[SubTmOnAdTm]
  {
    \typing{\Delta}{t}[A] \\
    \typing{\Delta}{a}[\adso{A}{B}]\\
    \typing{\Gamma}{\sigma}[\Delta]
  }{\conv{\Gamma}{\subs{\ad{t}{a}}{\sigma}}{\ad{\subs{t}{\sigma}}{\subs{a}{\sigma}}}[\subs{B}{\sigma}]} \\

  \inferdef[TransTy]
    {%\Gamma, \Delta \ty \Ctx \\
    % \sigma,\tau \ty \Sub(\Gamma,\Delta) \\
      \typing{\Delta}{A} \\
      \typing{\Gamma}{\mu}[\transso[\Delta]{\sigma}{\tau}] \\
    }{\typing{\Gamma}{\trans{A}{\mu}}[\adso{\subs{A}{\sigma}}{\subs{A}{\tau}}]} \and
    \inferdef[TransTyId]
    {
      \typing{\Delta}{A} \\
      \typing{\Gamma}{\sigma}[\Delta] \\
    }{\conv{\Gamma}{\trans{A}{\id_\sigma}}{\id_{A[\sigma]}}[\adso{\subs{A}{\sigma}}{\subs{A}{\sigma}}]} \and
  \inferdef[TransTyComp]
  { \typing{\Delta}{A}\\
    \typing{\Gamma}{\mu}[\transso[\Delta]{\sigma}{\tau}]\\
    \typing{\Gamma}{\nu}[\transso[\Delta]{\tau}{\xi}]}
  {\conv{\Gamma}{\trans{A}{\nu \circ \mu}}{\trans{A}{\nu} \circ \trans{A}{\mu}}[\adso{\subs{A}{\sigma}}{\subs{A}{\xi}}]} \and
    \inferdef[SubTyOnTransTy]
    {
      \typing{\Xi}{A} \\
      \typing{\Delta}{\mu}[\transso[\Xi]{\sigma}{\tau}] \\
      \typing{\Gamma}{\delta}[\Delta]
    }{\conv{\Gamma}{\subs{\trans{A}{\mu}}{\delta}}{\trans{A}{\mu\circ\delta}}[\adso{\subs{A}{\sigma\circ\delta}}{\subs{A}{\tau\circ\delta}}]} \and
    \inferdef[TransTyOnSubTy]
    {
      \typing{\Xi}{A} \\
      \typing{\Delta}{\xi}[\Xi] \\
      \typing{\Gamma}{\nu}[\transso[\Delta]{\sigma}{\tau}]
    }{\conv{\Gamma}{\trans{A[\xi]}{\nu}}{\trans{A}{\xi\circ\nu}}[\adso{\subs{A}{\xi\circ\sigma}}{\subs{A}{\xi\circ\tau}}]} \and
    \inferdef[TransTyNatural]
    {%\Gamma, \Delta \ty \Ctx \\
    % \sigma,\tau \ty \Sub(\Gamma,\Delta) \\
      \typing{\Delta}{f}[\adso{A}{B}] \\
      \typing{\Gamma}{\mu}[\transso[\Delta]{\sigma}{\tau}]
    }{\conv{\Gamma}{
        \trans{B}{\mu} \whisk \subs{f}{\sigma}
      }{\subs{f}{\tau} \whisk \trans{A}{\mu}
      }[\adso{\subs{A}{\sigma}}{\subs{B}{\tau}}]
    } \and
  \inferdef[TransTyAdTm]
    {%\Gamma, \Delta \ty \Ctx \\
    % \sigma,\tau \ty \Sub(\Gamma,\Delta) \\
      \typing{\Delta}{t}[A] \\
      \typing{\Gamma}{\mu}[\transso[\Delta]{\sigma}{\tau}]
    }{\conv{\Gamma}{
        \ad{\subs{t}{\sigma}}{\trans{A}{\mu}}
      }{\subs{t}{\tau}
      }[\subs{A}{\tau}]
    }

\end{mathparpagebreakable}

\subsection{Empty context and context dualisation}
\begin{mathparpagebreakable}
  \inferdef[CtxEmp]{ }{\ctxty{\emp}} \and
  \inferdef[SubEmp]{\ctxty{\Gamma}}{\typing{\Gamma}{\emp_{\impl{\Gamma}}}[\emp]} \and
  \inferdef[SubEmpExt]{\typing{\Gamma}{\sigma}[\emp]}
  {\conv{\Gamma}{\sigma}{\emp_{\Gamma}}[\emp]} \and
  % \inferdef{ }{+ \ty \Dir} \and \inferdef{ }{- \ty \Dir} \and
  % \inferdef{d, d' \ty \Dir}{d\ d' \ty \Dir} \\
  % \inferdef{ }{+ + \convop + \convop - - \\ + - \convop - \convop - +} \\
  \inferdef[CtxDual]
  {\ctxty{\Gamma}}{\ctxty{\dual{\Gamma}{-}}} \and
  \inferdef[SubDual]
  {
    % d \ty \Dir \\
    % \impl{\Gamma, \Delta \ty \Ctx} \\
    \typing{\Gamma}{\sigma}[\Delta]}{
    \typing{\dual{\Gamma}{-}}{\dual{\sigma}{-}}[\dual{\Delta}{-}]} \and
  \inferdef[TransDual]
  {
    % \impl{\Gamma, \Delta \ty \Ctx} \\
    % \impl{\sigma, \tau \ty \Sub(\Gamma,\Delta)} \\
    \typing{\Gamma}{\mu}[\transso[\Delta]{\sigma}{\tau}] \\
  }{
    \typing{\dual{\Gamma}{-}}{\dual{\mu}{-}}
      [\transso[\dual{\Delta}{-}]{\dual{\tau}{-}}{\dual{\sigma}{-}}]
  } \and
  \inferdef[CtxEmpDual]{ }{\conv{}{\dual{\emp}{-}}{\emp}} \and
  % \inferdef[SubEmpDual]
  %   {\ctxty{\Gamma}}
  %   {\conv{\dual{\Gamma}{-}}{\dual{\emp_\Gamma}{-}}{\emp_{\dual{\Gamma}{-}}}[\emp]}
  % This is derivable from SubEmpExt and CtxEmpDual
    \and
  \inferdef[SubIdDual]
    {\ctxty{\Gamma}}
    {\conv{\dual{\Gamma}{-}}{\dual{\id_\Gamma}{-}}{\id_{\dual{\Gamma}{-}}}[\dual{\Gamma}{-}]} \and
  \inferdef[SubCompDual]
    {\typing{\Delta}{\tau}[\Xi] \\
      \typing{\Gamma}{\sigma}[\Delta]
    }{\conv{\dual{\Gamma}{-}}{\dual{(\tau \circ \sigma)}{-}}{\dual{\tau}{-} \circ \dual{\sigma}{-}}[\dual{\Xi}{-}]} \and
  \inferdef[TransIdDual]
    {\typing{\Gamma}{\sigma}[\Delta]}
    {\conv{\dual{\Gamma}{-}}{\dual{\id_\sigma}{-}}{\id_{\dual{\sigma}{-}}}[\transso[\dual{\Delta}{-}]{\dual{\sigma}{-}}{\dual{\sigma}{-}}]} \and
  \inferdef[TransCompDual]{
    \typing{\Gamma}{\mu}[\transso[\Delta]{\rho}{\sigma}] \\
    \typing{\Gamma}{\nu}[\transso[\Delta]{\sigma}{\tau}] \\
    }{
      \conv{\dual{\Gamma}{-}}
      {\dual{(\mu \vcomp \nu)}{-}}
      {\dual{\nu}{-} \vcomp {\dual{\mu}{-}}}
      [\transso[\dual{\Delta}{-}]{\dual{\tau}{-}}{\dual{\rho}{-}}]} \and
  \inferdef[CtxPlus]
    {\ctxty{\Gamma}}
    {\conv{}{\dual{\Gamma}{+}}{\Gamma}}  \and
  \inferdef[SubPlus]
    {\typing{\Gamma}{\sigma}[\Delta]}
    {\conv{\Gamma}{\dual{\sigma}{+}}{\sigma}[\Delta]} \and
  \inferdef[TransPlus]
    {\typing{\Gamma}{\mu}[\transso[\Delta]{\sigma}{\tau}]}
    {\conv{\Gamma}{\dual{\mu}{+}}{\mu}[\transso[\Delta]{\sigma}{\tau}]} \and
  \inferdef[CtxDoubleDual]
    {\ctxty{\Gamma}}
    {\conv{}{\dual{{\dual{\Gamma}{-}}}{-}}{\Gamma}}  \and
  \inferdef[SubDoubleDual]
    {\typing{\Gamma}{\sigma}[\Delta]}
    {\conv{\Gamma}{\dual{{\dual{\sigma}{-}}}{-}}{\sigma}[\Delta]} \and
  \inferdef[TransDoubleDual]
    {\typing{\Gamma}{\mu}[\transso[\Delta]{\sigma}{\tau}]}
    {\conv{\Gamma}{\dual{{\dual{\mu}{-}}}{-}}{\mu}[\transso[\Delta]{\sigma}{\tau}]} \and
  \inferdef[TransWhiskerLeftDual]{
      \typing{\Delta}{\tau}[\Xi]\\
      \typing{\Gamma}{\mu}[\transso[\Delta]{\sigma}{\sigma'}]
    }{\conv{\Gamma}{\dual{(\tau \circ \mu)}{-}}{\dual{\tau}{-} \circ \dual{\mu}{-}}[\transso[\Xi]{\tau \circ \sigma}{\tau \circ \sigma'}]} \and
  \inferdef[TransWhiskerRightDual]{
      \typing{\Delta}{\nu}[\transso[\Xi]{\tau}{\tau'}] \\
      \typing{\Gamma}{\sigma}[\Delta]
    }{\conv{\Gamma}{\dual{(\nu \circ \sigma)}{-}}{\dual{\nu}{-} \circ \dual{\sigma}{-}}[\transso[\Xi]{\tau \circ \sigma}{\tau' \circ \sigma}]} \and
\end{mathparpagebreakable}

\subsection{Term variables in \AdapTTt}
\begin{mathparpagebreakable}
 \inferdef[CtxExtTm]
  {
    \ctxty{\Gamma} \\
    \typing{\dual{\Gamma}{d}}{A}
  }{
    \ctxty{\Gamma \ext_{\impl{d}} A}
  }[\(d:\Dir\)]
   \and
  \inferdef[SubExtTm]
  {
    % \impl{\Gamma, \Delta \ty \Ctx} \\
    \typing{\Gamma}{\sigma}[\Delta] \\
    \impl{\typing{\dual{\Delta}{d}}{A}} \\
    \typing{\dual{\Gamma}{d}}{t}[\subs{A}{\dual{\sigma}{d}}]
  }{\typing{\Gamma}{\sigma \ext_{\impl{d}} t}[\Delta \ext_{\impl{d}} A]
  }[\(d:\Dir\)] \and
  \inferdef[WkTm]
  {
    \typing{\dual{\Gamma}{d}}{A}
  }{
    \typing{\Gamma \ext_{\impl{d}} A}{\wk_{A}}[\Gamma]
  }[\(d : \Dir\)]
  \and
  \inferdef[VarZTm]{
    \impl{\typing{\Gamma}{A}}
  }{\typing{\Gamma \ext_{+} A}{\tmvz_{\impl{A}}}[\subs{A}{\wk}]}
  \and

  \inferdef[SubTL]{
    \typing{\Gamma}{\sigma}[\Delta] \\
    \typing{\dual{\Delta}{d}}{A} \\
    \typing{\Gamma}{t}[\subs{A}{\dual{\sigma}{d}}]
  }{
    \conv{\Gamma}
    {\wk \circ (\sigma \ext_d t)}
    {\sigma}[\Delta]
    }[\(d : \Dir\)]
  \and

  \inferdef[SubEta]{
    \typing{\Delta}{A} \\
    \typing{\Gamma}{\sigma}[\Delta \ext A] \\
  }{
    \conv{\Gamma}
    {(\wk \circ \sigma) \ext_d \subs{\tmvz_A}{\sigma}}
    {\sigma}[\Delta]
    }
  \and
  \inferdef[AdTmVarZ]{
    \typing{\Delta}{A} \\
    \typing{\Gamma}{\sigma}[\Delta \ext A] \\
    \typing{\Gamma}{\tau}[\Delta \ext A] \\
    \typing{\Gamma}{\mu}[\transso[\Delta\ext_+A]{\sigma}{\tau}]
  }{
    \conv{\Gamma}
    {\ad{\subs{\tmvz_A}{\sigma}}{\trans{A}{\wk \whisk \mu}}}
    {\subs{\tmvz_A}{\tau}}
    [\subs{A}{\tau}]
  }
  \and
  \inferdef[SubTmExtVarZ]{
    \typing{\Gamma}{\sigma}[\Delta] \\
    \typing{\Delta}{A} \\
    \typing{\Gamma}{t}[\subs{A}{\sigma}]
    }{
      \conv{\Gamma}
      {\subs{\tmvz_A}{\sigma \ext t} }
      {t}
      [\subs{A}{\sigma}]
    }
  \and

  \inferdef[TransTm+]{
    % \impl{\Gamma, \Delta \ty \Ctx} \\
    % \impl{\sigma, \tau \ty \Sub(\Gamma,\Delta)} \\
    \typing{\Gamma}{\mu}[\transso[\Delta]{\sigma}{\tau}] \\
    % \impl{A \ty \Ty(\Delta)} \\
    \typing{\Gamma}{t}[\subs{A}{\sigma}]}
  {
    \typing{\Gamma}{\mu \ext_{+} t}[\transso%[\Delta \ext_{+} A]
      {\sigma \ext_{+} t}{\tau \ext_{+} \had{t}{\trans{A}{\mu}}}]
  } \and
  \inferdef[TransTm-]{
    % \impl{\Gamma, \Delta \ty \Ctx} \\
    % \impl{\sigma, \tau \ty \Sub(\Gamma,\Delta)} \\
    \typing{\Gamma}{\mu}[\transso[\Delta]{\sigma}{\tau}] \\
    % \impl{A \ty \Ty_{-}(\Delta)} \\
    \typing{\dual{\Gamma}{-}}{t}[\subs{A}{\dual{\tau}{-}}]
  }{
    \typing{\Gamma}{\mu \ext_{-} t}[\transso%[\Delta \ext_{-} A]
      {\sigma \ext_{-} \had{t}{\trans{A}{\dual{\mu}{-}}}}{\tau \ext_{-} t}]
  } \and
  \inferdef[TransTl+]{
    % \impl{\Gamma, \Delta \ty \Ctx} \\
    \typing{\Gamma}{\mu}[\transso[\Delta]{\sigma}{\tau}] \\
    % \impl{A \ty \Ty(\Delta)} \\
    \typing{\Gamma}{t}[\subs{A}{\sigma}]
  }{
    \conv{\Gamma}{\wk \circ (\mu \ext_{+} t)}{\mu}[\transso[\Delta]{\sigma}{\tau}]}
  \and
    \inferdef[TransTl-]{
    % \impl{\Gamma, \Delta \ty \Ctx} \\
    \typing{\Gamma}{\mu}[\transso[\Delta]{\sigma}{\tau}] \\
    % \impl{A \ty \Ty(\Delta)} \\
    \typing{\dual{\Gamma}{-}}{t}[\subs{A}{\dual{\tau}{-}}]
  }{\conv{\Gamma}{\wk \circ (\mu \ext_{-} t)}{\mu}[\transso[\Delta]{\sigma}{\tau}]}
  \and
  \inferdef[TransEta]{
    % \impl{\Gamma, \Delta \ty \Ctx} \\
    % \impl{A \ty \Ty(\Delta)} \\
    % \impl{\sigma, \tau \ty \Sub(\Gamma,\Delta \ext A)} \\
    \typing{\Gamma}{\mu}[\transso[\Delta\ext_+A]{\sigma}{\tau}] \\
  }{
    \conv{\Gamma}{\mu}{(\wk \circ \mu) \ext_+ \subs{\tmvz}{\sigma}}[\transso[\Delta\ext_+A]{\sigma}{\tau}]} \and

  \inferdef[CtxExtTmDual]
  {
    \ctxty{\Gamma} \\
    \typing{\dual{\Gamma}{d}}{A}
    }{
    \conv{}
    {\dual{(\Gamma \ext_{\impl{d}} A)}{-}}
    {\dual{\Gamma}{-} \ext_{\impl{-d}} A}}[\(d : \Dir\)] \and
  \inferdef[WkTmDual]
  {
    \ctxty{\Gamma} \\
    \typing{\dual{\Gamma}{d}}{A}
    }{ \conv{\dual{(\Gamma \ext_{\impl{d}} A)}{-}}{\dual{\wk}{-}}{\wk}[\dual{\Gamma}{-}]}[\(d : \Dir\)] \and
  \inferdef[SubExtTmDual]
  {
    \typing{\Gamma}{\sigma}[\Delta] \\
    \typing{\dual{\Gamma}{d}}{A} \\
    \typing{\Gamma}{t}[\subs{A}{\dual{\sigma}{d}}]
  }{
    \conv{\dual{\Gamma}{-}}{\dual{(\sigma \ext_{\impl{d}} t)}{-}}
    {\dual{\sigma}{-} \ext_{\impl{-d}} t}
    [\dual{(\Delta \ext_{\impl{d}} A)}{-}]}[\(d : \Dir\)] \and
  \inferdef[TransTm+Dual]
  {
    % \impl{\Gamma, \Delta \ty \Ctx} \\
    % \impl{\sigma, \tau \ty \Sub(\Gamma,\Delta)} \\
    \typing{\Gamma}{\mu}[\transso[\Delta]{\sigma}{\tau}] \\
    \typing{\Delta}{A} \\
    \typing{\Gamma}{t}[\subs{A}{\sigma}]
  }
  {
    \conv{\Gamma}{\dual{(\mu \ext_{+} t)}{-}}
    {\dual{\mu}{-}  \ext_{-} t}
    [\transso[\Delta \ext A]
      {\dual{\tau}{-} \ext \had{t}{\trans{A}{\mu}}}
      {\dual{\sigma}{-} \ext t}] \\
    }
  \and
  \inferdef[TransTm-Dual]
  {
    \typing{\Gamma}{\mu}[\transso[\Delta]{\sigma}{\tau}] \\
    \typing{\dual{\Delta}{-}}{A} \\
    \typing{\dual{\Gamma}{-}}{t}[\subs{A}{\dual{\tau}{-}}]
  }
  { \conv{\dual{\Gamma}{-}}
    {\dual{(\mu \ext_{-} t)}{-}}
    {\dual{\mu}{-}  \ext_{+} t}
    [\transso[\Delta \ext_{-} A]
      {\dual{\tau}{-} \ext_{-} t}
      {\dual{\sigma}{-} \ext_{-} \had{t}{\trans{A}{\dual{\mu}{-}}}}]
  } \and
  \inferdef[TyTransSub]
  {
    \typing{\Delta}{A}\\
    \typing{\Gamma}{\mu}[\transso[\Delta]{\tau}{\xi}]\\
    \typing{\Xi}{\sigma}[\Gamma]\\
  }
  {
    \conv{\Xi}{\subs{\trans{A}{\mu}}{\sigma}}{\trans{A}{\mu\circ\sigma}}[\adso{\subs{A}{\tau\circ\sigma}}{\subs{A}{\xi\circ\sigma}}]
  } \and
  \inferdef[TySubTrans]
  {
    \typing{\Delta}{A}\\
    \typing{\Gamma}{\sigma}[\Delta]\\
    \typing{\Xi}{\mu}[\transso[\Gamma]{\tau}{\xi}]\\
  }
  {
    \conv{\Xi}{\trans{\subs{A}{\sigma}}{\mu}}{\trans{A}{\sigma\circ\mu}}[\adso{\subs{A}{\sigma\circ\tau}}{\subs{A}{\sigma\circ\xi}}]
  }
\end{mathparpagebreakable}

\subsection{Telescopes}

\begin{mathparpagebreakable}
  \inferdef[CtxExtTel]
  {
    \ctxty{\Gamma} \\ \typing{\dual{\Gamma}{d}}{\Theta}
  }{\ctxty{\Gamma \ext_{d} \Theta}}[\(d : \Dir\)] \and
  \inferdef[TelEmp]
  {\ctxty{\Gamma}}{\typing{\Gamma}{\emp}} \and
  \inferdef[TelExtTy]
  {
    \typing{\Gamma} \\
    \typing{\Gamma}{\Theta} \\
    \typing{\Gamma \ext_{+} \Theta}{A}
  }{\typing{\Gamma}{\Theta \ext A}} \\
  \inferdef[WkTel]
  {
    \ctxty{\Gamma} \\
    \typing{\dual{\Gamma}{d}}{\Theta}
  }{
    \typing{\Gamma \ext_{d} \Theta}{\wk_{\Theta}}[\Gamma]
  }[\(d : \Dir\)]
  \and
  \inferdef[SubTel]
  {\typing{\Gamma}{\sigma}[\Delta]\\
    \typing{\Delta}{\Theta}}
  {\typing{\Gamma}{\subs{\Theta}{\sigma}}}
  \and
  \inferdef[TransTel]
    {%\Gamma, \Delta \ty \Ctx \\
    % \sigma,\tau \ty \Sub(\Gamma,\Delta) \\
      \typing{\Delta}{\Theta} \\
      \typing{\Gamma}{\mu}[\transso[\Delta]{\sigma}{\tau}] \\
    }{\typing{\Gamma}{\trans{\Theta}{\mu}}[\adso{\subs{\Theta}{\sigma}}{\subs{\Theta}{\tau}}]}
  \and
  \inferdef[AdInst]
  {
    \typing{\Gamma}{a}[\adso{\Theta_1}{\Theta_2}] \\
    \typing{\Gamma}{\iota}[\Theta_1]
  }{\typing{\Gamma}{\ad{\iota}{a}}[\Theta_2]}
  \and
  \inferdef[VarInst]
  {
    \ctxty{\Gamma} \\
    \typing{\Gamma}{\Theta}
  }{
    \typing{\Gamma \ext_{+} \Theta}{\vinst}[\subs{\Theta}{\wk_{\Theta}}]
  } \and
  \inferdef[SubExtInst]
  {
    % \impl{\Gamma, \Delta \ty \Ctx} \\
    \typing{\Gamma}{\sigma}[\Delta] \\
    \impl{\typing{\dual{\Delta}{d}}{\Theta}} \\
    \typing{\dual{\Gamma}{d}}{\iota}[\subs{\Theta}{\dual{\sigma}{d}}]
  }{\typing{\Gamma}{\sigma \ext_{d} \iota}[\Delta \ext_{d} \Theta]}[\(d : \Dir\)]
  \\
  \inferdef[TransExt+Inst]
  {
    % \impl{\Gamma, \Delta \ty \Ctx} \\
    % \impl{\sigma, \tau \ty \Sub(\Gamma,\Delta)} \\
    \typing{\Gamma}{\mu}[\transso[\Delta]{\sigma}{\tau}] \\
    \impl{\typing{\Delta}{\Theta}} \\
    \typing{\Gamma}{\iota}[\subs{\Theta}{\sigma}]
  }{
    \typing{\Gamma}{\mu \ext_{+} \iota}[\transso{\sigma \ext \iota}
      {\tau \ext \had{\iota}{\trans{\Theta}{\mu}}}]
  } \and
  \inferdef[TransExt-Inst]
  {
    % \impl{\Gamma, \Delta \ty \Ctx} \\
    % \impl{\sigma, \tau \ty \Sub(\Gamma,\Delta)} \\
    \typing{\Gamma}{\mu}[\transso[\Delta]{\sigma}{\tau}] \\
    \impl{\typing{\dual{\Delta}{-}}{\Theta}} \\
    \typing{\dual{\Gamma}{-}}{\iota}[\subs{\Theta}{\dual{\tau}{-}}]
  }{
    \typing{\Gamma}{\mu \ext_{-} \iota}[\transso{\sigma \ext_{-}
      \had{\iota}{\trans{\Theta}{\dual{\mu}{-}}}}{\tau \ext \iota}]} \\

  \inferdef[SubTelId]
    {\typing{\Gamma}{\Theta}}
    {\conv{\Gamma}{\subs{\Theta}{\id_\Gamma}}{A}}
  \and
  \inferdef[SubTelComp]
    {\typing{\Gamma}{\Theta}\\
      \typing{\Xi}{\sigma}[\Gamma]\\
      \typing{\Delta}{\tau}[\Xi]}
    {\conv{\Delta}{\subs{\Theta}{\sigma \circ \tau}}{\subs{\subs{\Theta}{\sigma}}{\tau}}}
  \and
  \inferdef[SubTelOnEmp]
  {
    \typing{\Gamma}{\sigma}[\Delta]
  }{
    \conv{\Gamma}{\subs{\emp}{\sigma}}{\emp}
  }
  \and
  \inferdef[SubTelOnExt]
  {
    \typing{\Delta}{\Theta} \\
    \typing{\Delta \ext \Theta}{A} \\
    \typing{\Gamma}{\sigma}[\Delta]
  }{
    \conv{\Gamma}{\subs{(\Theta \ext A)}{\sigma}}{\subs{\Theta}{\sigma} \ext \subs{A}{(\sigma \circ \wk) \ext \vinst_\Theta}}
  }
  \inferdef[SubInst]
  {%\Gamma, \Delta \ty \Ctx \\
     \typing{\Delta}{\iota}[\Theta] \\
    \typing{\Gamma}{\sigma}[\Delta]
    % \typing{\Delta}{A,B}[\Ty] \\
    }{\typing{\Gamma}{\subs{\iota}{\sigma}}[\subs{\Theta}{\sigma}]} \\

\inferdef[SubInstId]
    {\typing{\Gamma}{\iota}[\Theta]}
    {\conv{\Gamma}{\subs{\iota}{\id_\Gamma}}{\iota}[\Theta]}
  \inferdef[SubInstComp]
    {\typing{\Gamma}{\iota}[\Theta]\\
      \typing{\Xi}{\sigma}[\Gamma]\\
      \typing{\Delta}{\tau}[\Xi]}
    {\conv{\Delta}{\subs{\iota}{\sigma \circ \tau}}{\subs{\subs{\iota}{\sigma}}{\tau}}
    [\subs{\Theta}{\sigma \circ \tau}]}

\inferdef[TelAdId]
    {%\Gamma \ty \Ctx \\
    \typing{\Gamma}{\Theta} \\
    }{\typing{\Gamma}{\id_{\Theta}}[\adso{\Theta}{\Theta}]} \and
  \inferdef[TelAdComp]
    {%\Gamma \ty \Ctx \\
      % \typing{\Gamma}{A,B,C}[\Ty] \\
    \typing{\Gamma}{g}[\adso{\Theta_2}{\Theta_3}] \\
    \typing{\Gamma}{f}[\adso{\Theta_1}{\Theta_2}]
    }{\typing{\Gamma}{g \circ f}[\adso{\Theta_1}{\Theta_3}]} \and

  \inferdef[SubTelAd]
    {%\Gamma, \Delta \ty \Ctx \\
    % \typing{\Delta}{A,B}[\Ty] \\
      \typing{\Delta}{f}[\adso{\Theta_1}{\Theta_2}]\\
      \typing{\Gamma}{\sigma}[\Delta]
    }{\typing{\Gamma}{\subs{f}{\sigma}}[\adso{\subs{\Theta_1}{\sigma}}{\subs{\Theta_2}{\sigma}}]} \and

  \inferdef[TelAdRightUnitality]
    {%\Gamma \ty \Ctx \\
    \typing{\Gamma}{a}[\adso{\Theta_1}{\Theta_2}]\\
    }{\conv{\Gamma}{a \circ \id_{\Theta_1}}{a}[\adso{\Theta_1}{\Theta_2}]} \and
  \inferdef[TelAdLeftUnitality]
    {%\Gamma \ty \Ctx \\
    \typing{\Gamma}{a}[\adso{\Theta_1}{\Theta_2}]\\
    }{\conv{\Gamma}{\id_{\Theta_2} \circ a}{a}[\adso{\Theta_1}{\Theta_2}]} \and
  \inferdef[TelAdAssociativity]
  {%\Gamma, \Delta \ty \Ctx \\
    \typing{\Gamma}{a}[\adso{\Theta_1}{\Theta_2}] \\
    \typing{\Gamma}{b}[\adso{\Theta_2}{\Theta_3}] \\
    \typing{\Gamma}{c}[\adso{\Theta_3}{\Theta_4}] \\
    }{\conv{\Gamma}{c \circ (b \circ a)}{(c \circ b) \circ a}[\adso{\Theta_1}{\Theta_4}]}
  \and
  \inferdef[SubTelAdId]
    {\typing{\Gamma}{a}[\adso{\Theta_1}{\Theta_2}]}
    {\conv{\Gamma}{\subs{a}{\id_\Gamma}}{a}[\adso{\Theta_1}{\Theta_2}]}
  \inferdef[SubTelAdComp]
    {\typing{\Gamma}{a}[\adso{\Theta_1}{\Theta_2}]\\
      \typing{\Xi}{\sigma}[\Gamma]\\
      \typing{\Delta}{\tau}[\Xi]}
    {\conv{\Delta}{\subs{a}{\sigma \circ \tau}}{\subs{\subs{a}{\sigma}}{\tau}}
      [\adso{\subs{\Theta_1}{\sigma \circ \tau}}{\subs{\Theta_2}{\sigma \circ \tau}}]}
  \and
  \inferdef[SubTelAdOnId]
    {
      \typing{\Delta}{\Theta} \\
      \typing{\Gamma}{\sigma}[\Delta]}
    {
      \conv{\Gamma}
        {\subs{\id_\Theta}{\sigma}}
        {\id_{\subs{\Theta}{\sigma}}}
        [\adso{\subs{\Theta}{\sigma}}{\subs{\Theta}{\sigma}}]
    }
  \and
  \inferdef[SubTelAdOnComp]
    { \typing{\Gamma}{a}[\adso{\Theta_2}{\Theta_3}]\\
      \typing{\Gamma}{b}[\adso{\Theta_1}{\Theta_2}]\\
      \typing{\Gamma}{\sigma}[\Delta]}
    {\conv{\Delta}{\subs{(a \circ b)}{\sigma}}{(\subs{a}{\sigma}) \circ (\subs{b}{\sigma})}
      [\adso{\subs{\Theta_1}{\sigma}}{\subs{\Theta_3}{\sigma}}]}
  \and
  \inferdef[TransTelNaturality]
    {
      \typing{\Delta}{a}[\adso{\Theta_1}{\Theta_2}] \\
      \typing{\Gamma}{\mu}[\transso[\Delta]{\sigma}{\tau}]
    }{\conv{\Gamma}{
        \trans{\Theta_2}{\mu} \whisk \subs{a}{\sigma}
      }{\subs{a}{\tau} \whisk \trans{\Theta_1}{\mu}
      }[\adso{\subs{\Theta_1}{\sigma}}{\subs{\Theta_2}{\tau}}]
    } \and
    \inferdef[TransTelId]
    {
      \typing{\Delta}{\Theta} \\
      \typing{\Gamma}{\sigma}[\Delta] \\
    }{\typing{\Gamma}{\trans{\Theta}{\id_\sigma} \convop \id_{\Theta[\sigma]}}[\adso{\subs{A}{\sigma}}{\subs{A}{\sigma}}]} \and
    \inferdef[SubTelAdTransTel]
    {
      \typing{\Delta}{\Theta} \\
      \typing{\Gamma}{\mu}[\transso[\Delta]{\sigma}{\tau}] \\
      \typing{\Xi}{\xi}[\Gamma]
    }{\typing{\Gamma}{\trans{\Theta}{\mu}[\xi] \convop \trans{\Theta}{\mu\circ\xi}}[\adso{\subs{\Theta}{\sigma\circ\xi}}{\subs{\Theta}{\tau\circ\xi}}]} \and
    \inferdef[TransTelComp]
    { \typing{\Delta}{\Theta}\\
      \typing{\Gamma}{\mu}[\transso[\Delta]{\sigma}{\tau}]\\
      \typing{\Gamma}{\nu}[\transso[\Delta]{\tau}{\xi}]}
    {\conv{\Gamma}{\trans{\Theta}{\nu \circ \mu}}{\trans{\Theta}{\nu} \circ \trans{\Theta}{\mu}}[\adso{\subs{\Theta}{\sigma}}{\subs{\Theta}{\xi}}]} \and
    \inferdef[TransTelSubTel]
    {
      \typing{\Delta}{\Theta} \\
      \typing{\Xi}{\sigma}[\Delta] \\
      \typing{\Gamma}{\nu}[\transso[\Xi]{\tau}{\xi}]
    }{\typing{\Gamma}{\trans{\Theta[\sigma]}{\nu} \convop \trans{\Theta}{\sigma\circ\nu}}[\adso{\subs{\Theta}{\sigma\circ\tau}}{\subs{\Theta}{\sigma\circ\xi}}]} \and
  \inferdef[InstTransTel]
    { \typing{\Delta}{\iota}[\Theta_1] \\
      \typing{\Gamma}{\mu}[\transso[\Delta]{\sigma}{\tau}]
    }{\conv{\Gamma}{
        \ad{\subs{\iota}{\sigma}}{\trans{\Theta_1}{\mu}}
      }{\subs{\iota}{\tau}
      }[\subs{\Theta_1}{\tau}]
    } \and
  \inferdef[AdInstId]
  {
    \typing{\Gamma}{\iota}[\Theta] \\
  }{\conv{\Gamma}{\ad{\iota}{\id_\Theta}}{\iota}[\Theta]} \\
  \inferdef[AdInstComp]
  {
    \typing{\Gamma}{\iota}[\Theta_1] \\
    \typing{\Gamma}{a}[\adso{\Theta_2}{\Theta_3}]\\
    \typing{\Gamma}{b}[\adso{\Theta_1}{\Theta_2}]\\
  }{\conv{\Gamma}{\ad{\iota}{a \circ b}}{\ad{\ad{\iota}{b}}{a}}[\Theta_3]} \\
  \inferdef[SubInstOnAdInst]
  {
    \typing{\Delta}{\iota}[\Theta_1] \\
    \typing{\Delta}{a}[\adso{\Theta_1}{\Theta_2}]\\
    \typing{\Gamma}{\sigma}[\Delta]
  }{\conv{\Gamma}{\subs{\ad{\iota}{a}}{\sigma}}{\ad{\subs{\iota}{\sigma}}{\subs{a}{\sigma}}}[\subs{\Theta_2}{\sigma}]} \\
  \inferdef[CtxExtTelDual] {
    \typing{\dual{\Gamma}{d}}{\Theta} \\
  } {
    \conv{}{\dual{(\Gamma \ext_d \Theta)}{-}}{\dual{\Gamma}{-} \ext_{-d} \Theta}
  } \and
  \inferdef[SubExtInstDual] {
    \typing{\Gamma}{\sigma}[\Delta] \\
    \typing{\dual{\Delta}{d}}{\Gamma} \\
    \typing{\dual{\Gamma}{d}}
  } {
    \conv{\dual{\Gamma}{-}}{\dual{(\sigma \ext_d \iota)}{-}}{ \dual{\iota}{-} \ext_{-d} \iota}[\dual{\Delta \ext_d \Theta}{-}]
  } \and
  \inferdef[WkTelDual] {
    \typing{\dual{\Gamma}{d}}{\Theta}
  } {
    \conv{\dual{\Gamma \ext_d \Theta}{-}}{\dual{\wk}{-}}{\wk}[\dual{\Gamma}{-}]
  } \and
  \inferdef[VarInstAdTrans] {
    \typing{\Delta}{\Theta} \\
    \typing{\Gamma}{\sigma}[\Delta \ext \Theta] \\
    \typing{\Gamma}{\tau}[\Delta \ext \Theta] \\
    \typing{\Gamma}{\mu}[\transso[\Delta\ext_+\Theta]{\sigma}{\tau}]
  }{
    \conv{\Gamma}
    {\ad{\subs{\vinst_\Theta}{\sigma}}{\trans{\Theta}{\wk \whisk \mu}}}
    {\subs{\vinst_\Theta}{\tau}}
    [\subs{\Theta}{\tau}]
  } \and
  \inferdef[InstEmp] {
    \ctxty{\Gamma}
  } {
    \typing{\Gamma}{\emp}[\emp]
  } \and
  \inferdef[InstExtTm] {
    \typing{\Gamma}{\Theta} \\
    \typing{\Gamma}{\iota}[\Theta] \\
    \typing{\Gamma \ext \Theta }{T} \\
    \typing{\Gamma}{t}[\subs{T}{\id \ext \iota}]
  } {
    \typing{\Gamma}{\iota \ext t}[\Theta \ext T]
  } \and
  \inferdef[InstTelEmp] {
    \typing{\Gamma}{\iota}[\emp]
  } {
    \conv{\Gamma}{\iota}{\emp}[\emp]
  } \and
  \inferdef[SubOnInstEmp] {
    \typing{\Gamma}{\sigma}[\Delta]
  } {
    \conv{\Gamma}{\subs{\emp}{\sigma}}{\emp}[\emp]
  } \and
  \inferdef[SubExtInstTl]{
    \typing{\Gamma}{\sigma}[\Delta] \\
    \typing{\dual{\Delta}{d}}{\Theta} \\
    \typing{\Gamma}{\iota}[\subs{\Theta}{\dual{\sigma}{d}}]
  }{
    \conv{\Gamma}
    {\wk \circ (\sigma \ext_d \iota)}
    {\sigma}[\Delta]
    }[\(d : \Dir\)]
  \and
  \inferdef[SubExtInstEta]{
    \typing{\Delta}{\Theta} \\
    \typing{\Gamma}{\sigma}[\Delta \ext \Theta] \\
  }{
    \conv{\Gamma}
    {(\wk \circ \sigma) \ext_d \subs{\vinst_\Theta}{\sigma}}
    {\sigma}[\Delta]
  } \and
  \inferdef[SubInstExtVarInst]{
    \typing{\Gamma}{\sigma}[\Delta] \\
    \typing{\Delta}{\Theta} \\
    \typing{\Gamma}{\iota}[\subs{\Theta}{\sigma}]
    }{
      \conv{\Gamma}
      {\subs{\vinst_\Theta}{\sigma \ext \iota} }
      {\iota}
      [\subs{\Theta}{\sigma}]
    }
  \and
  \inferdef[CtxExtTelEmp] {
    \ctxty{\Gamma}
  } {
    \conv{}{\Gamma \ext_d \emp}{\Gamma}
  } \and
  \inferdef[CtxExtTelExt] {
    \typing{\dual{\Gamma}{d}}{\Theta} \\
    \typing{\dual{\Gamma}{d} \ext \Theta}{A}
  } {
    \conv{}{\Gamma \ext_d (\Theta \ext A)}{(\Gamma \ext_d \Theta) \ext A}
  }
  \inferdef[WkTelWkTy]{
    \typing{\dual{\Gamma}{d}}{\Theta} \\
    \typing{\dual{\Gamma}{d} \ext_{+} \Theta}{A} \\
  }{
    \conv{\Gamma \ext_{d} \Theta \ext_{+} A}{\wk_{\Theta} \circ \wk_{A}}{\wk_{\Theta \ext A}}[\Gamma]
  }[\(d: \Dir\)]
  \and
  \inferdef[VarInstExtVarZ]{
    \typing{\dual{\Gamma}{d}}{\Theta} \\
    \typing{\dual{\Gamma}{d} \ext_{+} \Theta}{A} \\
  }{
    \conv{\Gamma \ext_{d} (\Theta \ext A)}{\vinst_{\Theta \ext A}}{\subs{\vinst_{\Theta}}{\wk}
      \ext \tmvz}[\Theta \ext A]
  }[\(d: \Dir\)]
  \and
  \inferdef[SubTelAd]{
    % \impl{\Gamma,\Delta \ty \Ctx} \\
    \typing{\Gamma}{\sigma}[\Delta] \\
    \typing{\dual{\Gamma}{d}}{\Theta} \\
    \typing{{\Delta}{d}}{\Theta'} \\
    \typing{\dual{\Gamma}{d}}{\alpha}[\adso{\Theta}{\subs{\Theta'}{\dual{\sigma}{d}}}]
  }{
    \defconv{\Gamma \ext_{d} \Theta}{\sigma \pext_{d} \alpha}
      {\dual{\left((\dual{\sigma}{d} \circ \wk) \ext \ad{\vinst}{\subs{\alpha}{\wk}}\right)}{d}}
      [\Delta \ext_{d} \Theta']
  }[\(d: \Dir\)] \and
  \inferdef[SubExtInstEmp]{
    \typing{\Gamma}{\sigma}[\Delta]
  }{\conv{\Gamma}{\sigma \ext_{d} \emp}{\sigma}[\Delta]}[\(d: \Dir\)]
  \and
  \inferdef[SubExtInstExt]{
    \typing{\Gamma}{\sigma}[\Delta] \\
    \typing{\Delta}{\Theta} \\
    \typing{\Gamma}{\iota}[\subs{\Theta}{\sigma}] \\
    \typing{\Delta \ext \Theta}{A} \\
    \typing{\Gamma}{t}[\subs{A}{\sigma \ext \iota}]
  }{
    \conv{\Gamma}{\sigma \ext_{d} (\iota \ext t)}{(\sigma \ext \iota) \ext t}
    [\Delta \ext \Theta \ext A]
  }[\(d: \Dir\)] \and
  \inferdef[TelAdExt]{
    \typing{\Gamma}{\alpha}[\adso{\Theta}{\Theta'}] \\
    \typing{\Gamma \ext \Theta}{f}[\adso{A}{\subs{A'}{\id \pext \alpha}}]
  }{
    \typing{\Gamma}{\alpha \ext f}[\adso{\Theta \ext A}{\Theta' \ext A'}]
  } \and
  \inferdef[TelAdExtId]{
    \typing{\Gamma}{\Theta} \\
    \typing{\Gamma \ext \Theta}{A}
  }{
    \conv{\Gamma}{\id_{\Theta} \ext \id_{A}}{\id_{\Theta \ext A}}
      [\adso{\Theta \ext A}{\Theta \ext A}]
  } \and
  \inferdef[TelAdExtComp]{
    \typing{\Gamma}{\alpha}[\adso{\Theta}{\Theta'}] \\
    \typing{\Gamma}{\beta}[\adso{\Theta'}{\Theta''}] \\
    \typing{\Gamma \ext \Theta}{f}[\adso{A}{\subs{A'}{\id \pext \alpha}}] \\
    \typing{\Gamma \ext \Theta'}{g}[\adso{A'}{\subs{A''}{\id \pext \beta}}]
  }{
    \conv{\Gamma}{(\beta \ext b) \circ (\alpha \ext a)}
      {(\beta \circ \alpha) \ext (\subs{b}{\id \pext \alpha} \circ A)}
      [\adso{\Theta \ext A}{\Theta'' \ext A''}]
  } \and
  \inferdef[TransInst+]{
    \typing{\Gamma}{\mu}[\transso[\Delta]{\sigma}{\tau}] \\
    \typing{\Gamma}{\iota}[\subs{\Theta}{\sigma}]}
  {
    \typing{\Gamma}{\mu \ext \iota}[\transso
      {\sigma \ext \iota}{\tau \ext \ad{\iota}{\trans{\Theta}{\mu}}}]
  } \and
  \inferdef[TransInst-]{
    \typing{\Gamma}{\mu}[\transso[\Delta]{\sigma}{\tau}] \\
    \typing{\dual{\Gamma}{-}}{\iota}[\subs{\Theta}{\dual{\tau}{-}}]
  }{
    \typing{\Gamma}{\mu \ext_{-} \iota}[\transso
      {\sigma \ext_{-} \ad{\iota}{\trans{A}{\dual{\mu}{-}}}}{\tau \ext_{-} t}]
  } \and
  \inferdef[TransInst+Tl]{
    \typing{\Gamma}{\mu}[\transso[\Delta]{\sigma}{\tau}] \\
    \typing{\Gamma}{\iota}[\subs{\Theta}{\sigma}]
  }{
    \conv{\Gamma}{\wk \circ (\mu \ext_{+} \iota)}{\mu}[\transso{\sigma}{\tau}]}
  \and
    \inferdef[TransInst-Tl]{
    \typing{\Gamma}{\mu}[\transso[\Delta]{\sigma}{\tau}] \\
    \typing{\dual{\Gamma}{-}}{\iota}[\subs{A}{\dual{\tau}{-}}]
  }{\conv{\Gamma}{\wk \circ (\mu \ext_{-} \iota)}{\mu}[\transso{\sigma}{\tau}]}
  \and
  \inferdef[TransInstEta]{
    \typing{\Gamma}{\mu}[\transso[\Delta \ext_+ \Theta]{\sigma}{\tau}] \\
  }{
    \conv{\Gamma}{\mu}{(\wk \circ \mu) \ext_+ \subs{\vinst}{\sigma}}[\transso{\sigma}{\tau}]}
  \and
  \inferdef[TransInst+Dual]
  {
    \typing{\Gamma}{\mu}[\transso[\Delta]{\sigma}{\tau}] \\
    % \typing{\Gamma}{A} \\
    \typing{\Gamma}{\iota}[\subs{\Theta}{\sigma}]
  }
  {
    \conv{\Gamma}{\dual{(\mu \ext_{+} \iota)}{-}}
    {\dual{\mu}{-} \ext_{-} \iota}
    [\transso[\Delta \ext \Theta]
      {\dual{\tau}{-} \ext \ad{\iota}{\trans{A}{\mu}}}
      {\dual{\sigma}{-} \ext \iota}] \\
    }
  \and
  \inferdef[TransInst-Dual]
  {
    \typing{\Gamma}{\mu}[\transso[\Delta]{\sigma}{\tau}] \\
    % \typing{\dual{\Gamma}{-}}{\Theta} \\
    \typing{\dual{\Gamma}{-}}{\iota}[\subs{\Theta}{\dual{\tau}{-}}]
  }
  { \conv{\dual{\Gamma}{-}}
    {\dual{(\mu \ext_{-} \iota)}{-}}
    {\dual{\mu}{-} \ext_{+} \iota}
    [\transso[\Delta \ext \Theta]
      {\dual{\tau}{-} \ext_{-} \iota}
      {\dual{\sigma}{-} \ext_{-} \ad{\iota}{\trans{\Theta}{\dual{\mu}{-}}}}]
  } \\
  \inferdef[TelExtTel]
  {
    \typing{\Gamma}{\Theta_1} \\
    \typing{\Gamma\ext_+\Theta_1}{\Theta_2}
  }{
    \typing{\Gamma}{\Theta_1\ext\Theta_2}
  }
  \and
   \inferdef[InstExtInst]
  {
    \typing{\Gamma}{\iota_1}[\Theta_1] \\
    \typing{\Gamma\ext_+\Theta_1}{\Theta_2}\\
    \typing{\Gamma}{\iota_2}[\subs{\Theta_2}{\id_\Gamma \ext_+ \iota_1}]
  }{
    \typing{\Gamma}{\iota_1\ext\iota_2}[\Theta_1\ext\Theta_2]
  }
  \\
  \inferdef[CtxExtTelExtTel]
  {
    \typing{\Gamma}{\Theta_1} \\
    \typing{\Gamma\ext_+\Theta_1}{\Theta_2}
  }{
    \typing{}{\Gamma\ext_+(\Theta_1\ext\Theta_2) \convop (\Gamma\ext_+\Theta_1)\ext_+\Theta_2}
  }
  \and
  \inferdef[TelExtTelEmp]
  {
    \typing{\Gamma}{\Theta} \\
   }{
    \typing{\Gamma}{\Theta\ext\emp \convop \Theta}
  }
  \and
  \inferdef[TelExtTelExtTy]
  {
    \typing{\Gamma}{\Theta_1} \\
    \typing{\Gamma\ext_+\Theta_1}{\Theta_2} \\
    \typing{(\Gamma\ext_+\Theta_1)\ext_+\Theta_2}{A}
   }{
    \typing{\Gamma}{\Theta_1\ext(\Theta_2\ext A) \convop (\Theta_1 \ext\Theta_2)\ext A}
  } \and
   \inferdef[TelExtTelExtTel]
  {
    \typing{\Gamma}{\Theta_1} \\
    \typing{\Gamma\ext_+\Theta_1}{\Theta_2} \\
    \typing{(\Gamma\ext_+\Theta_1)\ext_+\Theta_2}{\Theta_3}
   }{
    \typing{\Gamma}{\Theta_1\ext(\Theta_2\ext \Theta_3) \convop (\Theta_1 \ext\Theta_2)\ext \Theta_3}
  } \and
  \inferdef[SubTelOnExtTel]
  {
    \typing{\Delta}{\sigma}[\Gamma]\\
    \typing{\Gamma}{\Theta_1} \\
    \typing{\Gamma\ext_+\Theta_1}{\Theta_2}
  }{
    \typing{\Delta}{(\Theta_1\ext\Theta_2)[\sigma] \convop \Theta_1[\sigma] \ext \Theta_2[\sigma\pext \id_{\Theta_1[\sigma]}] }
  }
  \\
  \inferdef[WkTelWkTel]
  {
    \typing{\Gamma}{\Theta_1} \\
    \typing{\Gamma\ext_+\Theta_1}{\Theta_2}
  }{
    \typing{(\Gamma\ext_+\Theta_1)\ext_+\Theta_2}{\wk_{\Gamma,\Theta_1}\circ\wk_{\Gamma\ext_+\Theta_1,\Theta_2} \convop \wk_{\Gamma,\Theta_1\ext\Theta_2}}[\Gamma]
  }
  \and
  \inferdef[SubExtInstExtInst]
  { \typing{\Gamma}{\sigma}[\Delta]\\
    \typing{\Delta}{\Theta_1} \\
    \typing{\Gamma}{\iota_1}[\subs{\Theta_1}{\sigma}]\\
    \typing{\Delta \ext_+ \Theta_1}{\Theta_2}\\
    \typing{\Gamma}{\iota_2}[\subs{\Theta_2}{\Theta_1 \ext_+ \iota_1}]
  }
  {\conv{\Gamma}{\sigma \ext_+ (\iota_1 \ext \iota_2)}{(\sigma \ext \iota_1)
      \ext \iota_2}[\Delta \ext_+ (\Theta_1 \ext \Theta_2)]}
  \\
    \inferdef[TelAdExtTelAd]
  {
    \typing{\Gamma}{a}[\adso{\Theta_1}{\Theta_1'}] \\
    \typing{\Gamma\ext\Theta_1}{\Theta_2}\\
    \typing{\Gamma\ext\Theta_1'}{\Theta_2'}\\
    \typing{\Gamma\ext_+\Theta_1}{b}[\adso{\Theta_2}{\Theta_2'[\id\pext a]}]
  }{
    \typing{\Gamma}{a\ext b}[\adso{\Theta_1\ext\Theta_2}{\Theta_1'\ext\Theta_2'}]
  } \\
  \inferdef[TelAdExtTelAdEmp]
  {
    \typing{\Gamma}{a}[\adso{\Theta_1}{\Theta_1'}]
  }{
    \typing{\Gamma}{a\ext\emp\convop a}[\adso{\Theta_1}{\Theta_1'}]
  } \and
  \inferdeft[SubTelAdOnExtAd]
  {
    \typing{\Gamma}{a}[\adso{\Theta_1}{\Theta_1'}] \\
    \typing{\Gamma\ext\Theta_1}{A}\\
    \typing{\Gamma\ext\Theta_1'}{A'}\\
    \typing{\Gamma\ext_+\Theta_1}{b}[\adso{A}{A'[\id\pext a]}] \\
    \typing{\Delta}{\sigma}[\Gamma]
  }{
    \conv{\Gamma}{\subs{(a\ext b)}{\sigma}}{\subs{a}{\sigma}\ext \subs{b}{(\sigma \circ \wk) \ext \vinst}}[\adso{\subs{\Theta_1}{\sigma}\ext\subs{A}{(\sigma \circ \wk) \ext \vinst}}{\subs{\Theta_1'}{\sigma}\ext\subs{A'}{(\sigma \circ \wk) \ext \vinst}}]
  } \and
    \inferdeft[SubTelAdOnExtAd]
  {
    \typing{\Gamma}{a}[\adso{\Theta_1}{\Theta_1'}] \\
    \typing{\Gamma\ext\Theta_1}{\Theta_2}\\
    \typing{\Gamma\ext\Theta_1'}{\Theta_2'}\\
    \typing{\Gamma\ext_+\Theta_1}{b}[\adso{\Theta_2}{\Theta_2'[\id\pext a]}] \\
    \typing{\Delta}{\sigma}[\Gamma]
  }{
    \conv{\Gamma}{\subs{(a\ext b)}{\sigma}}{\subs{a}{\sigma}\ext \subs{b}{(\sigma \circ \wk) \ext \vinst}}[\adso{\subs{\Theta_1}{\sigma}\ext\subs{\Theta_2}{(\sigma \circ \wk) \ext \vinst}}{\subs{\Theta_1'}{\sigma}\ext\subs{\Theta_2'}{(\sigma \circ \wk) \ext \vinst}}]
  } \and
  \inferdef[TelTransSub]
  {
    \typing{\Delta}{\Theta}\\
    \typing{\Gamma}{\mu}[\transso[\Delta]{\tau}{\xi}]\\
    \typing{\Xi}{\sigma}[\Gamma]\\
  }
  {
      \conv{\Xi}{\subs{\trans{\Theta}{\mu}}{\sigma}}{\trans{\Theta}{\mu\circ\sigma}}[\adso{\subs{\Theta}{\tau\circ\sigma}}{\subs{\Theta}{\xi\circ\sigma}}]
  } \and
  \inferdef[TelSubTrans]
  {
    \typing{\Delta}{\Theta}\\
    \typing{\Gamma}{\sigma}[\Delta]\\
    \typing{\Xi}{\mu}[\transso[\Gamma]{\tau}{\xi}]\\
  }
  {
    \conv{\Xi}{\trans{\subs{\Theta}{\sigma}}{\mu}}{\trans{\Theta}{\sigma\circ\mu}}[\adso{\subs{\Theta}{\sigma\circ\tau}}{\subs{\Theta}{\sigma\circ\xi}}]
  }
  \and
  \inferdef[VarInstTelExtTy]
  { \typing{\Gamma}{\Theta} \\
    \typing{\Gamma \ext_+ \Theta}{A}
  }
  {\conv{\Gamma \ext_+ (\Theta \ext A)}
    {\vinst_{\Theta \ext A}}
    {\subs{\vinst_\Theta}{\wk} \ext \tmvz}
    [\subs{(\Theta \ext A)}{\wk}]}
  \and
  \inferdef[VarInstTelExtTel]
  {
    \typing{\Gamma}{\Theta_1} \\
    \typing{\Gamma \ext_+ \Theta_1}{\Theta_2}
  }
  {\conv{\Gamma \ext_+ (\Theta_1 \ext \Theta_2)}
    {\vinst_{\Theta_1 \ext \Theta_2}}
    {\subs{\vinst_{\Theta_1}}{\wk} \ext \vinst_{\Theta_2}}
    [\subs{(\Theta_1 \ext \Theta_2)}{\wk}]
  }
  \and
  \inferdef[TelAdExtTelAdExtAd]
  { \typing{\Gamma}{a_1}[\adso{\Theta_1}{\Theta'_1}]\\
    \typing{\Gamma \ext_+ \Theta_1}{\Theta_2}\\
    \typing{\Gamma \ext_+ \Theta'_1}{\Theta'_2}\\
    \typing{\Gamma \ext_+ \Theta_1}{a_2}[\adso{\Theta_2}{\subs{\Theta'_2}{\id
        \pext a_1}}]\\
    \typing{\Gamma \ext_+ (\Theta_1 \ext \Theta_2)}{A} \\
    \typing{\Gamma \ext_+ (\Theta'_1 \ext \Theta'_2)}{A'} \\
    \typing{\Gamma \ext_+ (\Theta_1 \ext \Theta_2)}{a}[\adso{A}{\subs{A'}{\id
        \pext (a_1 \ext a_2)}}]
  }
  {\conv{\Gamma}{(a_1 \ext a_2) \ext a}{a_1 \ext (a_2 \ext a)}[\adso{\Theta_1 \ext
      \Theta_2 \ext A}{\Theta'_1 \ext \Theta'_2 \ext A'}]}
  \and
  \inferdef[TelAdExtTelAdExtTelAd]
  { \typing{\Gamma}{a_1}[\adso{\Theta_1}{\Theta'_1}]\\
    \typing{\Gamma \ext_+ \Theta_1}{\Theta_2}\\
    \typing{\Gamma \ext_+ \Theta'_1}{\Theta'_2}\\
    \typing{\Gamma \ext_+ \Theta_1}{a_2}[\adso{\Theta_2}{\subs{\Theta'_2}{\id
        \pext a_1}}]\\
    \typing{\Gamma \ext_+ (\Theta_1 \ext \Theta_2)}{\Theta_3} \\
    \typing{\Gamma \ext_+ (\Theta'_1 \ext \Theta'_2)}{\Theta'_3} \\
    \typing{\Gamma \ext_+ (\Theta_1 \ext \Theta_2)}{a_3}[\adso{\Theta_3}{\subs{\Theta'_3}{\id \pext (a_1 \ext a_2)}}]
  }
  {\conv{\Gamma}{(a_1 \ext a_2) \ext a_3}{a_1 \ext (a_2 \ext a_3)}[\adso{\Theta_1 \ext
      \Theta_2 \ext \Theta_3}{\Theta'_1 \ext \Theta'_2 \ext \Theta'_3}]}
  % \and
  % \inferdef[WkTelExtTelAdapt]
  % {\typing{\Gamma}{a_1}[\adso{\Theta_1}{\Theta'_1}]\\
  %   \typing{\Gamma \ext_+ \Theta_1}{\Theta_2}\\
  %   \typing{\Gamma \ext_+ \Theta'_1}{\Theta'_2}\\
  %   \typing{\Gamma \ext_+ \Theta_1}{a_2}[\adso{\Theta_2}{\subs{\Theta'_2}{\id
  %       \pext a_1}}]}
  % {\conv{}{}{}}
\end{mathparpagebreakable}

\subsection{Type variables}
\label{sec:adapTTt-full-rules-type-variables}

\begin{mathparpagebreakable}
  \inferdef[CtxTy]
  {\ctxty{\Gamma}\\
    \typing{\dual{\Gamma}{d'}}{\Theta}}
  {\ctxty{{\Gamma \ext*_{d'} (\tyext{\Theta}[d])}}}
  [\(d, d' : \Dir\)]
  \and
  \inferdef[SubTy]
  {
    % \impl{\Gamma, \Delta \ty \Ctx} \\
    \typing{\Gamma}{\sigma}[\Delta] \\
    \typing{\dual{\Delta}{d'}}{\Theta} \\
    \typing{\dual{\left(\Gamma \ext_{d'} \subs{\Theta}
      {\dual{\sigma}{d'}}\right)}{d}}{A}
  }{\typing{\Gamma}{\sigma \ext*_{d} A}[\Delta \ext*_{d'} (\tyext{\Theta}[d])]}
  [\(d, d' : \Dir\)]
  \and
  \inferdef[TransAd++]{
    % \impl{\Gamma, \Delta \ty \Ctx} \\
    % \impl{\sigma, \tau \ty \Sub(\Gamma,\Delta)} \\
    \typing{\Gamma}{\mu}[\transso[\Delta]{\sigma}{\tau}] \\
    \typing{\Delta}{\Theta} \\
    \typing{\Gamma \ext_{+} \hsubs{\Theta}{\sigma}}{A} \\
    \typing{\Gamma \ext_{+} \hsubs{\Theta}{\tau}}{B} \\
    \typing{\Gamma \ext_{+} \hsubs{\Theta}{\sigma}}{f}[
      \adso{A}{\hsubs{B}{\id_{\Gamma} \pext_{+} \trans{\Theta}{\mu}}}]
  }{
    \typing{\Gamma}{\mu \ext*_{+} f}
      [\transso[\Delta \ext*_{+} (\tyext{\Theta}[+])]{\sigma \ext*_{+} A}{\tau \ext*_{+} B}]
  }
  \and
  \inferdef[TransAd+-]{
    % \impl{\Gamma, \Delta \ty \Ctx} \\
    % \impl{\sigma, \tau \ty \Sub(\Gamma,\Delta)} \\
    \typing{\Gamma}{\mu}[\transso[\Delta]{\sigma}{\tau}] \\
    \typing{\Delta}{\Theta} \\
    \typing{\dual{\left(\Gamma \ext_{+} \hsubs{\Theta}{\sigma}\right)}{-}}{A} \\
    \typing{\dual{\left(\Gamma \ext_{+} \hsubs{\Theta}{\tau}\right)}{-}}{B} \\
    \typing{\dual{\left(\Gamma \ext_{+} \hsubs{\Theta}{\sigma}\right)}{-}}{f}[
      \adso{\hsubs{B}{\id_{\Gamma} \pext_{+} \trans{\Theta}{\mu}}}{A}]
  }{
    \typing{\Gamma}{\mu \ext*_{-} f}
      [\transso[\Delta \ext*_{+} (\tyext{\Theta}[-])]{\sigma \ext*_{-} A}{\tau \ext*_{-} B}]
  }
  \and
  \inferdef[TransAd-+]{
    % \impl{\Gamma, \Delta \ty \Ctx} \\
    % \impl{\sigma, \tau \ty \Sub(\Gamma,\Delta)} \\
    \typing{\Gamma}{\mu}[\transso[\Delta]{\sigma}{\tau}] \\
    \typing{\dual{\Delta}{-}}{\Theta} \\
    \impl{\typing{\Gamma \ext_{-} \hsubs{\Theta}{\dual{\sigma}{-}}}{A}} \\
    \impl{\typing{\Gamma \ext_{-} \hsubs{\Theta}{\dual{\tau}{-}}}{B}} \\
    \typing{\Gamma \ext_{\impl{-}} \hsubs{\Theta}{\dual{\tau}{-}}}{f}[
      \adso{\hsubs{A}{\id_{\Gamma} \pext_{-} \trans{\Theta}{\dual{\mu}{-}}}}{B}]
  }{
    \typing{\Gamma}{\mu \ext*_{+} f}
      [\transso[\Delta \ext*_{-} (\tyext{\Theta}[+])]{\sigma \ext*_{+} A}{\tau \ext*_{+} B}]
  }
  \and
  \inferdef[TransAd-{}-]{
    % \impl{\Gamma, \Delta \ty \Ctx} \\
    % \impl{\sigma, \tau \ty \Sub(\Gamma,\Delta)} \\
    \typing{\Gamma}{\mu}[\transso[\Delta]{\sigma}{\tau}] \\
    \typing{\dual{\Delta}{-}}{\Theta} \\
    \impl{\typing{\dual{\left(\Gamma \ext_{-} \hsubs{\Theta}{\dual{\sigma}{-}}\right)}{-}}{A}} \\
    \impl{\typing{\dual{\left(\Gamma \ext_{-} \hsubs{\Theta}{\dual{\tau}{-}}\right)}{-}}{B}} \\
    \typing{\dual{\left(\Gamma \ext_{\impl{-}} \hsubs{\Theta}{\dual{\tau}{-}}\right)}{-}}{f}[
      \adso{B}{\hsubs{A}{\dual{\left(\id_{\Gamma} \pext_{-} \trans{\Theta}{\dual{\mu}{-}}\right)}{d}}}]
  }{
    \typing{\Gamma}{\mu \ext*_{-} f}
      [\transso[\Delta \ext*_{-} (\tyext{\Theta}[-])]{\sigma \ext*_{-} A}{\tau \ext*_{-} B}]
  }
  \and
  \inferdef[WkTy]{
    \ctxty{\Gamma} \\
    \typing{\dual{\Gamma}{d'}}{\Theta}}{
    \typing{\Gamma \ext*_{d'} (\tyext{\Theta}[d])}{\wk}[\Gamma]
  }[\(d,d' : \Dir\)]
  \and
  \inferdef[VarZTy]{
    \ctxty{\Gamma} \\
    \typing{\dual{\Gamma}{d}}{\Theta}
  }{\typing{\Gamma \ext*_{d} (\tyext{\Theta}[+]) \ext_{d} \subs{\Theta}{\dual{\wk}{d}}}{\tyvz}
  }[\(d \ty \Dir\)] \and
  \inferdef[CtxTyDual]{
    \typing{\dual{\Gamma}{d'}}{\Theta}
  }{
    \ctxconv{\dual{(\Gamma \ext*_{d'} \tyext{\Theta}[d])}{-}}{\dual{\Gamma}{-} \ext*_{-d'} \tyext{\Theta}[-d]}
  }[\(d,d' : \Dir\)] \and
  \inferdef[WkTyDual]{
    \typing{\dual{\Gamma}{d'}}{\Theta}
  }{
    \conv{\dual{(\Gamma \ext*_{d'} \tyext{\Theta}[d])}{-}}{\dual{\wk}{-}}{\wk}[\dual{\Gamma}{-}]
  }[\(d,d' : \Dir\)] \and
  \inferdef[SubTyDual]{
    \typing{\Gamma}{\sigma}[\Delta] \\
    \typing{\dual{\Delta}{d'}}{\Theta} \\
    \typing{\dual{(\Gamma \ext_{d'} \subs{\Theta}{\dual{\sigma}{d'}})}{d}}{A}
  }{
    \conv{\Gamma}{\dual{(\sigma \ext*_{d} A)}{-}}{\dual{\sigma}{-} \ext*_{-d} A}
      [\dual{(\Delta \ext*_{d'} \tyext{\Theta}[d])}{-}]
  }[\(d,d' : \Dir\)] \and
  \inferdef[TransAd++Dual]{
    \typing{\Gamma}{\mu}[\transso[\Delta]{\sigma}{\tau}] \\
    \typing{\Delta}{\Theta} \\
    \typing{\Gamma \ext_{+} \subs{\Theta}{\sigma}}{A} \\
    \typing{\Gamma \ext_{+} \subs{\Theta}{\tau}}{B} \\
    \typing{\Gamma \ext_{+} \subs{\Theta}{\sigma}}{a}[\adso{A}{\subs{B}{\id \pext \trans{\Theta}{\mu}}}]
  }{
    \conv{\dual{\Gamma}{-}}{\dual{(\mu \ext*_{+} a)}{-}}{\dual{\mu}{-} \ext*_{-} a}
      [\transso[\dual{(\Delta \ext*_{+} \tyext{\Theta}[+])}{-}]{\dual{(\tau \ext*_{-} B)}{-}}{\dual{(\sigma \ext*_{-} A)}{-}}]
  }\and
  \inferdef[TransAd+-Dual]{
    \typing{\Gamma}{\mu}[\transso[\Delta]{\sigma}{\tau}] \\
    \typing{\Delta}{\Theta} \\
    \typing{\dual{\left(\Gamma \ext_{+} \subs{\Theta}{\sigma}\right)}{-}}{A} \\
    \typing{\dual{\left(\Gamma \ext_{+} \subs{\Theta}{\tau}\right)}{-}}{B} \\
    \typing{\dual{\left(\Gamma \ext_{+} \subs{\Theta}{\sigma}\right)}{-}}{a}[\adso{\subs{B}{\dual{\left(\id \pext \trans{\Theta}{\mu}\right)}{-}}}{A}]
  }{
    \conv{\dual{\Gamma}{-}}{\dual{(\mu \ext*_{-} a)}{-}}{\dual{\mu}{-} \ext*_{+} a}
      [\transso[\dual{(\Delta \ext*_{+} \tyext{\Theta}[-])}{-}]{\dual{(\tau \ext*_{+} B)}{-}}{\dual{(\sigma \ext*_{-} A)}{-}}]
  } \and
  \inferdef[TransAd-+Dual]{
    \typing{\Gamma}{\mu}[\transso[\Delta]{\sigma}{\tau}] \\
    \typing{\dual{\Delta}{-}}{\Theta} \\
    \typing{\Gamma \ext_{-} \subs{\Theta}{\dual{\sigma}{-}}}{A} \\
    \typing{\Gamma \ext_{-} \subs{\Theta}{\dual{\tau}{-}}}{B} \\
    \typing{\Gamma \ext_{-} \subs{\Theta}{\dual{\tau}{-}}}{a}
      [\adso{\subs{A}{\id \pext \trans{\Theta}{\dual{\mu}{-}}}}{B}]
  }{
    \conv{\dual{\Gamma}{-}}{\dual{(\mu \ext*_{+} a)}{-}}{\dual{\mu}{-} \ext*_{-} a}
      [\transso[\dual{(\Delta \ext*_{-} \tyext{\Theta}[+])}{-}]{\dual{(\tau \ext*_{+} B)}{-}}{\dual{(\sigma \ext*_{+} A)}{-}}]
  }
  \and
  \inferdef[TransAd-{}-Dual]{
    \typing{\Gamma}{\mu}[\transso[\Delta]{\sigma}{\tau}] \\
    \typing{\dual{\Delta}{-}}{\Theta} \\
    \typing{\dual{(\Gamma \ext_{-} \subs{\Theta}{\dual{\sigma}{-}})}{-}}{A} \\
    \typing{\dual{(\Gamma \ext_{-} \subs{\Theta}{\dual{\tau}{-}})}{-}}{B} \\
    \typing{\dual{(\Gamma \ext_{-} \subs{\Theta}{\dual{\tau}{-}})}{-}}{a}
      [\adso{B}{\subs{A}{\dual{\left(\id \pext \trans{\Theta}{\dual{\mu}{-}}\right)}{-}}}]
  }{
    \conv{\dual{\Gamma}{-}}{\dual{(\mu \ext*_{-} a)}{-}}{\dual{\mu}{-} \ext*_{+} a}
      [\transso[\dual{(\Delta \ext*_{-} \tyext{\Theta}[-])}{-}]{\dual{(\tau \ext*_{-} B)}{-}}{\dual{(\sigma \ext*_{-} A)}{-}}]
  }
  \and
  \inferdef[SubTlTy]{
     % \impl{\Gamma, \Delta \ty \Ctx} \\
    \typing{\Gamma}{\sigma}[\Delta] \\
    \typing{\dual{\Delta}{d'}}{\Theta} \\
    \typing{\dual{\left(\Gamma \ext_{d'} \subs{\Theta}
      {\dual{\sigma}{d'}}\right)}{d}}{A}
  }{\conv{\Gamma}{\wk \circ (\sigma \ext*_{d} A)}{\sigma}[\Delta]}
  [\(d, d' : \Dir\)] \and
  \inferdef[SubTyExtVarZ]{
     % \impl{\Gamma, \Delta \ty \Ctx} \\
    \typing{\Gamma}{\sigma}[\Delta] \\
    \typing{\dual{\Delta}{d}}{\Theta} \\
    \typing{\Gamma \ext_{d} \subs{\Theta}{\dual{\sigma}{d}}}{A} \\
    \typing{\dual{\Gamma}{d}}{\iota}[\subs{\Theta}{\dual{\sigma}{d}}]
  }{\conv{\Gamma}{\subs{\tyvz}{\sigma \ext*_{+} A \ext_{d} \iota}}
    {\subs{A}{\id_{\Gamma} \ext \iota}}}
  [\(d : \Dir\)] \and
  \inferdef[SubEtaTy]{
    \ctxty{\Gamma,\Delta} \\
    \typing{\dual{\Delta}{d}}{\Theta} \\
    \typing{\Gamma}{\sigma}[\Delta \ext*_{d} \tyext{\Theta}{+}]
  }{\conv{\Gamma}{\sigma}{(\wk \circ \sigma) \ext* \subs{\tyvz}{\sigma \pext \id}}
    [\Delta \ext*_{d} \tyext{\Theta}{+}]
  }[\(d : \Dir\)] \and
  \inferdef[TransTlAd++]{
    \typing{\Gamma}{\mu}[\transso[\Delta]{\sigma}{\tau}] \\
    \typing{\Delta}{\Theta} \\
    \typing{\Gamma \ext_{+} \subs{\Theta}{\sigma}}{A} \\
    \typing{\Gamma \ext_{+} \subs{\Theta}{\tau}}{B} \\
    \typing{\Gamma \ext_{+} \subs{\Theta}{\sigma}}{f}[
      \adso{A}{\subs{B}{\id_{\Gamma} \pext_+ \trans{\Theta}{\mu}}}]
  }{
    \conv{\Gamma}{\wk \circ (\mu \ext*_{+} f)}{\mu}
      [\transso[\Delta]{\sigma}{\tau}]
    }
  \and
  \inferdef[TransTlAd+-]{
    \typing{\Gamma}{\mu}[\transso[\Delta]{\sigma}{\tau}] \\
    \typing{\Delta}{\Theta} \\
    \typing{\dual{\left(\Gamma \ext_{+} \subs{\Theta}{\sigma}\right)}{-}}{A} \\
    \typing{\dual{\left(\Gamma \ext_{+} \subs{\Theta}{\tau}\right)}{-}}{B} \\
    \typing{\dual{\left(\Gamma \ext_{+} \subs{\Theta}{\sigma}\right)}{-}}{f}[
      \adso{\subs{B}{\dual{\left(\id_{\Gamma} \pext_+ \trans{\Theta}{\mu}\right)}{-}}}{A}]
  }{
    \conv{\Gamma}{\wk \circ (\mu \ext*_{d} f)}{\mu}
      [\transso[\Delta]{\sigma}{\tau}]
    }
  \and
  \inferdef[TransTlAd-+]{
    \typing{\Gamma}{\mu}[\transso[\Delta]{\sigma}{\tau}] \\
    \typing{\dual{\Delta}{-}}{\Theta} \\
    \impl{\typing{\Gamma \ext_{-} \subs{\Theta}{\dual{\sigma}{-}}}{A}} \\
    \impl{\typing{\Gamma \ext_{-} \subs{\Theta}{\dual{\tau}{-}}}{B}} \\
    \typing{\Gamma \ext_{-} \subs{\Theta}{\dual{\tau}{-}}}{f}[
      \adso{\subs{A}{\dual{\left(\id_{\dual{\Gamma}{-}} \pext_- \trans{\Theta}{\dual{\mu}{-}}\right)}{d}}}{B}]
  }{
    \conv{\Gamma}{\wk \circ (\mu \ext*_{+} f)}{\mu}
      [\transso[\Delta]{\sigma}{\tau}]
  }
  \and
  \inferdef[TransTlAd-{}-]{
    \typing{\Gamma}{\mu}[\transso[\Delta]{\sigma}{\tau}] \\
    \typing{\dual{\Delta}{-}}{\Theta} \\
    \impl{\typing{\dual{\left(\Gamma \ext_{-} \subs{\Theta}{\dual{\sigma}{-}}\right)}{-}}{A}} \\
    \impl{\typing{\dual{\left(\Gamma \ext_{-} \subs{\Theta}{\dual{\tau}{-}}\right)}{-}}{B}} \\
    \typing{\dual{\left(\Gamma \ext_{-} \subs{\Theta}{\dual{\tau}{-}}\right)}{-}}{f}[
      \adso{B}{\subs{A}{\dual{\left(\id_{\dual{\Gamma}{-}} \pext_- \trans{\Theta}{\dual{\mu}{-}}\right)}{d}}}]
  }{
    \conv{\Gamma}{\wk \circ (\mu \ext*_{-} f)}{\mu}
      [\transso[\Delta]{\sigma}{\tau}]
  }
  \and
  \inferdef[TransAdEta+]{
    \typing{\Delta}{\Theta} \\
    \typing{\Gamma}{\mu}[\transso[\Delta \ext*_{+} \tyext{\Theta}{+}]{\sigma}{\tau}]
  }{
    \conv{\Gamma}{\mu}{(\wk \circ \mu) \ext* \trans{\tyvz}{\mu \pext_{+} \id}}
      [\transso[\Delta \ext*_{+} \tyext{\Theta}{+}]{\sigma}{\tau}]
    }[\(d : \Dir\)]
    \and
    \inferdef[TransAdEta-]{
    \typing{\dual{\Delta}{-}}{\Theta} \\
    \typing{\Gamma}{\mu}[\transso[\Delta \ext*_{-} \tyext{\Theta}{+}]{\sigma}{\tau}]
  }{
    \conv{\Gamma}{\mu}{(\wk \circ \mu) \ext* \trans{\tyvz}{\mu \pext_{-} \id}}
      [\transso[\Delta \ext*_{-} \tyext{\Theta}{+}]{\sigma}{\tau}]
    }[\(d : \Dir\)]
  \and
  \inferdef[TransHdAd+]{
    \typing{\Gamma}{\mu}[\transso[\Delta]{\sigma}{\tau}] \\
    \typing{\Delta}{\Theta} \\
    \typing{\Gamma \ext_{+} \subs{\Theta}{\sigma}}{A} \\
    \typing{\Gamma \ext_{+} \subs{\Theta}{\tau}}{B} \\
    \typing{\Gamma \ext_{\impl{+}} \subs{\Theta}{\sigma}}{f}[
      \adso{A}{\subs{B}{\id_{\Gamma} \pext_{+} \trans{\Theta}{\mu}}}]
  }{
    \conv{\Gamma}{\trans{\tyvz}{\mu \ext*_{+} f \ext \iota}}
      {\subs{f}{\id \ext \iota}}
      [\adso{\subs{A}{\id \ext \iota}}{\subs{B}{\id \ext \ad{\iota}{\trans{\Theta}{\mu}}}}]
    }[\(d : \Dir\)]
    \and
    \inferdef[TransHdAd-]{
    \typing{\Gamma}{\mu}[\transso[\Delta]{\sigma}{\tau}] \\
    \typing{\dual{\Delta}{-}}{\Theta} \\
    \typing{\Gamma \ext_{-} \subs{\Theta}{\dual{\sigma}{-}}}{A} \\
    \typing{\Gamma \ext_{-} \subs{\Theta}{\dual{\tau}{-}}}{B} \\
    \typing{\Gamma \ext_{\impl{-}} \subs{\Theta}{\dual{\tau}{-}}}{f}[
      \adso{\subs{A}{\id_{\Gamma} \pext_{-} \trans{\Theta}{\dual{\mu}{-}}}}{B}]
  }{
    \conv{\Gamma}{\trans{\tyvz}{\mu \ext*_{+} f \ext \iota}}
      {\subs{f}{\id \ext \iota}}
      [\adso{\subs{A}{\id \ext \iota}}{\subs{B}{\id \ext \ad{\iota}{\trans{\Theta}{\mu}}}}]
    }[\(d : \Dir\)]
  \end{mathparpagebreakable}

\subsection{Rules for Pi}

\begin{mathparpagebreakable}
  \inferdef[PiTy]{
    \typing{\dual{\Gamma}{-}}{A} \\
    \typing{\Gamma \ext_- A}{B}
  }{
    \typing{\Gamma}{\Pi A. B }
  } \and
  \inferdef[LamTm]{
    \typing{\dual{\Gamma}{-}}{A} \\
    \typing{\Gamma \ext_- A}{B} \\
    \typing{\Gamma \ext_- A}{t}[B]
  }{
    \typing{\Gamma}{\l t}[\Pi A. B]
  } \and
  \inferdef[AppTm]{
    \typing{\dual{\Gamma}{-}}{A} \\
    \typing{\Gamma \ext_- A}{B} \\
    \typing{\Gamma }{f}[\Pi A. B] \\
    \typing{\Gamma }{x}[A] \\
  }{
    \typing{\Gamma}{\app f x}[\subs{B}{\id_\Gamma \ext t}]
  } \and
  \inferdef[PiSub]
  {
    \typing{\Gamma}{\sigma}[\Delta] \\
    \typing{\dual{\Delta}{-}}{A} \\
    \typing{\Delta \ext_- A}{B}
  }{
    \conv{\Gamma}{\subs{\Pi A. B}{\sigma}}{\Pi \subs{A}{\sigma}.\subs{B}{(\sigma \circ \wk) \ext\tmvz_{\subs{A}{\sigma}}}}[\Delta]
  } \and
  \inferdef[LamSub]
  {
    \typing{\Gamma}{\sigma}[\Delta] \\
    \typing{\dual{\Delta}{-}}{A} \\
    \typing{\Delta \ext_- A}{B} \\
    \typing{\Delta \ext_- A}{t}[B]
  }{
    \conv{\Gamma}{\subs{\l t}{\sigma}}{\l (\subs{t}{(\sigma \circ \wk) \ext\tmvz_{\subs{A}{\sigma}}})}[\Delta]
  } \and
  \inferdef[AppSub]
  {
    \typing{\Gamma}{\sigma}[\Delta] \\
    \typing{\dual{\Delta}{-}}{A} \\
    \typing{\Delta \ext_- A}{B} \\
    \typing{\Delta }{f}[\Pi A. B] \\
    \typing{\Delta }{x}[A] \\
  }{
    \conv{\Gamma}
      {\subs{(\app f x)}{\sigma}}
      {\app{\subs{f}{\sigma}}{\subs{x}{\sigma}}}
      [\Delta]
  } \and
  \inferdef[$\beta$]
  {
    \typing{\Gamma}{\sigma}[\Delta] \\
    \typing{\dual{\Delta}{-}}{A} \\
    \typing{\Delta \ext_- A}{B} \\
    \typing{\Delta \ext_- A}{b}[B] \\
    \typing{\Delta }{x}[A] \\
  }{
    \conv{\Gamma}
      {\app{\l b}{x}}
      {\subs{b}{(\sigma \circ \wk) \ext\tmvz_{\subs{A}{\sigma}}}}
  } \and
  \inferdef[$\eta$]
  {
    \typing{\Gamma}{\sigma}[\Delta] \\
    \typing{\dual{\Delta}{-}}{A} \\
    \typing{\Delta \ext_- A}{B} \\
    \typing{\Delta }{f}[\Pi A. B] \\
  }{
    \conv{\Gamma}
      {f}
      {\l (\app{\subs{f}{\wk}}{\tmvz_A})}
  } \and
  \inferdef[PiAd]
  { \typing{\Gamma}{a}[\adso{A'}{A}] \\\\
    \typing{\Gamma \ext A'}{b}[\adso{\subs{B}{\ad{\tmvz}{a}}}{B'}]}
  {\typing{\Gamma}{\trans{\P}{a \ext* b}}[\adso{\P A. B}{\P A'. B'}]}
  \and
  \inferdef[AppPiAd]
  { \typing{\Gamma}{f}[\P A.B] \\ \typing{\Gamma}{u}[A'] \\\\
  \typing{\Gamma}{a}[\adso{A'}{A}] \\
  \typing{\Gamma \ext A'}{b}[\adso{\subs{B}{\ad{\tmvz}{a}}}{B'}]
  }
  {\conv{\Gamma}{(\ad{f}{\trans{\P}{a \ext* b}})\;u}
    {\ad{(f\,(\ad{u}{a}))}{\subs{b}{u}}}[\subs{B'}{u}]
  } \and
  \inferdef[TelToPiTy] {
    \typing{\dual{\Gamma}{-}}{\Theta} \\
    \typing{\Gamma \ext_- \Theta}{A}
  }{
    \typing{\Gamma}{\Pi \Theta.A}
  } \and
  \inferdef[TelToPiEmp] {
    \typing{\Gamma }{A}
  }{
    \conv{\Gamma}{\Pi \emp.A}{A}
  } \and
  \inferdef[TelToPiExt] {
    \typing{\dual{\Gamma}{-}}{\Theta} \\
    \typing{\dual{\Gamma}{-} \ext_+ \Theta}{A} \\
    \typing{\Gamma \ext_- (\Theta \ext A)}{B}
  }{
    \conv{\Gamma}{\Pi (\Theta \ext A).B}{\Pi \Theta. \Pi A. B}
  }\and
  \inferdef[TelToPiSub] {
    \typing{\Gamma}{\sigma}{\Delta} \\
    \typing{\dual{\Delta}{-}}{\Theta} \\
    \typing{\Delta \ext_- \Theta}{A}
  }{
    \conv{\Gamma}{\subs{\Pi \Theta.A}{\sigma}}{\Pi \subs{\Theta}{\sigma}.\subs{A}{\sigma \ext \vinst}}
  }
\end{mathparpagebreakable}

\subsection{Inductive types}
\begin{mathparpagebreakable}

  \inferdef[DataDescDef]{
    \ctxty{\Gamma_{par}} \\
    \typing{\Gamma_{par}}{\Theta_{ind}}
  }{
    \IndDesc(\Gamma_{par},\Theta_{ind}) \defconvop \mList{\ConDesc(\Gamma_{par},\Theta_{ind})}
  } \and
  \inferdef[ConDescDef]{
    \ctxty{\Gamma_{par}} \\
    \typing{\Gamma_{par}}{\Theta_{ind}} \\
    \typing{\Gamma_{par}}{\Theta_{nr}} \\
    r \ty \mList{\RecDesc(\Gamma_{par},\Theta_{ind},\Theta_{nr})} \\
    \typing{\Gamma_{par} \ext \Theta_{nr}}{\iota}[\subs{\Theta_{ind}}{\wk}]
  }{
    \{\rfield{nrec} \coloneq \Theta_{nr};\rfield{rec} \coloneq r;\rfield{ind} \coloneq \iota\}
      \ty \ConDesc(\Gamma_{par},\Theta_{ind})
  } \\
  \inferdef[RecDescDef]{
    \ctxty{\Gamma_{par}} \\
    \typing{\Gamma_{par}}{\Theta_{ind}} \\
    \typing{\Gamma_{par}}{\Theta_{nr}} \\
    \typing{\dual{(\Gamma_{par} \ext \Theta_{nr})}{-}}{\Theta_{ar}}\\
    \typing{(\Gamma \ext_{+} \Theta_{nr}) \ext_{-} \Theta_{ar}}{\iota}[\subs{\Theta_{ind}}{\wk}]
  }{
    \{\rfield{arit} \coloneq \Theta_{ar} ;\rfield{rind} \coloneq \iota \}
    \ty \RecDesc(\Gamma_{par},\Theta_{ind},\Theta_{nr})
  }

  \inferdef[IndTy]{
      % \impl{\Gamma \ty \Ctx(E)} \\ \impl{\Theta \ty \Tel(E,\Gamma)} \\
      \impl{I \ty \IndDesc(\Gamma_{par},\Theta_{ind})} \\
    }{
      \typing{\Gamma_{par}\ext \Theta_{ind}}{\ind(I)}
    } \and
  \inferdef[IndCstr]
  {
    I \ty \IndDesc(\Gamma_{par},\Theta_{ind}) \\
    \impl{c \ty \ConDesc(\impl{\Gamma_{par},\Theta_{ind}})} \\
    ic \ty c \in I \\
    % \ctxty{\Delta} \\
    % \typing{\Delta}{p}[\Gamma_{par}] \\
    % \typing{\Delta}{arg}[\conData(c,p,\subs{\ind(I)}{p \pext \id})]
  }{
    \typing{\Gamma_{par} \ext \subs{\conData(c)}{\id \ext \ind(I)}}
      {\constr(ic)}[\subs{\ind(I)}{\id \ext \subs{\proj{c}{ind}}{\wk}}]
  } \and
  \inferdef[ConDataDef]{
    % \ctxty{\Gamma_{par}} \\
    % \typing{\Gamma_{par}}{\Theta_{ind}} \\
    c \ty \ConDesc(\Gamma_{par},\Theta_{ind}) \\
    % \ctxty{\Delta} \\
  }{
    \defconv{\Gamma_{par} \ext*_{+} (\tyext{\Theta_{ind}}[+])}
    {\conData(c)}
    {\left(\subs{\proj{c}{nrec}}{\wk} \ext \recDatas(\proj{c}{rec})\right)}
    % [\Tel]
  } \and
  \inferdef[RecDatas]{
    \ctxty{\Gamma_{par}} \\
    \typing{\Gamma_{par}}{\Theta_{ind}} \\
    \typing{\Gamma_{par}}{\Theta_{nr}} \\
    c \ty \mList{\RecDesc(\Gamma_{par},\Theta_{ind},\Theta_{nr})} \\
  }{
    \typing{\Gamma_{par} \ext*_{+} (\tyext{\Theta_{ind}}[+]) \ext \subs{\Theta_{nr}}{\wk}}
    {\recDatas(c)}% [\Tel]
  } \and
  \inferdef[RecDatasEmp]{
    \ctxty{\Gamma_{par}} \\
    \typing{\Gamma_{par}}{\Theta_{ind}} \\
    \typing{\Gamma_{par}}{\Theta_{nr}} \\
  }{
    \conv{\Gamma_{par} \ext*_{+} (\tyext{\Theta_{ind}}[+]) \ext \subs{\Theta_{nr}}{\wk}}
    {\recDatas(\emp)}{\emp}%[\Tel]
  } \and
  \inferdef[RecDatasExt]{
    % \ctxty{\Gamma_{par}} \\
    % \typing{\Gamma_{par}}{\Theta_{ind}} \\
    % \typing{\Gamma_{par}}{\Theta_{nr}} \\\\
    c \ty \mList{\RecDesc(\Gamma_{par},\Theta_{ind},\Theta_{nr})} \\
    r \ty \RecDesc(\Gamma_{par},\Theta_{ind},\Theta_{nr}) \\
  }{
    \conv{\Gamma_{par} \ext*_{+} (\tyext{\Theta_{ind}}[+]) \ext \subs{\Theta_{nr}}{\wk}}
    {\recDatas(c \ext r)}{\recDatas(c) \ext \subs{\recData(r)}{\wk}}%[\Tel]
  } \and
  \inferdef[RecDataDef]{
    % \ctxty{\Gamma_{par}} \\
    % \typing{\Gamma_{par}}{\Theta_{ind}} \\
    % \typing{\Gamma_{par}}{\Theta_{nr}} \\
    r \ty \RecDesc(\Gamma_{par},\Theta_{ind},\Theta_{nr}) \\
    % \ctxty{\Delta} \\
    % \typing{\Delta}{\sigma}[\Gamma_{par}] \\
    % \typing{\Delta \ext \subs{\Theta_{ind}}{p}}{A}
  }{
    \Gamma_{par} \ext*_{+} (\tyext{\Theta_{ind}}[+]) \ext \subs{\Theta_{nr}}{\wk} \vdash
      \recData(r) \defconvop \hspace{10em} \\
      \hspace{10em}
      \P \subs{\proj{r}{arit}}{\wk \pext \id}.
      \left(\hsubs{\tyvz}{(\wk \circ \wk) \ext (\subs{\proj{r}{rind}}{(\wk \circ \wk \circ \wk ) \ext \subs{\vinst}{\wk} \ext \vinst}\right)})
     % \ty \Ty
  }

  \inferdef[IndAdEqPar]{
    I \ty \IndDesc(\Gamma_{par},\emp) \\
    \impl{c \ty \ConDesc(\impl{\Gamma_{par},\emp})} \\
    ic \ty c \in I \\\\
    % \ctxty{\Delta} \\
    % \typing{\Delta}{p,p'}[\Gamma_{par}] \\
    \typing{\Delta}{\mu}[\transso[\Gamma_{par}]{p}{p'}] \\
    \typing{\Delta}{arg}[\subs{\conData}{p \ext* \subs{\ind(I)}{p}}]
}{
    \Delta \vdash
    \ad{\subs{\constr(ic)}{p \ext arg}}{\trans{\ind(I)}{\mu}}
        \convop \hspace{11.5em}\\
    \hspace{3em} \subs{\constr(ic)}{p' \ext
      \ad{arg}{
      \trans{\left(\subs{\conData(c)}{\id \ext* \ind(I)}\right)}{\mu}}}
      \ty \subs{\ind(I)}{p'}
} \and
\inferdef[IndAdEq]{
    I \ty \IndDesc(\Gamma_{par},\Theta_{ind}) \\
    \impl{c \ty \ConDesc(\impl{\Gamma_{par},\Theta_{ind}})} \\
    ic \ty c \in I \\
    % \ctxty{\Delta} \\
    % \typing{\Delta}{p,p'}[\Gamma_{par}] \\
    \typing{\Delta}{\mu}[\transso[\Gamma_{par}]{p}{p'}] \\
    \typing{\Delta}{argn}[\subs{\proj{c}{nrec}}{p}] \\
    \typing{\Delta}{argr}[\subs{\recDatas(\proj{c}{rec})}
      {p \ext* \subs{\ind(I)}{p \pext \id} \ext argn}]
}{
    \Delta \vdash
    \ad{\subs{\constr(ic)}{p \ext argn \ext argr}}
      {\trans{\ind(I)}{\mu \pext \subs{\proj{c}{ind}}{p \ext argn}}}
        \convop \hspace{3em} \\
    \subs{\constr(ic)}{p' \ext
      \ad{(argn \ext argr)}{
      \trans{\left(\subs{\conData(c)}{\id \ext* \ind(I)}\right)}{\mu}}} \\
    \hspace{7em}
      \ty \subs{\ind(I)}{p' \ext \ad{\subs{\proj{c}{ind}}{p \ext argn}}{{\trans{\Theta_{ind}}{\mu}}}}
}

\end{mathparpagebreakable}

\end{document}